\documentclass[preprint,10pt,1p]{elsarticle}
\usepackage{jcomp}
\usepackage{framed,multirow}
\usepackage[utf8]{inputenc}
\usepackage{bm}
\usepackage{amsmath}
\usepackage{mathrsfs}
\usepackage{graphicx}
\usepackage{epsfig, setspace}
\usepackage{url}
\usepackage{graphicx, transparent, color}
\usepackage{algorithm} 
\usepackage{algorithmicx}
\usepackage{etex}
\usepackage[bookmarks=true]{hyperref}
\usepackage{amsmath}
\usepackage{xcolor, soul}
\usepackage{rotating} 
\usepackage{pdflscape} 
\usepackage{silence}
\usepackage{ulem,cancel}
\newtheorem{remark}{\bf Remark}[section]
\newtheorem{example}{Example}[section]
\usepackage{graphicx}
\usepackage{float}
\usepackage{subfigure}
\usepackage{subfigmat}

\usepackage{tikz}
\usepackage{capt-of}
\usepackage{setspace,lipsum}
\usepackage{lineno}
\newcommand{\half}{\frac{1}{2}}
\usepackage{listings}
\usepackage{xcolor}
\definecolor{aquamarine}{rgb}{0.5, 1.0, 0.83}
\definecolor{OliveGreen}{rgb}{0,0.6,0}
\sethlcolor{aquamarine}
\definecolor{codegreen}{rgb}{0,0.6,0}
\definecolor{codegray}{rgb}{0.5,0.5,0.5}
\definecolor{codepurple}{rgb}{0.58,0,0.82}
\definecolor{backcolour}{rgb}{0.95,0.95,0.92}
\lstdefinestyle{mystyle}{
    backgroundcolor=\color{backcolour},   
    commentstyle=\color{codegreen},
    keywordstyle=\color{magenta},
    numberstyle=\tiny\color{codegray},
    stringstyle=\color{codepurple},
    basicstyle=\ttfamily\footnotesize,
    breakatwhitespace=false,         
    breaklines=true,                 
    captionpos=b,                    
    keepspaces=true,                 
    numbers=left,                    
    numbersep=5pt,                  
    showspaces=false,                
    showstringspaces=false,
    showtabs=false,                  
    tabsize=2
}
\begin{document}

\begin{frontmatter}
\title{Gradient Based Reconstruction: Inviscid and viscous flux discretizations, shock capturing, and its application to single and multicomponent flows}

\author[AA_address]{Amareshwara Sainadh Chamarthi \cortext[cor1]{Corresponding author. \\ 
E-mail address: skywayman9@gmail.com (Amareshwara Sainadh  Ch.).}}

\address[AA_address]{Faculty of Mechanical Engineering, Technion - Israel Institute of Technology, Haifa, Israel}

\begin{abstract}
This paper presents a gradient-based reconstruction approach for simulations of compressible single and multi-species Navier-Stokes equations. The novel feature of the proposed algorithm is the efficient reconstruction via derivative sharing between the inviscid and viscous schemes: highly accurate explicit and implicit gradients are used for the solution reconstruction expressed in terms of derivatives. The higher-order accurate gradients of the \textcolor{black}{ velocity components} are reused to compute the viscous fluxes for efficiency and significantly improve the solution and gradient quality, as demonstrated by several viscous-flow test cases. The viscous schemes are fourth-order accurate and carefully designed with a high-frequency damping property, which has been identified as a critically important property for stable compressible-flow simulations with shock waves [Chamarthi et al., JCP, 2022]. Shocks and material discontinuities are captured using a monotonicity-preserving (MP) scheme, which is also improved by reusing the gradients. \textcolor{black}{For inviscid test cases, The proposed scheme is fourth-order for linear and second-order accurate for non-linear problems.} Several numerical results obtained for simulations of complex viscous flows are presented to demonstrate the accuracy and robustness of the proposed methodology. 
\end{abstract}


\end{frontmatter}
\section{Introduction}\label{sec-1}

Solving Navier-Stokes equations involves computing gradients required for convective and viscous fluxes and even turbulent statistics like enstrophy. The gradients are also used in shock-capturing techniques like weighted essentially non-oscillatory (WENO) schemes \cite{jiang1995}. The primary motivation of this paper is to compute the gradients with high accuracy and superior spectral properties \textit{once} and reuse them wherever necessary to reduce the computational costs of the simulations yet resolve flow features with superior resolution.

To compute viscous fluxes, one has to compute the gradients of velocities and temperature. The viscous fluxes can be discretized either in conservative or non-conservative form. It is important to evaluate the viscous fluxes in such a way that it prevents the odd-even decoupling problem. Sandham et al. \cite{sandham2002entropy} evaluated the viscous fluxes by rewriting them in Laplacian form. Such an approach was utilized to avoid odd-even decoupling, which will cause oscillations, especially in practical simulations. They directly evaluated the second-order derivatives in the Laplacian form by an explicit scheme. They indicated that evaluating them by two successive operations of the first derivatives will lead to odd-even decoupling. Their approach may be considered a non-conservative form. On the other hand, viscous discretizations in a conservative framework are susceptible to the well-known odd-even decoupling problem. To avoid it, we construct our viscous schemes with an efficient approach discussed in our previous paper \cite{chamarthi2022} by defining a numerical flux with low-order consistent and damping terms involving free parameters that determine the order of accuracy and spectral properties. In this paper, the framework is extended to compact finite-difference schemes \cite{lele1992compact} and high-order explicit finite differences with superior damping and spectral properties.

In many finite-difference-type schemes, the inviscid and viscous schemes are constructed independently, but it would be more efficient if some quantities common to both schemes were identified and shared. Such is not very simple for general difference schemes, especially those based on flux reconstruction. On the other hand, in the implicit gradient schemes previously developed in Ref.\cite{chamarthi2021implicit}, the sharing is possible and done because we employ solution reconstructions, expressed in terms of gradients and Hessians, for both inviscid and viscous discretizations. Such a construction is common in unstructured-grid discretizations, but often the solution gradients need to be computed by two different methods for nonlinear-solver stability (unweighted least-squares method for inviscid fluxes) \cite{diskin2010comparison}. However, it is not an issue in the explicit time-stepping schemes employed in this work, where there is no need to solve a global system of residual equations. It is important to note that the implicit gradient schemes are similar to that of compact finite volume schemes proposed by Sengupta et al. \cite{sengupta2005new} in the year 2005. It is similar in that it uses the kappa-reconstruction scheme, Equation (\ref{eqn:3linear}) in this paper, with implicit gradients (see their Equation 1). One of the key advantages of the implicit gradient approach is the ability to reuse the same gradients for convective and viscous fluxes, which will improve the viscous flow simulations' solution quality and overall computational efficiency. Sengupta et al. \cite{sengupta2005new} have used upwind compact finite difference schemes to compute the first derivatives, whereas, in this paper, the first derivatives are computed by central compact or explicit finite difference schemes. The derivatives thus computed with central schemes are more suitable and can be reused for viscous fluxes. Implicit gradients are also proposed for unstructured grids \cite{nishikawa2018green}. This paper presents efficient, robust\textcolor{black}{, and} low dissipation linear schemes based on the gradients and a straightforward algorithm for compressible viscous flows involving discontinuities and material interfaces.

Compressible flow simulations involving discontinuities, such as shock waves, material interfaces, and \textcolor{black}{turbulence,} are challenging. Even for smooth initial data, the flow may develop sudden changes in pressure and density mathematically treated as discontinuities. These discontinuities will result in spurious oscillations (Gibbs phenomenon), which can be mitigated by introducing additional dissipation. The development of high-resolution schemes for capturing discontinuities dates back to the pioneering work of the Godunov scheme \cite{godunov1959}, where a piece-wise constant cell reconstruction was used. The high-order accurate weighted essentially non-oscillatory (WENO) scheme, introduced by Liu et al. \cite{Liu1994} and improved by Jiang and Shu \cite{jiang1995}, is the most widely used approach for capturing discontinuities, as well as smooth, complex flow features. Wide variety of WENO schemes are proposed in literature \cite{martin2006bandwidth, Hu2010, balsara2016efficient, Fan2014a, Henrick2005, fu2017targeted, Borges2008, Deng2000, Nonomura2013, wang2017compact}. Despite the popularity of the WENO schemes, other shock-capturing approaches have been extensively studied in the literature. The Monotonicity-Preserving (MP) scheme proposed by Suresh and Hyunh \cite{suresh1997accurate} can effectively suppress numerical oscillations across discontinuities and also preserve accuracy. Li and Jaberi \cite{li2012high} studied both MP and WENO schemes and concluded that the MP schemes had less numerical dissipation and better grid-convergence properties than the WENO scheme. Recently, Zhao et al. \cite{zhao2019general} have evaluated various shock-capturing schemes and found that the fifth-order MP scheme is an excellent choice for wave propagation and is the most efficient of all the evaluated fifth-order accurate schemes. Recently, Li et al. \cite{li2021low} improved the TENO schemes by filtering the non-smooth stencils by an extra nonlinear limiter such as MP limiter \cite{suresh1997accurate} or a total variation diminishing (TVD) limiters \cite{vanleer1979,arora1997well} instead of being completely discarded. Such an approach is necessary to remove oscillations observed in the TENO scheme. This paper uses the MP limiting procedure for shock-capturing, which is also improved by reusing the gradients. The contributions of the paper are as follows:

\begin{enumerate}
\item Extend the $\alpha$-damping viscous scheme proposed in \cite{chamarthi2022} for high order explicit and compact finite-difference schemes (implicit gradients) with superior spectral \textcolor{black}{properties}.
\item Higher-order gradients of the \textcolor{black}{velocity components} are reused for the inviscid fluxes, which will improve the efficiency of the numerical scheme and obtain high-resolution solutions.
\item Improve the MP limiting approach of \cite{suresh1997accurate} by using high order gradients in estimating the curvature terms.
\end{enumerate}

The rest of the paper is organized as follows. Section \ref{sec-2} introduces the governing equations of single- and multi-species flows. The discretization of viscous fluxes with the \textcolor{black}{$\alpha$-damping} approach, gradient-based reconstruction approach for inviscid fluxes, and shock-capturing approach for flows involving discontinuities are presented in Section \ref{sec-3}. Numerical results and discussion are presented in Section \ref{sec-4}, and finally, in Section \ref{sec-5}, we provide conclusions.

\section{Governing equations}\label{sec-2}

\subsection{Single component flows}
The compressible Navier–Stokes (NS) equations governing single-component fluid flows in a two-dimensional (2-D) Cartesian coordinate system can be expressed as:
\textcolor{black}{
\begin{equation}\label{CNS-base}
\frac{\partial \mathbf{Q}}{\partial t}+\frac{\partial \mathbf{F^c}}{\partial x}+\frac{\partial \mathbf{G^c}}{\partial y}+\frac{\partial \mathbf{H^c}}{\partial z}=\frac{\partial \mathbf{F^v}}{\partial x}+\frac{\partial \mathbf{G^v}}{\partial y}+\frac{\partial \mathbf{H^v}}{\partial z},
\end{equation}}
where  $\textbf{Q}$ is the conservative variable vector, $\mathbf{F^c}$, $\mathbf{G^c}$,  $\mathbf{H^c}$ and $\mathbf{F^v}$, $\mathbf{G^v}$, $\mathbf{H^v}$, are the convective (superscript $c$) and viscous (superscript $v$) flux vectors in each coordinate direction, respectively. The compressible Euler equations are obtained as a subset by excluding the viscous fluxes. The conservative variable, convective, and viscous flux vectors are given as:
\begin{equation}
\mathbf{Q}=\left[\begin{array}{c}
\rho \\
\rho u \\
\rho v \\
\rho w \\
E
\end{array}\right],\;
\mathbf{F^c}=\left[\begin{array}{c}
\rho u \\
\rho u^{2}+p \\
\rho u v \\
\rho u w \\
(E+p) u
\end{array}\right],\;
\mathbf{G^c}=\left[\begin{array}{c}
\rho v \\
\rho v u \\
\rho v^{2}+p \\
\rho v w \\
(E+p) v
\end{array}\right],
\mathbf{H^c}=\left[\begin{array}{c}
\rho w \\
\rho u w \\
\rho v w \\
\rho w^{2}+p \\
(E+p) w
\end{array}\right],
\end{equation}


\begin{equation}
\begin{array}{l}\label{eqn-visc}
\mathbf{F^v}=\left[0, \tau_{x x}, \tau_{x y}, \tau_{x z}, u \tau_{x x}+v \tau_{x y}+w \tau_{x z}-q_{x}\right]^{T}, \\
\mathbf{G^v}=\left[0, \tau_{x y}, \tau_{y y}, \tau_{y z}, u \tau_{y x}+v \tau_{y y}+w \tau_{y z}-q_{y}\right]^{T}, \\
\mathbf{H^v}=\left[0, \tau_{x z}, \tau_{y z}, \tau_{z z}, u \tau_{z x}+v \tau_{z y}+w \tau_{z z}-q_{z}\right]^{T},
\end{array}
\end{equation}
where $\rho$ is the density and $u$, $v$ and $w$ are velocity components in the $x-$, $y-$ and $z-$ directions, respectively. The total energy per unit volume of the fluid is given as $E = \rho (e + (u^2+v^2+w^2)/2)$, where $e$ is the specific internal energy. The system of equations is closed with the ideal gas equation of state which relates the thermodynamic pressure $p$ and the total energy:
\begin{equation}\label{eqn:pressure}
p = (\gamma -1) (E - \rho \frac{(u^2+v^2+w^2)}{2}),
\end{equation}
where $\gamma$ is the ratio of specific heats of the fluid. The components of the viscous stress tensor $\tau$ and the heat flux $q$ are defined in tensor notations as:

\begin{equation}\label{eqn:5-stress}
\tau_{i j}=\frac{\mu}{\operatorname{Re}}\left(\frac{\partial u_{i}}{\partial x_{j}}+\frac{\partial u_{j}}{\partial x_{i}}-\frac{2}{3} \frac{\partial u_{k}}{\partial x_{k}} \delta_{i j}\right),
\end{equation}

\begin{equation}\label{eqn:6-heat}
\begin{aligned}
\mathrm{q}_{i}=-\frac{\mu}{\operatorname{Re Pr Ma}(\gamma-1)} \frac{\partial T}{\partial x_{i}}, \quad T= \text{Ma}^{2} \gamma \frac{p}{\rho},
\end{aligned}
\end{equation}
where $\mu$ is the dynamic viscosity, $T$ is the temperature, $Ma$ and $Re$ are the Mach number and Reynolds number, and Pr is the Prandtl number. 

\subsection{Multicomponent flows}

In this study, the compressible multi-component flows as described by the quasi-conservative five equation model of Allaire et al. \cite{allaire2002five} including viscous effects \cite{coralic2014finite} are considered. For a system consisting of two fluids, there are two continuity equations, one momentum and one energy equation. In addition, an advection equation for the volume fraction of one of the two fluids is considered as:
\begin{equation}\label{5eqn-base}
\frac{\partial \mathbf{Q}}{\partial t}+\frac{\partial \mathbf{F^c}}{\partial x}+\frac{\partial \mathbf{G^c}}{\partial y}+\frac{\partial \mathbf{F^v}}{\partial x}+\frac{\partial \mathbf{G^v}}{\partial y}=\mathbf{S},
\end{equation}
where the state vector, flux vectors and source term, $\mathbf{S}$, are given by:
\begin{equation}
\mathbf{Q}=\left[\begin{array}{c}
\alpha_{1} \rho_{1} \\
\alpha_{2} \rho_{2} \\
\rho u \\
\rho v \\
E \\
\alpha_{1}
\end{array}\right], \quad \mathbf{F^c}=\left[\begin{array}{c}
\alpha_{1} \rho_{1} u \\
\alpha_{2} \rho_{2} u \\
\rho u^{2}+p \\
\rho v u \\
(E+p) u \\
\alpha_{1} u
\end{array}\right], \quad \mathbf{G^c}=\left[\begin{array}{c}
\alpha_{1} \rho_{1} v \\
\alpha_{2} \rho_{2} v \\
\rho u v \\
\rho v^{2}+p \\
(E+p) v \\
\alpha_{1} v
\end{array}\right],  \quad \mathbf{S}=\left[\begin{array}{c}
0 \\
0 \\
0 \\
0 \\
0 \\
\alpha_{1} \nabla \cdot \mathbf{u}
\end{array}\right],
\end{equation}
\textcolor{black}{
\begin{equation}
\mathbf{F^v}=\left[\begin{array}{c}
0 \\
0 \\
-\tau_{x x} \\
-\tau_{y x} \\
-\tau_{x x} u-\tau_{x y} v\\
0 \\
\end{array}\right],
\mathbf{G^v}=\left[\begin{array}{c}
0 \\
0 \\
-\tau_{x y} \\
-\tau_{y y} \\
-\tau_{y x} u-\tau_{y y} v\\
0 \\
\end{array}\right],
\end{equation}}
where $\rho_1$ and $\rho_2$ correspond to the densities of fluids $1$ and $2$, $\alpha_{1}$ and $\alpha_{2}$ are the volume fractions of the fluids $1$ and $2$, $\rho$, $u$,$v$, $p$ and $E$ are the density, $x-$ and $y-$ velocity components, pressure, total energy per unit volume of the mixture, respectively. The five-equation model is incomplete in the vicinity of the material interface where the fluids are in a mixed state when using a diffuse interface approach for mathematical modeling. In order to complete the closure, a set of mixture rules for various properties of the fluids must be defined. The mixture rules for the volume fractions of the two fluids $\alpha_1$ and $\alpha_2$, as well as the density and mixture rules for the ratio of specific heats $\gamma$ of the mixture, are given by:

\begin{equation}
\alpha_{2}=1-\alpha_{1}
\end{equation}

\begin{equation}
\rho=\rho_{1} \alpha_{1}+\rho_{2} \alpha_{2}
\end{equation}

\begin{equation}
\frac{1}{\gamma-1}=\frac{\alpha_{1}}{\gamma_{1}-1}+\frac{\alpha_{2}}{{\gamma}_{2}-1}
\end{equation}
where $\gamma_1$ and $\gamma_2$ are the specific heat ratios of fluids $1$ and $2$, respectively.  Finally, under the isobaric assumption, the equation of state, as given by Equation (\ref{eqn:pressure}), is used to close the system. The viscous terms are defined as in single-component flows. The advection equation for the volume fraction is written in non-conservative form, given by Equation (\ref{eqn-source}), following Johnsen and Colonius \cite{johnsen2006implementation}. This choice ensures the advection equation is consistently coupled with the governing equations, leading to an oscillation-free solution at the material interface.

\begin{equation}\label{eqn-source}
\frac{\partial \alpha_{1}}{\partial t}+\nabla \cdot\left(\alpha_{1} \mathbf{u}\right)=\alpha_{1} \nabla \cdot \mathbf{u}
\end{equation}

\section{Numerical method}\label{sec-3}

We apply a conservative numerical method to solve the above presented equations. The time evolution of the vector of cell-centered conservative variables $\mathbf{\hat Q}$ is given by the following semi-discrete equation applied to a Cartesian grid cell $I_{i,j,k} = [x_{i-\frac{1}{2}}, x_{i+\frac{1}{2}}] \times [y_{j-\frac{1}{2}}, y_{j+\frac{1}{2}}] \times [z_{k-\frac{1}{2}}, z_{k+\frac{1}{2}}]$, expressed as an ordinary differential equation:

\textcolor{black}{
 \begin{equation}\label{eqn-differencing}
\frac{\mathrm{d}}{\mathrm{dt}} {\mathbf{\hat Q}}_{i,j,k} = \mathbf{Res}_{i,j,k} = - \frac{d \mathbf F}{dx}_{i,j,k} - \frac{d \mathbf G}{dy}_{i,j,k}- \frac{d \mathbf H}{dz}_{i,j,k} + \mathbf{S}_{i, j,k},
 \end{equation}}
 
\textcolor{black}{ where $ \mathbf{Res}_{i,j,k}$ denotes the residual. $\frac{d \mathbf F}{dx}_{i,j,k}$, $\frac{d \mathbf G}{dy}_{i,j,k}$ and $\frac{d \mathbf H}{dz}_{i,j,k}$ are approximations to the flux derivatives at the cell center, and we seek their high-order approximations in the conservative form:}
 \textcolor{black}{
 \begin{equation}\label{eqn-differencing_residual}
\begin{aligned}
\frac{d \mathbf F}{dx}_{i,j,k}  = &\frac{1}{\Delta x}\left[\left(\mathbf {\hat{F}^c}_{i+ \frac{1}{2}, j, k}-\mathbf {\hat{F}^c}_{i- \frac{1}{2}, j, k}\right) - \left(\mathbf {\hat{F}^v}_{i+ \frac{1}{2}, j, k}-\mathbf {\hat{F}^v}_{i-\frac{1}{2}, j, k}\right)\right] \\
\frac{d \mathbf G}{dy}_{i,j,k}=&\frac{1}{\Delta y}\left[\left(\mathbf {\hat{G}^c}_{i, j+ \frac{1}{2}, k}-\mathbf {\hat{G}^c}_{i, j- \frac{1}{2}, k}\right)-\left(\mathbf {\hat{G}^v}_{i, j+ \frac{1}{2}, k}-\mathbf {\hat{G}^v}_{i, j- \frac{1}{2}, k}\right)\right]\\
\frac{d \mathbf H}{dz}_{i,j,k}=&\frac{1}{\Delta z}\left[\left(\mathbf {\hat{H}^c}_{i, j, k+ \frac{1}{2}}-\mathbf {\hat{H}^c}_{i, j, k- \frac{1}{2}}\right)-\left(\mathbf {\hat{H}^v}_{i, j, k+ \frac{1}{2}}-\mathbf {\hat{H}^v}_{i, j, k- \frac{1}{2}}\right)\right],
\end{aligned}
\end{equation}}
where $\Delta x=x_{i+1 / 2}-x_{i-1 / 2}$, $\Delta y=y_{j+1 / 2}-y_{j-1 / 2}$, $\Delta z=z_{k+1 / 2}-z_{k-1 / 2}$, $\mathbf {\hat{F}^c}$, $\mathbf {\hat{G}^c}$, $\mathbf {\hat{H}^c}$  and $\mathbf {\hat{F}^v}$, $\mathbf {\hat{G}^v}$ and $\mathbf {\hat{H}^v}$ are interpreted as numerical approximation of the convective and viscous fluxes in the $x-$, $y-$ and $z-$directions, respectively, and $\mathbf{S}_{i, j, k}$ is the source term.  The conserved variables are integrated in time using the third-order TVD Runge–Kutta scheme \cite{jiang1995}:
\begin{eqnarray}\label{rk}
\mathbf{\hat Q}_{i, j, k}^{(1)}&=&\mathbf{\hat Q}_{i, j, k}^{n}+\Delta t \mathbf{Res}\left(\mathbf{\hat Q}_{i, j, k}^{n}\right) \\
\mathbf{\hat Q}_{i, j, k}^{(2)}&=&\frac{3}{4} \mathbf{\hat Q}_{i, j, k}^{n}+\frac{1}{4} \mathbf{\hat Q}_{i, j, k}^{(1)}+\frac{1}{4} \Delta t \mathbf{Res}\left(\mathbf{\hat Q}_{i, j, k}^{(1)}\right) \\
\mathbf{\hat Q}_{i, j, k}^{n+1}&=&\frac{1}{3} \mathbf{\hat Q}_{i, j, k}^{n}+\frac{2}{3} \mathbf{\hat Q}_{i, j, k}^{(2)}+\frac{2}{3} \Delta t \mathbf{Res}\left(\mathbf{\hat Q}_{i, j, k}^{(2)}\right)
\end{eqnarray}
{\textcolor{black}{where $\mathbf{\hat Q}$ denote vectors of numerical solutions, which defined as \textit{point values} at cell centers (not cell averages)} and $\mathbf{Res}$ denote residuals given by the right-hand side of Equation (\ref{eqn-differencing}), respectively.} The superscripts ${n}$ and ${n+1}$ denote the current and the next time-steps, and superscripts ${(1)-(2)}$ corresponds to intermediate steps. The time step $\Delta t$ is computed as:

\begin{equation}\label{eqn:cfl}
\Delta t= \text{CFL} \cdot \min \left(\Delta t_{viscous}, \Delta t_{inviscid}\right),
\end{equation}
where
\begin{equation}
\Delta t_{inviscid}=  \min _{i, j, k}\left(\frac{\Delta x_{i}}{\left|u_{i, j, k}\right|+c_{i, j, k}}, \frac{\Delta y_{j}}{\left|v_{i, j, k}\right|+c_{i, j, k}}, \frac{\Delta z_{k}}{\left|w_{i, j , k}\right|+c_{i, j, k}}\right),
\end{equation}
where $c$ is the speed of sound and given by $c=\sqrt{\gamma{p}/\rho}$.  \textcolor{black}{ The calculation of $\Delta t_{viscous}$ is discussed later.} The remaining step is to describe the details of the numerical approximation of fluxes on the RHS of Equation (\ref{eqn-differencing_residual}). In subsection (\ref{sec-3.2}), the computation of viscous fluxes is explained.  In subsection (\ref{sec-3.1}), the computation of convective fluxes is explained, which \textcolor{black}{explains} the idea of \textcolor{black}{gradient-based} reconstruction. In the later subsections, (\ref{sec-3.1.1}), the monotonicity-preserving implicit gradient shock-capturing algorithm is presented. 

\subsection{Spatial discretization of viscous fluxes}\label{sec-3.2}

In this section, the discretization of the viscous fluxes is presented. For simplicity, a one-dimensional scenario is considered. The grid is discretized on a uniform grid with $N$ cells on a spatial domain spanning $x \in \left[x_l, x_r \right]$. The cell center locations are at $x_i = x_l + (i - 1/2) \Delta x$, $\forall j \in \{1, \: 2, \: \dots, \: N\}$, where $\Delta x = (x_r - x_l)/N$. In one-dimension, the viscous flux at the interface is as follows:
\begin{equation}
\mathbf{\hat F^v}_{i+\frac{1}{2}}=\left[\begin{array}{c}
0 \\
-\tau_{i+\frac{1}{2}} \\
-\tau_{i+\frac{1}{2}} u_{i+\frac{1}{2}}+q_{i+\frac{1}{2}}
\end{array}\right]
\end{equation}
where
\begin{equation}\label{eqn:visc-interface}
\tau_{i+1 / 2}=\frac{4}{3} \mu_{i+1 / 2}\left(\frac{\partial u}{\partial x}\right)_{i+1 / 2}, \quad q_{i+1 / 2}=-\frac{\mu_{i+1 / 2}}{\operatorname{Pr}(\gamma-1)}\left(\frac{\partial T}{\partial x}\right)_{i+1 / 2}
\end{equation}
As it can be seen from Equation (\ref{eqn:visc-interface}) the viscous fluxes at cell interfaces $x_{i+\frac{1}{2}}$, $\forall i \in \{ 0, \: 1, \: 2, \: \dots, \: N \}$, has to be evaluated. The gradients of the velocity and temperature at the interface should be computed. For this purpose, we consider the $\alpha$-damping approach of Nishikawa \cite{Nishikawa2011a}, which is known to prevent odd-even decoupling, is considered and is briefly presented here. The velocity gradient at the cell-interface can be computed by the following equation,

\begin{equation}\label{eqn:alpha-damping}
\begin{aligned}
\left(\frac{\partial u}{\partial x}\right)_{i+\frac{1}{2}}=\underbrace{\frac{1}{2}\left[\left(\frac{\partial u}{\partial x}\right)_{i}+\left(\frac{\partial u}{\partial x}\right)_{i+1}\right]}_{\text{Consistent term}}+\underbrace{\frac{\alpha}{2 \Delta x}\left({u}_{i+ \frac{1}{2}}^{R}-{u}_{i+ \frac{1}{2}}^{L}\right)}_{\text{Damping term}},
\end{aligned}
\end{equation}
where,
\textcolor{black}{\begin{equation}\label{eqn:vel-visc}
\begin{aligned}
{u}_{i+ \frac{1}{2}}^{L} &={\hat {u}}_{i}+\frac{\Delta x}{2} \left(\frac{\partial u}{\partial x}\right)_{i},{u}_{i+ \frac{1}{2}}^{R} &={\hat {u}}_{i+1}-\frac{\Delta x}{2} \left(\frac{\partial u}{\partial x}\right)_{i+1}  \quad 
\end{aligned}
\end{equation}}
The first term in the Equation (\ref{eqn:alpha-damping}) is called the consistent term, which approximates the physical flux consistently, and the second term is called the damping term, which helps in high-frequency damping. By substituting the Equations (\ref{eqn:vel-visc}) in the Equation (\ref{eqn:alpha-damping}) we get the following equation:

\textcolor{black}{\begin{equation}
\begin{aligned}
\left(\frac{\partial u}{\partial x}\right)_{i+1 / 2}&=\frac{1}{2}\left( u_{i}^{\prime}+ u_{i+1}^{\prime}\right)+\frac{\alpha}{2 \Delta x}\left(\hat u_{i+1}-\frac{\Delta x}{2}  u_{i+1}^{\prime}-\hat u_{i}-\frac{\Delta x}{2}  u_{i}^{\prime}\right),
\end{aligned}
\end{equation}}
\textcolor{black}{where $u_{i}^{\prime}$ represents $\left(\frac{\partial u}{\partial x}\right)_{i}$. The interface velocity can be computed by the arithmetic average:}
\begin{equation}\label{eqn:vel-inter}
\begin{aligned}
u_{i+\frac{1}{2}} = \frac{1}{2}\left({u}_{i+ \frac{1}{2}}^{L} + {u}_{i+ \frac{1}{2}}^{R} \right)
\end{aligned}
\end{equation}

The gradient of velocity at the cell-centres, $\left(\frac{\partial u}{\partial x}\right)_{i}$, in the consistent terms, should be computed accurately, which is the objective of this work. Nishikawa has shown in Ref. \cite{Nishikawa2011a} that the \textcolor{black}{$\alpha$-damping} approach with $\alpha$ = $\frac{8}{3}$ will lead to fourth-order accuracy for the diffusion fluxes. Nishikawa has considered explicit second-order central differences in evaluating the gradients at the cell centres. Chamarthi et al.\cite{chamarthi2022} \textcolor{black}{used} fourth-order central differences and obtained a sixth-order scheme. In Ref. \cite{chamarthi2022} an additional parameter $\beta$ along with the curvature term in the damping term is used and is not considered here. This paper extends the approach to fourth-order implicit gradients (compact finite differences) and sixth-order explicit gradients computed at the cell centres. The formula for the sixth-order explicit derivative is as follows,
\begin{eqnarray}\label{eqn:six_visc}
u_{i}^{\prime} 
 =
 \frac{  1 }{60} \left[ 
         45  \frac{  u_{i+1} - u_{i-1}   }{\Delta x}
        -  9  \frac{  u_{i+2} - u_{i-2}   }{\Delta x}
     +   {\color{black}   \frac{  u_{i+3} - u_{i-3}   }{\Delta x} }
 \right].
 \end{eqnarray}
 The optimized fourth-order implicit gradients for the computation of the derivatives is as follows,

\textcolor{black}{
\begin{equation}\label{eqn:four_visc}
     \beta  u_{i-\textcolor{black}1}^{\prime}+u_{i}^{\prime}+\beta u_{i-1}^{\prime}=\frac{b}{4 \Delta x}\left(u_{i+2}-u_{i-2}\right)+\frac{a}{2 \Delta x}\left(u_{i+1}-u_{i-1}\right),
   \end{equation}}
where, \textcolor{black}{$ \beta$ = $\frac{5}{14}$,} $a=\frac{11}{7}$, $b=\frac{1}{7}$ and $i= 1, 2, 3, ....., N-1, N$. Implicit gradient approach require solving a system of equations which can be done easily by a tridiagonal solver. 

\begin{figure}[H]
\centering
 \includegraphics[width=0.5\textwidth]{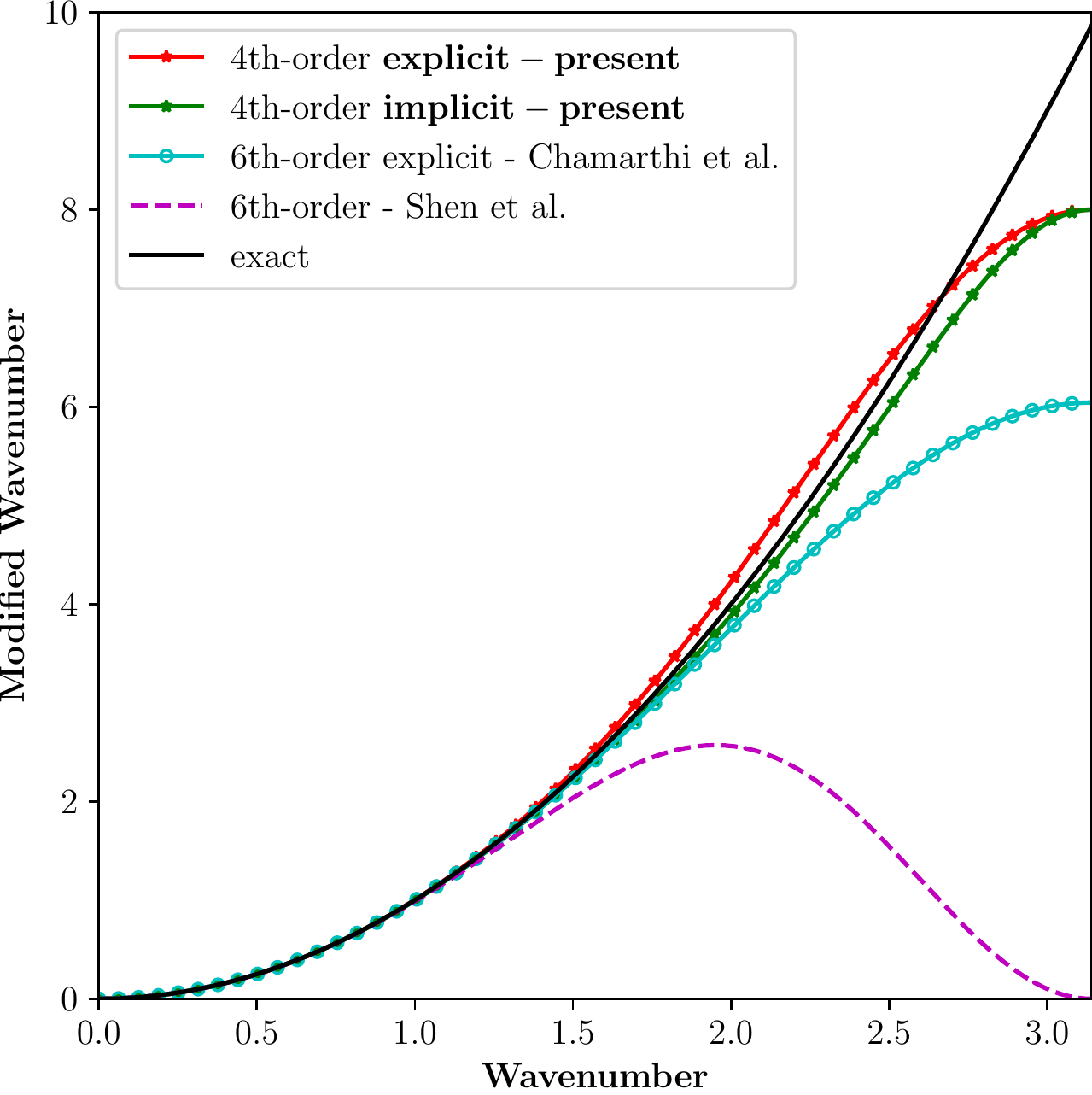}
\caption{Differencing error for second derivative vs wavenumber.}
\label{fig:second_deriv}
\end{figure}

 It was found through Fourier analysis that the present schemes, both explicit and implicit, will be fourth-order accurate by setting $\alpha = 4$, and it will be second-order accurate for any other value. The modified wave numbers for these approximate second derivatives are plotted in Fig. \ref{fig:second_deriv}. The fourth-order \textcolor{black}{$\alpha$-damping} scheme using implicit and explicit gradients can resolve the high-frequency modes and conform to the exact modified wave number much more closely despite being only fourth-order accurate. The spectral and damping properties are superior to that of the sixth-order scheme presented in \cite{chamarthi2022}. The finite-difference approximation for viscous fluxes proposed by Shen et al. \cite{shen2010large}, which is sixth-order accurate and does not resolve any of the high-frequency modes, is also shown in Fig. \ref{fig:second_deriv}. The Fourier analysis of  Shen et al. \cite{shen2010large} is presented in \cite{chamarthi2022}.

 For completeness, the Fourier error analysis of the proposed viscous flux discretization is given below \cite{Nishikawa2010,lele1992compact,moin2010fundamentals}. The Fourier analysis helps analyze the spectral properties of a scheme. Consider the Fourier form of a function 
\begin{equation}
  f(x) = e^{i k x} 
\end{equation}
where $i = \sqrt{-1}$ is the imaginary unit, $k = \frac{2 \pi}{L} n$ is the wave number, and $n = 1, 2, ..., N/2$. This function can then be differentiated twice to obtain a Fourier form of the second derivative. 
\begin{equation}
  f''(x) = (i k)^2 e^{i k x} = - k^2 e^{i k x} 
\end{equation}
This can be extended to a discrete form by plugging in values of for $x_j$ such that
\begin{equation}
  x_j = \frac{L}{N} j
\end{equation}
for $j = 0, 1, 2, ..., N-1$ grid points.  Using the discrete values of the function, our approximation to the second derivative can be represented in Fourier form as
\begin{equation}
  \frac{\delta^2 f}{\delta x^2}_{j}\vline = k' f_j
\end{equation}
where $k'$ is a modified wavenumber for the approximation. The exact form of the modified wave number is $k' = -k^2$ and is plotted as a reference in Fig. \ref{fig:second_deriv}. For the fourth-order implicit  gradient-based scheme, denoted as $\alpha$-IG in this paper, we find the modified wavenumber with the \textcolor{black}{$\alpha$-damping} approach to be 

\begin{equation}\label{mod-IG}
k_{\alpha-IG}' =\frac{\sin ^2\left(\frac{k}{2}\right) (-16 \cos (k)+\cos (2 k)-33)}{5 \cos (k)+7}
\end{equation}
\textcolor{black}{By expanding the above equation, we obtain the following equation,}
\begin{equation}
\begin{split}
k_{\alpha-IG}' = &-k^2+\frac{k^6}{240}-\frac{17 k^8}{24192}+\frac{17 k^{10}}{691200}+O\left(k^{11}\right)
\end{split}
\end{equation}

From the above equation, it can be seen that the $\alpha$-IG scheme is fourth-order accurate. Similarly, the Fourier analysis \textcolor{black}{carried} out for the sixth-order explicit gradients, denoted as $\alpha$-EG in this paper, is as follows:

\begin{equation}
  \begin{split}
k_{\alpha-EG}' = \frac{1}{15} \sin ^2\left(\frac{k}{2}\right) (29 \cos (k)-7 \cos (2 k)+\cos (3 k)-83)
\end{split}
\end{equation}
By expanding the above equation, we obtain the following equation,
\begin{equation}
  \begin{split}
k_{\alpha-EG}' = -k^2+\frac{k^6}{360}-\frac{73 k^8}{10080}+\frac{223 k^{10}}{86400}+O\left(k^{11}\right),
\end{split}
\end{equation}
which shows that the $\alpha$-EG scheme scheme is also fourth order accurate.

\begin{remark}\normalfont	By substituting $\alpha$ = 4 in the cell interface gradients, the second derivative can be explicitly written as follows:

\textcolor{black}{\begin{equation}
\begin{aligned}
\left(\frac{\partial u}{\partial x}\right)_{i+1 / 2}&=\frac{1}{2}\left( u_{i}^{\prime}+ u_{i+1}^{\prime}\right)+\frac{4}{2 \Delta x}\left(\hat u_{i+1}-\frac{\Delta x}{2}  u_{i+1}^{\prime}-\hat u_{i}-\frac{\Delta x}{2}  u_{i}^{\prime}\right),\\
\\
\left(\frac{\partial u}{\partial x}\right)_{i-1 / 2}&=\frac{1}{2}\left( u_{i-1}^{\prime}+ u_{i}^{\prime}\right)+\frac{4}{2 \Delta x}\left(\hat u_{i}-\frac{\Delta x}{2}  u_{i}^{\prime}-\hat u_{i-1}-\frac{\Delta x}{2}  u_{i-1}^{\prime}\right),
\end{aligned}
\end{equation}}

\textcolor{black}{\begin{equation}
\begin{aligned}
\left(\frac{\partial^2 u}{\partial x^2}\right)_{i} =  \frac{\left(\frac{\partial u}{\partial x}\right)_{i+1 / 2} - \left(\frac{\partial u}{\partial x}\right)_{i-1 / 2}} {\Delta x}
\end{aligned}
\end{equation}}

\textcolor{black}{\begin{equation}\label{eqn:adam}
\left(\frac{\partial^2 u}{\partial x^2}\right)_{i}=\left(\frac{2}{ \Delta x^{2}}\right)\left(\hat u_{i+1}-2 \hat u_{i}+\hat u_{i-1}\right)-(\frac{1}{ 2 \Delta x})\left(u_{i+1}^{\prime}-u_{i-1}^{\prime}\right)
\end{equation}}

During the literature survey, it was found that the paper published by Adam \cite{adam1977highly} in 1977, where the second derivatives are computed as a function of the solution values and the first derivatives, see his Equation 14, is the same as that of the Equation (\ref{eqn:adam}) in this paper. Adam used standard fourth-order compact finite differences to compute the first derivative, whereas this paper uses an optimized fourth-order compact scheme or explicit sixth-order scheme. It can be seen that the \textcolor{black}{$\alpha$-damping} scheme was reduced to the one obtained by Adam even though we have used a different compact finite difference scheme for the first derivative. 
\end{remark}

\begin{remark}
\normalfont The readers need to understand that viscous fluxes are implemented via the $\alpha$-damping approach and not directly using the Equation	 (\ref{eqn:adam}). It was done a) to show the similarity between the present scheme and that of \cite{adam1977highly} and b) the second derivatives thus obtained can be used in the reconstruction of inviscid fluxes, which will be shown in the next section.
\end{remark}

\textcolor{black}{Finally, the maximum time step allowable for explicit time-stepping for the $\alpha$-damping approach is as follows:}
\textcolor{black}{\begin{equation}
 \Delta t_{viscous}=  \min _{i, j, k}\left(\frac{1}{\alpha} \frac{\Delta x_{i}^{2}}{\nu_{i, j, k}}, \frac{1}{\alpha}  \frac{\Delta y_{j}^{2}}{\nu_{i, j, k}}, \frac{1}{\alpha}  \frac{\Delta z_{k}^{2}}{\nu_{i, j, k}}\right),
\end{equation}}
where, $\alpha$ = 4 and $\nu$ is the kinematic viscosity defined as $\nu = \mu / \rho$. This time step is restriction is derived by Nishikawa, see Equation 4.22 in \cite{Nishikawa2010}, and is applicable here as well. The value of $\alpha$ is slightly more restrictive than the value 38/15 obtained in \cite{chamarthi2022}; nevertheless, it is important for stable computations.

\subsection{Spatial discretization of convective fluxes}\label{sec-3.1}

 In this section, the spatial discretization of the convective fluxes in Equation (\ref{eqn-differencing_residual}), $\mathbf {\hat{F^c}}_{i- 1 / 2}$ and $\mathbf {\hat{F^c}}_{i+ 1 / 2}$, is presented.  These are computed by a Riemann solver  \cite{deledicque2007exact, ivings1998riemann,roe1981approximate, batten1997choice, einfeldt1988godunov, toro1994restoration, osher1982upwind} and can be written in canonical form as follows:
\begin{equation}\label{Numerical Flux}
\mathbf {\hat{F}^c}_{i+ 1 / 2}={F}^{\rm Riemann}_{i+\frac{1}{2}}(\mathbf{Q}_{i+\frac{1}{2}}^{L},\mathbf{Q}_{i+\frac{1}{2}}^{R}), 
\end{equation}

\begin{equation}
\mathbf{F}^{\rm Riemann}_{i+\frac{1}{2}}
= \frac{1}{2}
\left[
{\mathbf{F}}(\mathbf{Q}_{i+\frac{1}{2}}^{L}) 
+ 
{\mathbf{F}}(\mathbf{Q}_{i+\frac{1}{2}}^{R})
\right]
-
 \frac{1}{2} | {\mathbf{A}_{i+\frac{1}{2}}}|({\mathbf{Q}^R_{i+\frac{1}{2}}}-{\mathbf{Q}^L_{i+\frac{1}{2}}}),
\label{eqn:Riemann}
\end{equation}
where ``L'' and ``R'' \textcolor{black}{denotes} the reconstructed states from the left and right side of a cell interface, respectively, and $|{\mathbf{A}_{i+\frac{1}{2}}}|$ denotes the inviscid Jacobian of the Euler equations.  In this paper, the HLLC Riemann solver is used unless otherwise \textcolor{black}{stated,} and implementation details are provided in the Appendix \ref{sec-3.3}. The procedure to obtain the values at the left and right interfaces, $L$ and $R$, is presented below. Details are presented for a one-dimensional case for simplicity since it is easily extended to multi-dimensional problems via a dimension-by-dimension approach.   We consider the following general interpolation \cite{van1977towards} valid for  $x_{i-1 / 2} \leq x \leq x_{i+1 / 2}$: 

\begin{equation}\label{eqn:legendre}
\mathbf{U}(x)=\mathbf{\hat{U}}_{i}+\mathbf{U}'_i\left(x-x_{i}\right)+\textcolor{black}{\frac{3}{2}} \kappa \mathbf{U}''_i \left[\left(x-x_{i}\right)^{2}-\frac{\Delta x_i^{2}}{12}\right]
\end{equation}
where $\mathbf{\hat{U}}_{i}$ is the primitive variable vector, for single component flows $\mathbf{U}$ = $(\rho, u, v, w, p)^T$, at the cell-centers and ${ \mathbf{U}'_i}$, ${\mathbf{U}''_i}$ are the corresponding first- and second-derivatives within cell $i$.  For the numerical approximations of the Riemann problem, we need the values at the cell interfaces only. By setting $x = x_i \pm \Delta x/2$ and within a cell $i$ and $\kappa=\frac{1}{3}$ the following expression results:
\begin{equation}\label{eqn:3linear}
\begin{aligned}
\mathbf{U}_{i+ \frac{1}{2}}^{L} &=\mathbf{\hat {U}}_{i}+\frac{\Delta x}{2} \mathbf{U}'_i+\frac{\Delta x^2}{12} \mathbf{U}''_i \\
\mathbf{U}_{i+ \frac{1}{2}}^{R} &=\mathbf{\hat {U}}_{i+1}-\frac{\Delta x}{2}  \mathbf{U}'_{i+1}+\frac{\Delta x^2}{12}  \mathbf{U}''_{i+1} \quad 
\end{aligned}
\end{equation}
By choosing primitive variables for reconstruction for the inviscid \textcolor{black}{fluxes,} the gradients can be reused for the viscous fluxes (or vice versa), which is the main motivation behind this work.  For the multi-component flows the primitive variables in two-dimensions are, $\mathbf{U}$ = $(\rho_{1} \alpha_{1},\rho_{2} \alpha_{2}, u, v, p,\alpha_1)^T$. The values of $\mathbf{Q}_{i+\frac{1}{2}}^{L}$  are therefore obtained from $\mathbf{U}_{i+\frac{1}{2}}^{L}$.

Therefore, the velocity gradients in Equation (\ref{eqn:3linear}) are readily available as they are already computed earlier for the viscous fluxes by using Equation (\ref{eqn:six_visc}), sixth-order explicit derivatives, or by using Equation (\ref{eqn:four_visc}), fourth-order implicit gradients. The second derivative (Hessians), ${\mathbf{U}''_i}$ in Equation (\ref{eqn:3linear}) can also be easily computed by using Equation (\ref{eqn:adam}),
which means the second derivatives are computed from the first-derivatives and the cell center values. Apart from reusing the velocity gradients, it is also possible to obtain the temperature gradients necessary for the viscous fluxes from the pressure and density gradients \cite{wang2017compact}, and such an approach is not considered here. Therefore, for single and multi-component flows, only velocity gradients are shared between the inviscid and viscous fluxes. As the interface values are obtained by using the gradients of the variables, the approach is called {\textcolor{black}{Gradient-based} reconstruction  (GRB)}.  The final expression for the left interface can  be written as

\begin{equation}\label{eqn:IG2}
 {\mathbf{U}}_{i+\frac{1}{2}}^{L,GRB} ={\hat {\mathbf{U}}}_{i}+\frac{\Delta x}{2} {\mathbf{U}}'_i+\frac{\Delta x^2}{12} \left[\left(\frac{2}{ \Delta x^{2}}\right)\left({\hat {\mathbf{U}}}_{i+1}-2 {\hat {\mathbf{U}}}_i+{\hat {\mathbf{U}}}_{i-1}\right)-(\frac{1}{ 2 \Delta x})\left({ {\mathbf{U}}}_{i+1}^{\prime}-{ {\mathbf{U}}}_{i-1}^{\prime}\right)\right],
\end{equation}
The explicit gradient-based reconstruction approach, which uses explicit sixth-order finite differences, is denoted as \textbf{EG6} and implicit gradient-based reconstruction, using optimized fourth-order compact finite differences, is denoted as \textbf{IG4H} (H for Hermitian) in this paper. It can be noticed that the only difference between the EG6 and IG4H schemes is in the computation of the first-order derivatives and the rest of the approach, including viscous flux computation, the parameter $\alpha$ in damping term, the formula for the computation of second-derivatives by Equation (\ref{eqn:adam}), is the same.

A Fourier analysis is performed, similar to that of viscous fluxes presented in the earlier section, but for the first derivative to understand the spectral properties of the proposed schemes. First, we consider the EG6 gradient-based reconstruction approach with sixth-order explicit scheme, Equation (\ref{eqn:six_visc}), substituted into the (\ref{eqn:3linear}). Substituting the Fourier mode into the residual computed with the explicit gradients, we obtain

\begin{align}\label{fourier:EG6_res}
\begin{split}
k_{EG6}'=\frac{1}{360} i e^{-\frac{1}{2} (i k)} \sin \left(\frac{k}{2}\right) \biggl[\biggl(540 i \sin (k)-108 i \sin (2 k)+12 i \sin (3 k)\biggr)+\\
	 \biggl(231 \cos (k)-44 \cos (2 k)+9 \cos (3 k)-\cos (4 k)+525\biggr)\biggr]
\end{split}
\end{align}

which can be expanded as

\begin{equation}\label{fourier:EG6_acc}
k_{EG6}'= i k+\frac{i k^5}{720}+\frac{k^6}{1440}+\frac{i k^7}{30240}+\frac{7 k^8}{1920}\cdots
\end{equation}

Similarly, we consider the IG4H gradient-based reconstruction approach with fourth-order implicit scheme, Equation (\ref{eqn:four_visc}), substituted into the Equation (\ref{eqn:3linear}). Substituting the Fourier mode into the residual computed with the implicit gradients, we obtain

\begin{equation}\label{fourier:IG4H_res}
k_{IG4H}'=\frac{e^{-4 i k}+1070 i \sin (k)+34 i \sin (2 k)+6 i \sin (3 k)-56 \cos (k)+28 \cos (2 k)-8 \cos (3 k)+35}{96 (5 \cos (k)+7)}
\end{equation}

which can be expanded as

\begin{equation}\label{fourier:IG4H_acc}
k_{IG4H}'=i k+\frac{i k^5}{720}+\frac{i k^7}{2016}+\frac{k^8}{2304}+\cdots
\end{equation}

\begin{figure}[H]
\centering
\subfigure[]{\includegraphics[width=0.45\textwidth]{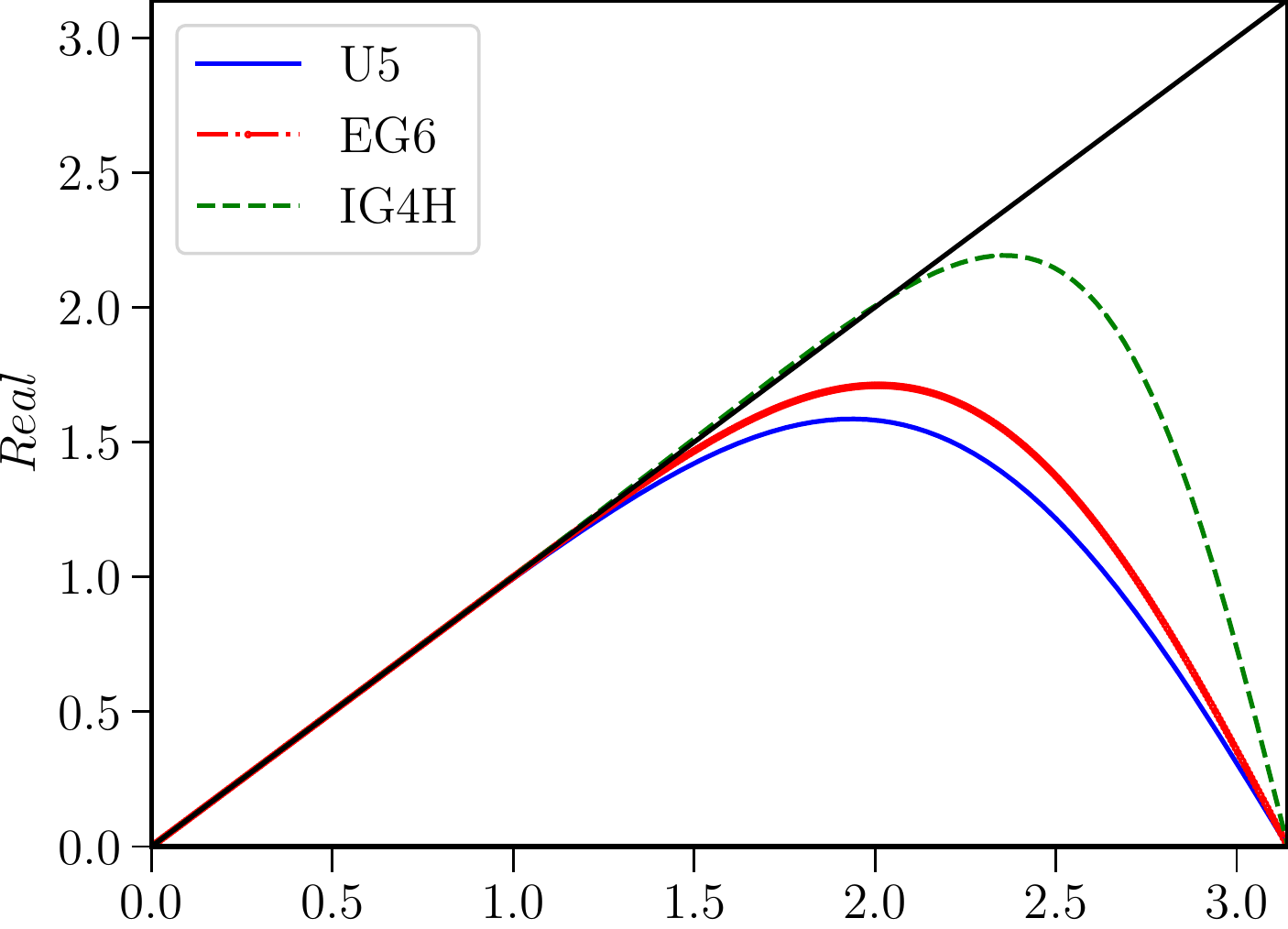}
\label{fig:dispersion_u}}
\subfigure[]{\includegraphics[width=0.47\textwidth]{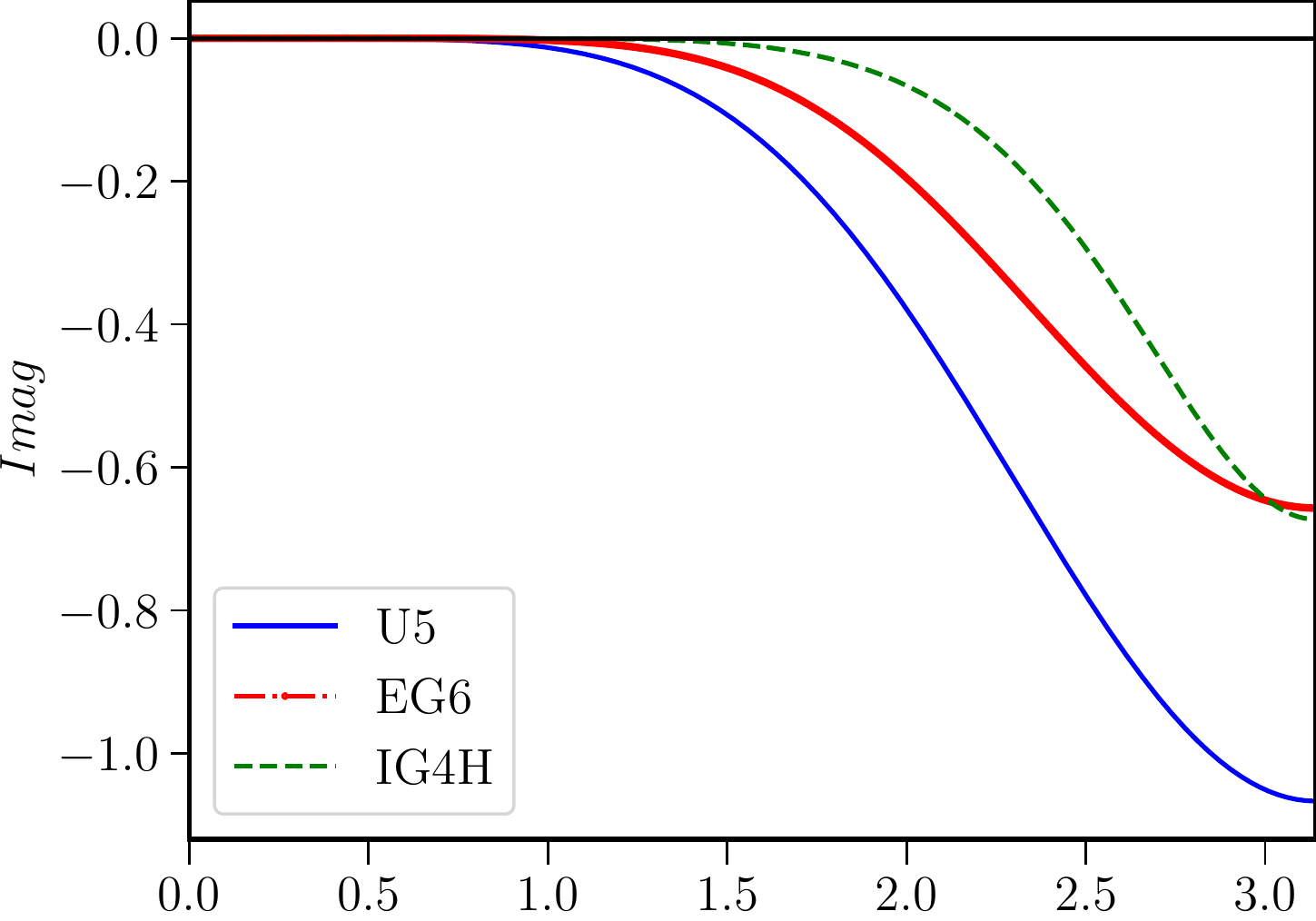}
\label{fig:dissipation_u}}
\caption{Dispersion and Dissipation properties of the unlimited schemes.}
\label{fig_disp_unlim}
\end{figure}

From Equations (\ref{fourier:EG6_acc}) and (\ref{fourier:IG4H_acc}) one can see that both the \textcolor{black}{gradient-based} reconstructions, EG6 and IG4H, are \textcolor{black}{fourth-order} accurate. Fourier analysis is also carried out for the fifth order linear upwind reconstruction scheme, denoted as U5, and is as follows:

\begin{equation}\label{eqn:linear5}
{\mathbf{U}}^{L,U5}_{i+\frac{1}{2}}=(2 { \mathbf{\hat U}}_{i-2} - 13{\mathbf{\hat U}}_{i-1} + 47{\mathbf{\hat U}}_i + 27{\mathbf{\hat U}}_{i+1} - 3{\mathbf{\hat U}}_{i+2})/60
\end{equation}

\begin{equation}\label{fourier:U5_acc}
k_{U5}' = \frac{1}{30} \left(-e^{-3 i k}+45 i \sin (k)-9 i \sin (2 k)-15 \cos (k)+6 \cos (2 k)+10\right)
\end{equation}

which can be expanded as
\begin{equation}
k_{U5}' = i k+\frac{k^6}{60}-\frac{i k^7}{140}-\frac{k^8}{240}+\cdots,
\end{equation}
which indicates the reconstruction formula given by Equation (\ref{eqn:linear5}) is fifth-order accurate. Fig. \ref{fig_disp_unlim} shows the dispersion and dissipation properties of the IG4H, EG6, and the U5 schemes,  i.e., imaginary and real parts of  Equations (\ref{fourier:EG6_res}), (\ref{fourier:IG4H_res}), and (\ref{fourier:U5_acc}) respectively. Figure \ref{fig:dispersion_u} shows that the dispersion property of the IG4H is better than EG6 and U5 schemes. Also, as can be seen in Figure \ref{fig:dissipation_u}, the proposed schemes have superior dissipation in comparison with the U5 scheme. Fourth-order accuracy achieved here, contrary to expectations that the $\kappa$-scheme can be only at most third accurate as shown in \cite{van2021towards}, is not by design but an unexpected outcome. \textcolor{black}{Also, The fourth-order accuracy is possible because the numerical solution at the cell center is a point value approximation.}

 \begin{remark}\label{eqn:kappa}
\normalfont  An observation in the literature is that optimizing the linear upwind or central schemes for superior spectral properties reduces formal-order accuracy. Weirs and Candler constructed a six-point (not \textit{sixth order}) low dissipation scheme, which is only third-order accurate \cite{weirs1997optimization}.    Sun et al. \cite{sun2011class} have proposed optimized linear schemes that are fourth-order accurate using a six-point stencil for DNS for compressible turbulent flows. Later Sun et al. \cite{sun2014sixth} have developed an optimized sixth-order scheme using an eight-point stencil. Fu et al. \cite{Fu2016} also optimized the linear upwind and central schemes in their paper for superior spectral properties, where the optimized schemes had larger absolute error than the linear base schemes (see Fig. 10). The optimized schemes were able to resolve the flow features significantly better than the base schemes. In Ref. \cite{fernandez2018posteriori} Fern{\'a}ndez-Fidalgo has used \textit{13 point} fourth-order central finite difference scheme of Bailly and Bogey \cite{bogey2004family} as their linear scheme, see their Table. 4. The order of accuracy of the improved five and six-point TENO schemes in Ref. \cite{fu2019improved} is only third and fourth-order; see their conclusions. These observations from the literature show that researchers have optimized the linear schemes for superior spectral properties at the expense of formal order accuracy. Furthermore, In Ref. \cite{fu2017targeted} Fu et al. proposed optimal schemes with tailored resolution properties, and it is interesting to note that in Fig. 3 of their paper, they obtained a linear scheme that has superior dispersion property, which is only second-order accurate.
The point is that the authors in Ref. \cite{{weirs1997optimization},{sun2011class},{sun2014sixth},{Fu2016},{fernandez2018posteriori},{fu2017targeted},{bogey2004family},{fu2019improved}} gave up high order of accuracy even for the smooth flow for other properties. In the current paper, the implicit gradient scheme, despite being formally fourth-order, gave better results than the fifth-order upwind scheme for all the test cases. Using a value of $1/2$ instead of $1/3$ for $\kappa$ in Equation (\ref{eqn:3linear}) will lead to \textcolor{black}{anti-dissipation} properties (not shown here), which means the scheme would be unstable and is not considered. It may be possible to tune or adjust the value of $\kappa$ for better accuracy, if required or possible, or superior spectral properties. Such optimization is beyond the scope of this work, and readers can refer to \cite{sengupta2005new}.

 \end{remark}

\begin{remark}\label{eqn:choice}
\normalfont It is possible to compute the gradients of the conservative variables, $\mathbf{Q}$ = ($\rho, \rho u, \rho v,\rho w, E)^T$, and substitute them	in the Equation (\ref{eqn:3linear}). In such a case, the velocity gradients required for viscous flux computation are to be recomputed or should be obtained from the conservative variable gradients. Such an approach will lead to additional computations. It is also the case if the fluxes were used in Equation (\ref {eqn:legendre}) for reconstruction. The choice of reconstruction variables ($\mathbf{U}$, $\mathbf{Q}$ or $\mathbf{F}$) will lead to significantly different results depending on the test case.  As explained earlier, the motivation of the present approach is to compute the gradients of the primitive variables once and reuse them in both the inviscid and viscous flux discretization.
\end{remark} 

The presented schemes so far are \textit{linear high-order} schemes. As is well-known, linear high-order methods will generate oscillations near a shock or discontinuity \cite{godunov1959}, and the shock-capturing procedure is presented in the following section, Section \ref{sec-3.1.1}.

\subsection{Shock-capturing approach}\label{sec-3.1.1}

In this section, the shock-capturing approach for the linear gradient-based reconstruction schemes described in the earlier section by using the monotonicity-preserving (MP) scheme of Suresh and Huynh \cite{suresh1997accurate}. The primary advantage of the MP limiter is that any high-order linear scheme can be used. Fang et al. \cite{fang2013optimized} have used the seventh-order upwind formula as the linear scheme and also optimized the upwind scheme for improved resolution. In the original MP paper \cite{suresh1997accurate}, Suresh and Huynh also suggested that the fifth-order accurate implicit scheme, also known as the compact reconstruction scheme in literature \cite{Pirozzoli2002,ghosh2012compact}, can be used in the place of the explicit fifth-order scheme, see their Equation (2.3). The MP scheme can preserve accuracy at the extrema and monotonicity near discontinuities. Since the $R$ interface values can be obtained via symmetry, only the $L$ interface values are discussed. The implementation of the MP scheme consists of four components. In the first component, calculate the interface value by an accurate upwind formula here it is the gradient-based reconstruction approach, 

\begin{equation}\label{eqn:IG}
 {\mathbf{U}}_{i+\frac{1}{2}}^{L,GRB} ={\hat {\mathbf{U}}}_{i}+\frac{\Delta x}{2} {\mathbf{U}}'_i+\frac{\Delta x^2}{12} \left[\left(\frac{2}{ \Delta x^{2}}\right)\left({\hat {\mathbf{U}}}_{i+1}-2 {\hat {\mathbf{U}}}_i+{\hat {\mathbf{U}}}_{i-1}\right)-(\frac{1}{ 2 \Delta x})\left({ {\mathbf{U}}}_{i+1}^{\prime}-{ {\mathbf{U}}}_{i-1}^{\prime}\right)\right],
\end{equation}
In the second component of the procedure, the following criterion will be checked to determine the necessity of applying the limiter to the linear upwind scheme:
\begin{equation} \label{eqn:hocus/mp5Condition}
    \left( {\mathbf{U}}^{L,GRB}_{i+1/2} - \hat{{\mathbf{U}}}_i \right) \left( {\mathbf{U}}^{L,GRB}_{i+1/2} - {\mathbf{U}}^{MP} \right) \leq 10^{-20}
\end{equation}
where ${\mathbf{U}^{MP}}$ is given by the following equation:
\begin{equation} \label{eqn:alpha}
\begin{aligned}
 &{\mathbf{U}}^{M P} ={\mathbf{\hat U}}_{i}+\operatorname{minmod}\left[{\mathbf{\hat U}}_{i+1}-{\mathbf{\hat U}}_{i}, {\beta}\left({\mathbf{\hat U}}_{i}-{\mathbf{\hat U}}_{i-1}\right)\right]\,\,,\\
\text{and,} &\operatorname{minmod}(a,b) = \frac{1}{2} \left(\operatorname{sign}(a)+\operatorname{sign}(b)\right)\min(|a|,|b|)\,\,.
\end{aligned}
\end{equation}
The parameter ${\beta}$ in the Equation (\ref{eqn:alpha}) is a constant and is taken as $7$ in this paper as in \cite{chamarthi2021high}. If the linear scheme satisfies the condition given by Equation (\ref{eqn:hocus/mp5Condition}), there is no need to apply the limiter and ${\mathbf{U}}^{L}_{i+\frac{1}{2}}$ = ${\mathbf{U}}^{L,GRB}_{i+\frac{1}{2}}$. \textcolor{black}{Otherwise, the MP limiter, the third component of the procedure,} is applied. The procedure of the MP limiter is described through the following set of equations:
\begin{equation}\label{eqn:curvature} 
 d_{i} ={\hat{\mathbf{U}}}_{i-1}-2 {\hat{\mathbf{U}}}_{i}+{\hat{\mathbf{U}}}_{i+1}
 \end{equation}
 
 \begin{equation}\label{eqn:min}
 d_{i+1 / 2}^{M} =\operatorname{minmod}\left(4 d_{i}-d_{i+1}, 4 d_{i+1}-d_i, d_{i},d_{i+1}\right)\\
 \end{equation}
 
\begin{equation}\label{eqn:mp-procedure}
\begin{aligned}
{\mathbf{U}}_{i+\frac{1}{2}}^{M D} &=\frac{1}{2}\left({\hat{\mathbf{U}}}_{i}+{\hat{\mathbf{U}}}_{i+1}\right)-\frac{1}{2} d_{i+\frac{1}{2}}^{M} \\
{\mathbf{U}}_{i+\frac{1}{2}}^{U L} &={\hat{\mathbf{U}}}_{i}+\alpha \left({\hat{\mathbf{U}}}_{i}-{\hat{\mathbf{U}}}_{i-1}\right) \\
{\mathbf{U}}_{i+\frac{1}{2}}^{L C} &=\frac{1}{2}\left(3 {\hat{\mathbf{U}}}_{i}-{\hat{\mathbf{U}}}_{i-1}\right)+\frac{4}{3} d_{i-\frac{1}{2}}^{M}\\
{\mathbf{U}}_{i+\frac{1}{2}}^{\min } &=\max \left[\min \left({\hat{\mathbf{U}}}_{i}, {\hat{\mathbf{U}}}_{i+1}, {U}_{i+\frac{1}{2}}^{M D}\right), \min \left({\hat{\mathbf{U}}}_{i}, {\mathbf{U}}_{i+1 / 2}^{U L}, {\mathbf{U}}_{i+1 / 2}^{L C}\right)\right] \\
{\mathbf{U}}_{i+\frac{1}{2}}^{\max } &=\min \left[\max \left({\hat{\mathbf{U}}}_{i}, {\hat{\mathbf{U}}}_{i+1}, {U}_{i+\frac{1}{2}}^{M D}\right), \max \left({\hat{\mathbf{U}}}_{i}, {\mathbf{U}}_{i+\frac{1}{2}}^{U L}, {\mathbf{U}}_{i+\frac{1}{2}}^{L C}\right)\right]
\end{aligned}
\end{equation}

In the final step, the new limited interface value at the interface is given by the following equation:

\begin{equation}\label{mp-final}
{\mathbf{U}}_{i+\frac{1}{2}}^{\text {MIG }} ={\mathbf{U}}^{L,GRB}_{i+\frac{1}{2}}+\operatorname{minmod}\left({\mathbf{U}}_{i+\frac{1}{2}}^{\min }-{\mathbf{U}}^{L,GRB}_{i+\frac{1}{2}}, {\mathbf{U}}_{i+\frac{1}{2}}^{\max }-{\mathbf{U}}^{L,GRB}_{i+\frac{1}{2}}\right).
\end{equation}

The schemes thus obtained from the above procedure are defined as Monotonicity-preserving explicit or implicit gradient reconstruction and are denoted as \textbf{MEG} or \textbf{MIG} in this paper. The non-linear scheme that uses implicit gradients, IG4H, as its base linear scheme is denoted as \textbf{MIG4} and the non-linear scheme that uses explicit gradients, EG6, as its base linear scheme is denoted as \textbf{MEG6}.

\textbf{Improvement of MP limiting:} Continuing with the idea of ``reusing'' the gradients the curvature terms $ d_{i}$ given by Equation (\ref{eqn:curvature}) are replaced with high-order second-derivatives (curvature terms) that are readily available from Equation (\ref{eqn:adam}),

\begin{equation}\label{eqn:curv}
d_i=2\left({\hat {\mathbf{U}}}_{i+1}-2 {\hat {\mathbf{U}}}_i+{\hat {\mathbf{U}}}_{i-1}\right)-0.5\Delta x\left({ {\mathbf{U}}}_{i+1}^{\prime}-{ {\mathbf{U}}}_{i-1}^{\prime}\right).
\end{equation}
Also, the curvature measurement $d_{i+1 / 2}^{M}$, given by Equation (\ref{eqn:min}), is defined as follows:
 \begin{equation}\label{eqn:m_curv}
 d_{i+1 / 2}^{M} =\operatorname{minmod}\left(d_{i},d_{i+1}\right)\\
 \end{equation}
  In the original paper by Suresh and Huynh \cite{suresh1997accurate} both approaches i.e. Equation (\ref{eqn:m_curv}) and (\ref{eqn:min}) are considered for $d_{i+1 / 2}^{M}$. In the current paper Equation (\ref{eqn:m_curv}) is considered.

\begin{figure}[H]
\centering
\subfigure[]{\includegraphics[width=0.45\textwidth]{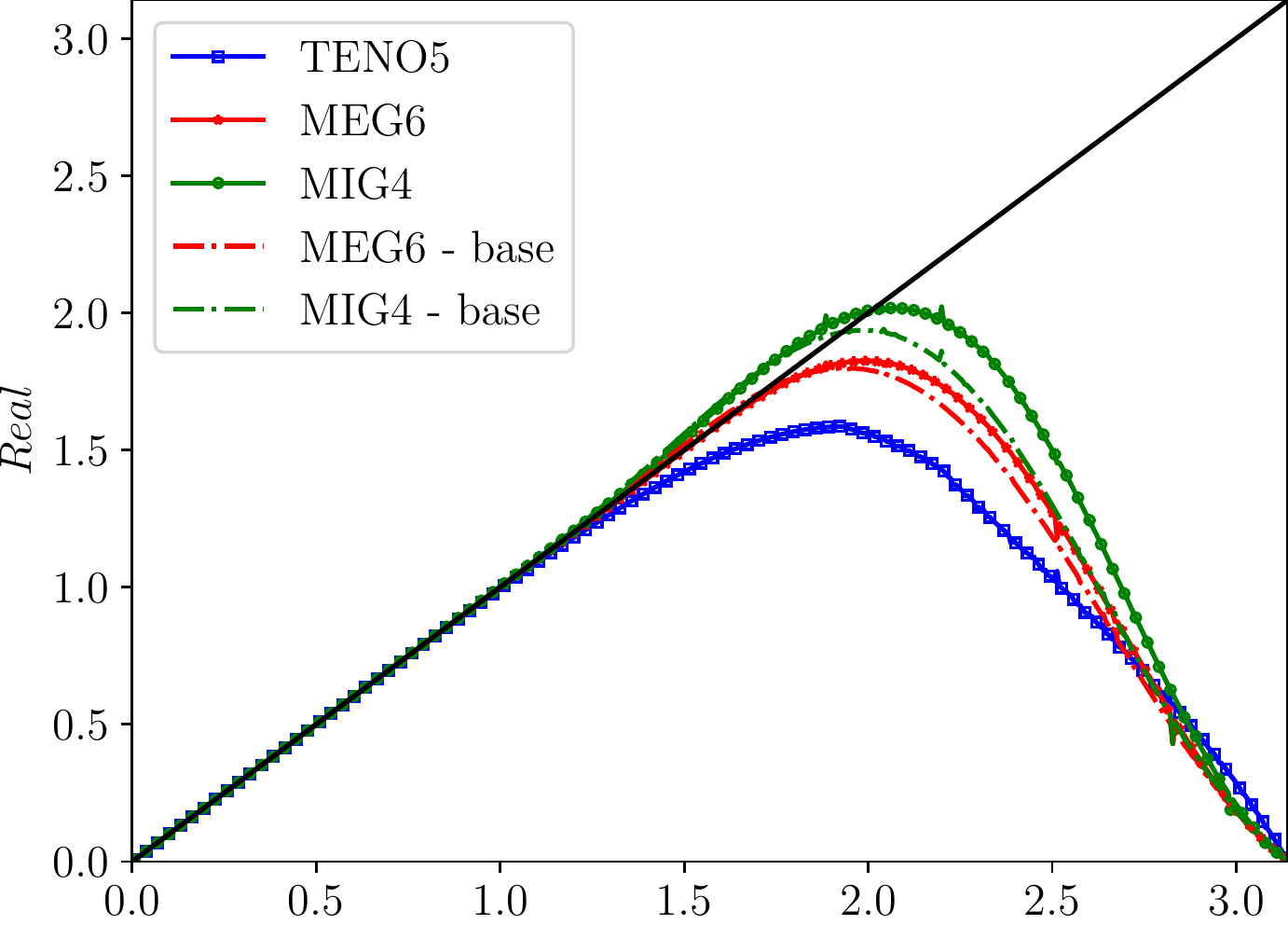}
\label{fig:dispersion}}
\subfigure[]{\includegraphics[width=0.47\textwidth]{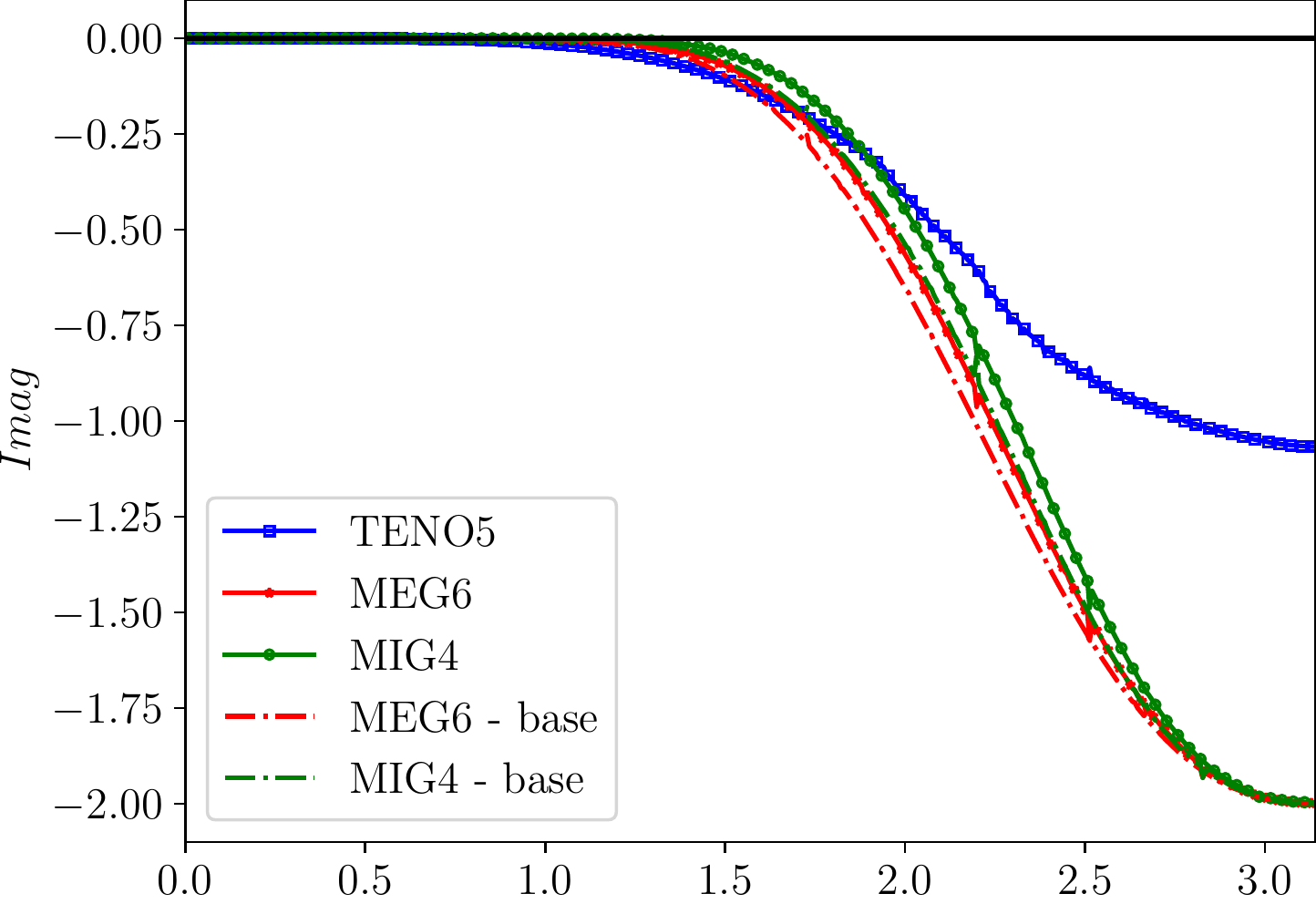}
\label{fig:dissipation}}
\caption{Dispersion and Dissipation properties of the base and improved nonlinear monotonicity preserving schemes.}
\label{fig_disp}
\end{figure}
The spectral properties of the developed nonlinear schemes are studied by the Approximate Dispersion Relation (ADR) analysis \cite{zhao2019general}. Following the analysis, Fig. \ref{fig_disp} shows the comparison of the dissipation and dispersion properties of the proposed schemes. Fig. \ref{fig:dispersion} shows the dispersion properties of the proposed schemes, and it can be seen that MIG4, with implicit gradients, has superior dispersion properties. Reusing the gradients for the curvature terms improved the dispersion properties of the MIG4 scheme (comparing MIG4 - base with standard $d_i$ and MIG4 with $d_i$ using Equation (\ref{eqn:curv})). Dispersion improvement in the MEG6 scheme is marginal, but it also improved. In this paper, we also compared the results using the fifth-order TENO scheme of Fu et al. \cite{fu2017targeted} along with the MIG scheme presented in this paper. The main advantage of the proposed schemes is the reusability of the gradients for both inviscid and viscous fluxes, which is a clear distinction from the TENO scheme. Nevertheless, we carried out simulations with the TENO scheme. It is noted that MIG4 \textcolor{black}{has} superior dispersion and dissipation properties than the MEG6 and TENO5 scheme. 

For coupled hyperbolic equations like the Euler equations, shock-capturing should be carried out using characteristic variables for \textit{cleanest} results \cite{van2006upwind}. It is fairly easy to implement characteristic variable transformation for the proposed schemes and summarized below:

\begin{enumerate}
\item Compute the first-order gradients of variables that are to be used for the reconstruction.
\item Compute the arithmetic or Roe averages at the interface $(x_{i+\frac{1}{2}})$ by using neighbouring cells, $(x_i)$ and $(x_{i+1})$. Compute the left $\bm{L_{n}}$ and right $\bm{R_{n}}$ eigenvectors. Transform the variables, $\bm{\hat \mathbf{U}}$, and their corresponding gradients, $\bm{\mathbf{U}}'$, into characteristic space by multiplying with the left eigenvectors.

\begin{equation}
	\hat{\bm{W}}_{m} = \bm{L}_{\bm{n}_{i+\frac{1}{2}}} \bm{\hat \mathbf{U}}_{m},
	\bm{W}'_{m} = \bm{L}_{\bm{n}_{i+\frac{1}{2}}} \bm{\mathbf{U}}'_{m},
\end{equation}
The transformed variables are denoted as $\hat{\bm{W}}_{m}$ and ${\bm{W}}'_{m}$, respectively, where $m$  is $\{i-2, i-1, i, i+1, i+2, i+3 \}$.

\item Obtain the left and right reconstructed states by using the transformed variables.
\begin{equation}\label{eqn:grab}
\begin{aligned}
 \mathbf{W}_{i+\frac{1}{2}}^{L,GRB} &=\mathbf{\hat {W}}_{i+0}+\frac{\Delta x}{2} \mathbf{W}'_{i}+\frac{\Delta x^2}{12} \left[\left(\frac{2}{ \Delta x^{2}}\right)\left(\mathbf{\hat {W}}_{i+1}-2 \mathbf{\hat {W}}_i+\mathbf{\hat {W}}_{i-1}\right)-(\frac{1}{ 2 \Delta x})\left(\mathbf{ {W}}_{i+1}^{\prime}-\mathbf{ {W}}_{i-1}^{\prime}\right)\right],\\
  \mathbf{W}_{i+\frac{1}{2}}^{R,GRB} &=\mathbf{\hat {W}}_{i+1}-\frac{\Delta x}{2} \mathbf{W}'_{i+1}+\frac{\Delta x^2}{12} \left[\left(\frac{2}{ \Delta x^{2}}\right)\left(\mathbf{\hat {W}}_{i+2}-2 \mathbf{\hat {W}}_{i+1}+\mathbf{\hat {W}}_{i}\right)-(\frac{1}{ 2 \Delta x})\left(\mathbf{ {W}}_{i+2}^{\prime}-\mathbf{ {W}}_{i}^{\prime}\right)\right].
\end{aligned}
\end{equation}

\item Carry out MP limiting procedure, through Equations (\textcolor{black}{\ref{eqn:hocus/mp5Condition}})-(\ref{mp-final}), and obtain left- and right-interface values denoted by ${\bm{W}}_{i+\frac{1}{2}}^{L, MIG}$ and ${\bm{W}}_{i+\frac{1}{2}}^{R, MIG}$. For improved MP limiting approach Equations (\ref{eqn:curv}) and (\ref{eqn:m_curv}) should be used. \\
\item After obtaining ${\bm{W}}_{i+\frac{1}{2}}^{L, MIG}$ and ${\bm{W}}_{i+\frac{1}{2}}^{R, MIG}$ the reconstructed states are then recovered by projecting the characteristic variables back to physical fields: \\
\textcolor{black}{
\begin{equation}
\begin{aligned}\label{IG-right-transform}
	{{\bm{\mathbf{U}}}}_{i+\frac{1}{2}}^{L,MIG} &= \bm{R}_{\bm{n}_{i+\frac{1}{2}}} {\bm{W}}_{i+\frac{1}{2}}^{L, MIG}, \\
  {{\bm{\mathbf{U}}}}_{i+\frac{1}{2}}^{R,MIG} &= \bm{R}_{\bm{n}_{i+\frac{1}{2}}} {\bm{W}}_{i+\frac{1}{2}}^{R, MIG}.
\end{aligned}
\end{equation}}
\end{enumerate}

\section{Results and discussion}\label{sec-4}

In this section, the proposed spatial discretization schemes are tested for a set of benchmark cases to assess the performance of both single- and multi-species flows. For the MEG6 scheme, the viscous fluxes are discretized using $\alpha$-EG, and for the MIG4 scheme, the $\alpha$-IG approach is used unless otherwise stated. The value of CFL in Equation (\ref{eqn:cfl}) is taken as 0.2 for all the simulations unless otherwise stated.

\textcolor{black}{The test cases are organized so that the important aspects and advantages of the proposed numerical methods are highlighted in the first eight examples. Then the standard one-dimensional test cases are presented. Example \ref {ex:vs} shows the importance of the viscous flux discretization, Examples \ref{ex:ssl}, \ref{shock-entropy}, and \ref{ex:rm} show the advantage of the implicit gradient approach, and Example \ref{ex:dmr} shows that the proposed approach is free of oscillations, unlike the TENO scheme. Finally, Examples \ref{shock-entropy} and \ref{ex:rp} show the improvement of the MP scheme by using the high-order curvature terms. Examples \ref{leblanc} and \ref{ex:triple} also show the advantage of the implicit gradient approach.}

\subsection{Multi-dimensional single species inviscid and viscous test cases}

\begin{example}\label{ex:vs}{Viscous Shock tube} (Importance of viscous flux discretization)
\end{example}

\begin{figure}[H]
\begin{onehalfspacing}
\centering\offinterlineskip
\subfigure[\textcolor{black}{TENO5, Shen}]{\includegraphics[width=0.25\textwidth]{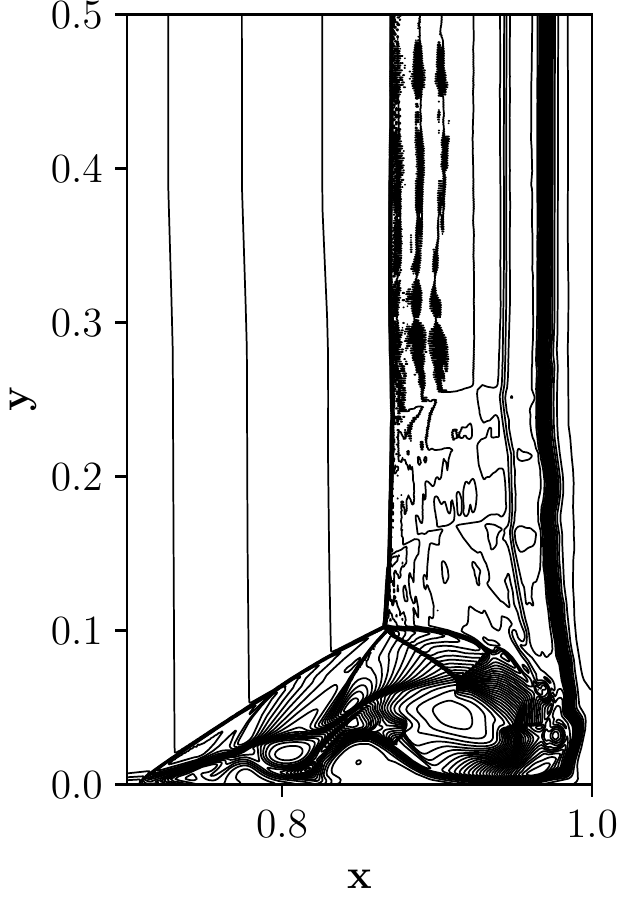}
\label{fig:1000_shen_nodamp_50}}
\subfigure[\textcolor{black}{TENO5, $\alpha$-EG}]{\includegraphics[width=0.25\textwidth]{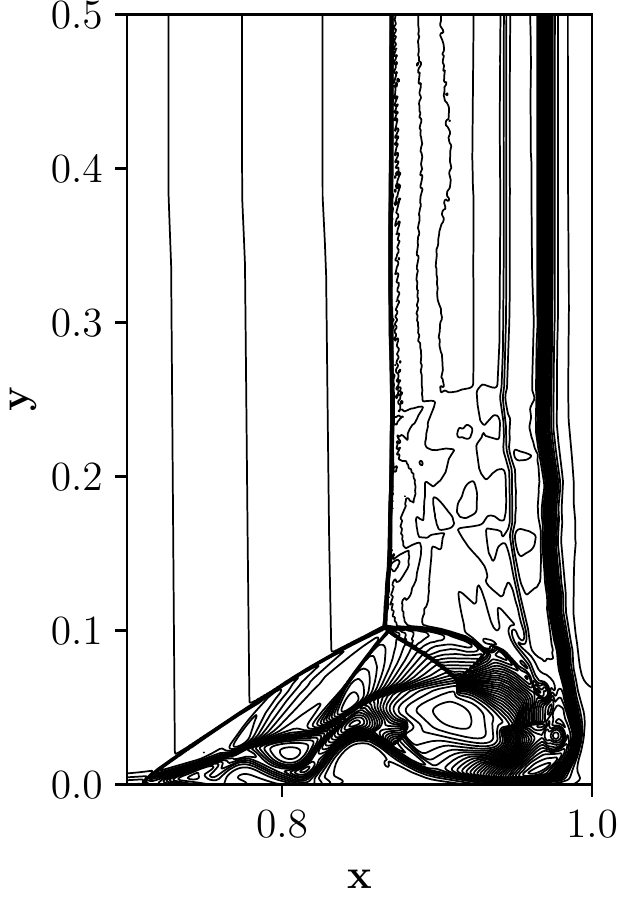}
\label{fig:1000_alpha_eg_teno}}
\subfigure[\textcolor{black}{TENO5, $\alpha$-IG}]{\includegraphics[width=0.25\textwidth]{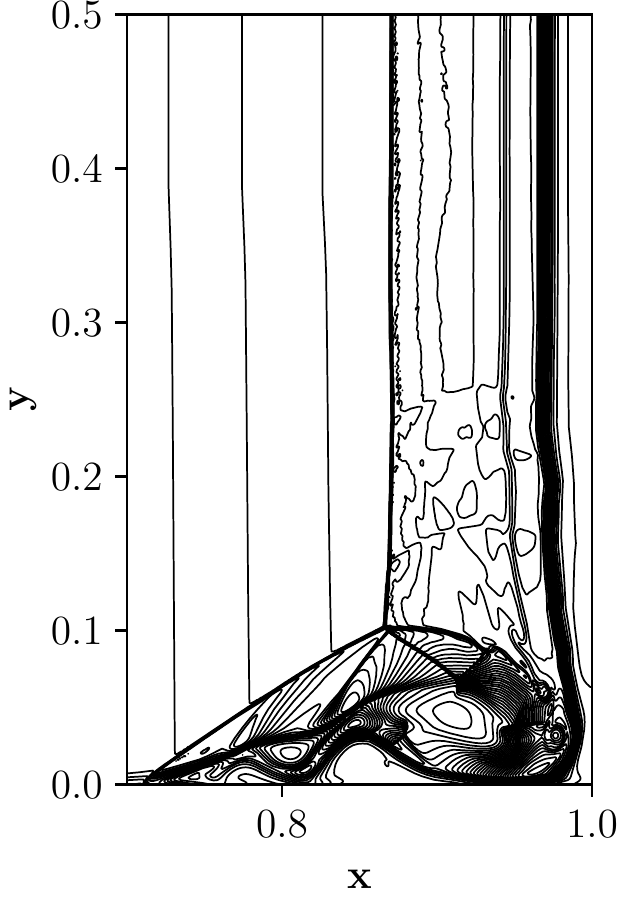}
\label{fig:1000_alpha_ig_teno}}
\subfigure[\textcolor{black}{MEG6, $\alpha$-EG}]{\includegraphics[width=0.25\textwidth]{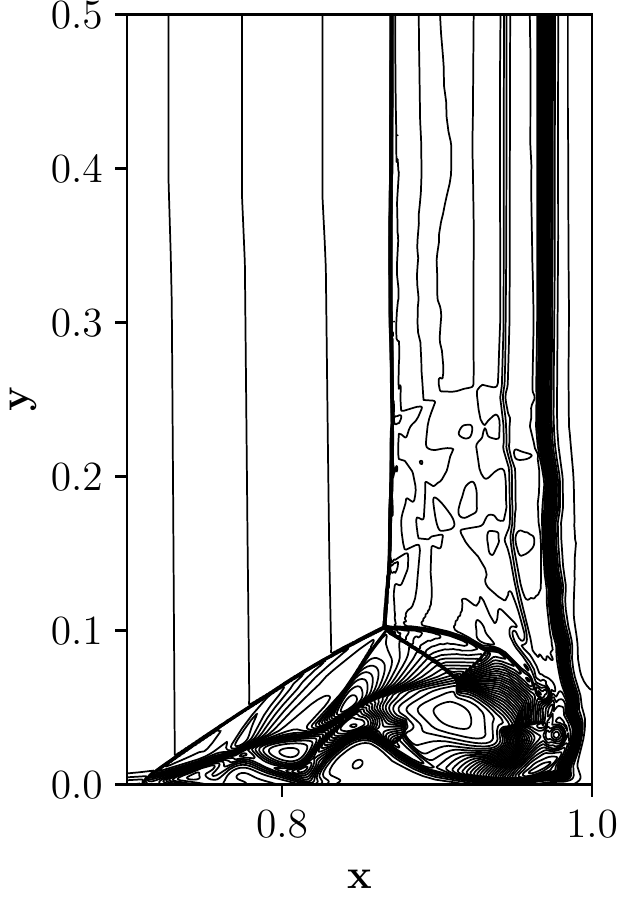}
\label{fig:1000_meg6_50}}
\subfigure[\textcolor{black}{MEG4, $\alpha$-IG}]{\includegraphics[width=0.25\textwidth]{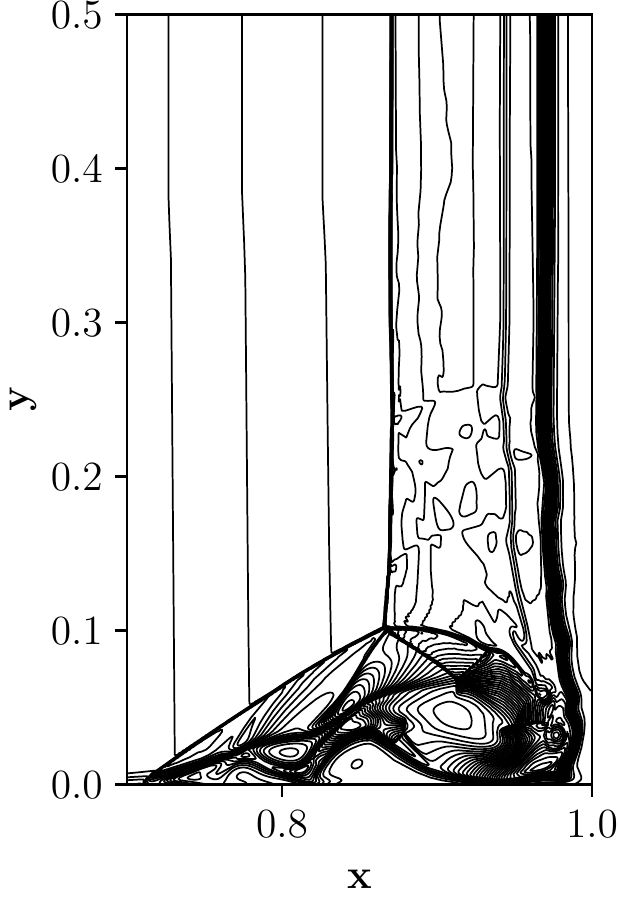}
\label{fig:1000_mig4_50}}
\subfigure[\textcolor{black}{TENO5, Shen's scheme}]{\includegraphics[width=0.45\textwidth]{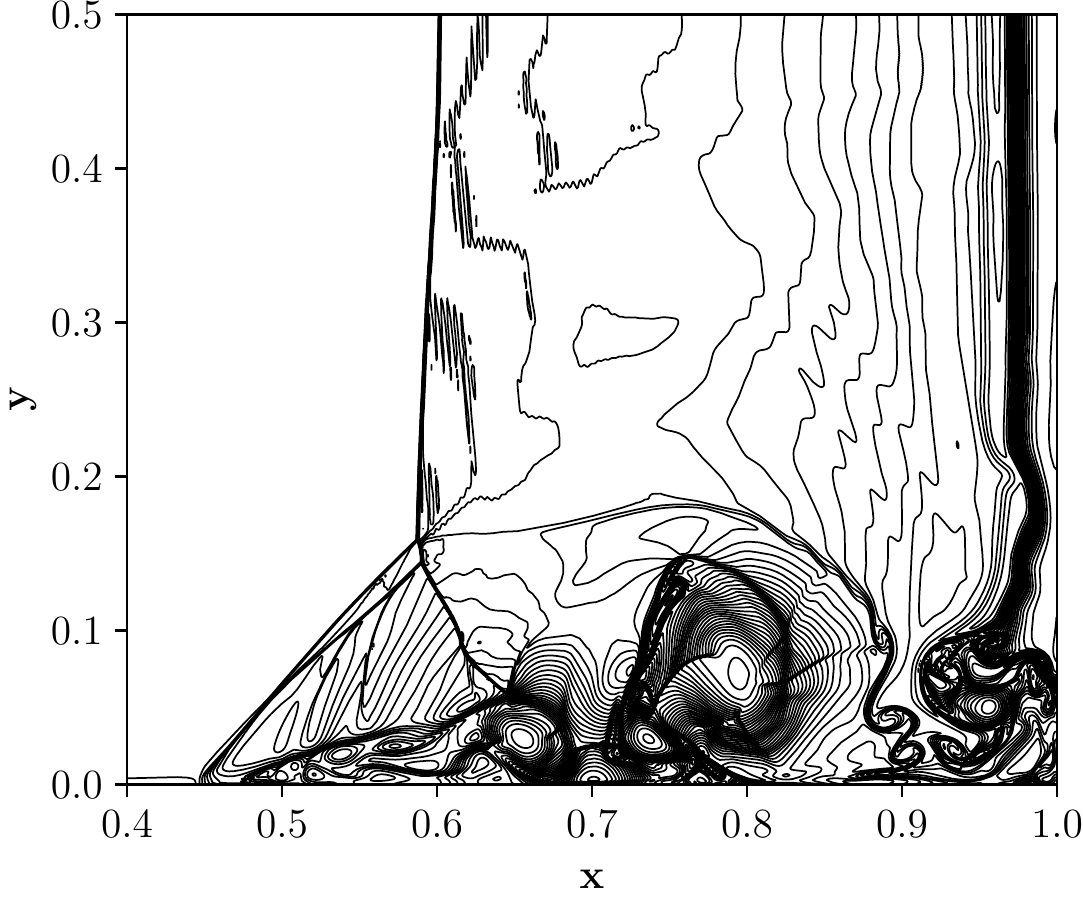}
\label{fig:1000_teno_useless}}
\caption{Figs. \ref{fig:1000_shen_nodamp_50} - \ref{fig:1000_mig4_50} show density contours obtained with various schemes for the Example \ref{ex:vs} at $t$ = 0.50 for $Re=1000$ on a grid size of 1280 $\times$ 640. Fig. \ref{fig:1000_teno_useless} is at the time instance of $t$=1.0. These figures are drawn with 38 density contours.}
\label{fig_damp_1000}
\end{onehalfspacing}
\end{figure}

\begin{figure}[H]
\begin{onehalfspacing}
\centering\offinterlineskip
\subfigure[\textcolor{black}{TENO5, $\alpha$-EG}]{\includegraphics[width=0.45\textwidth]{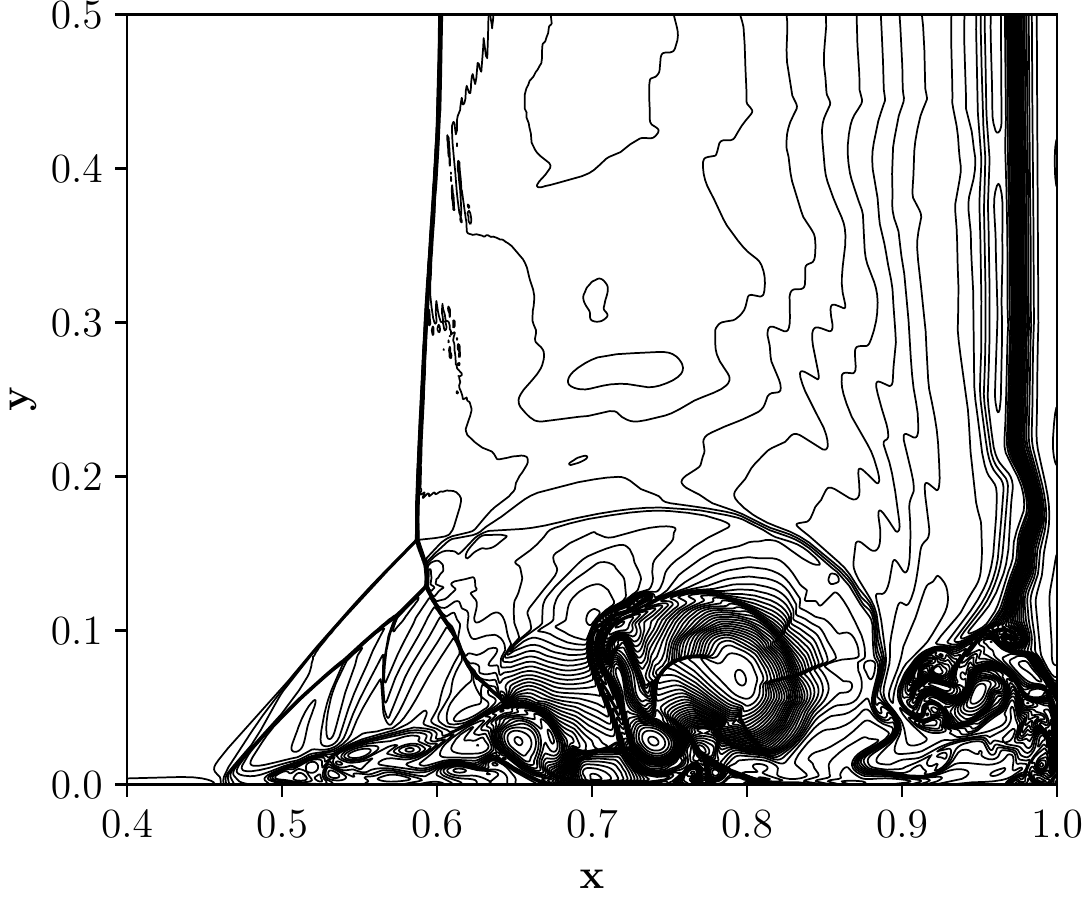}
\label{fig:VST_TENO-EG}}
\subfigure[\textcolor{black}{TENO5, $\alpha$-IG}]{\includegraphics[width=0.45\textwidth]{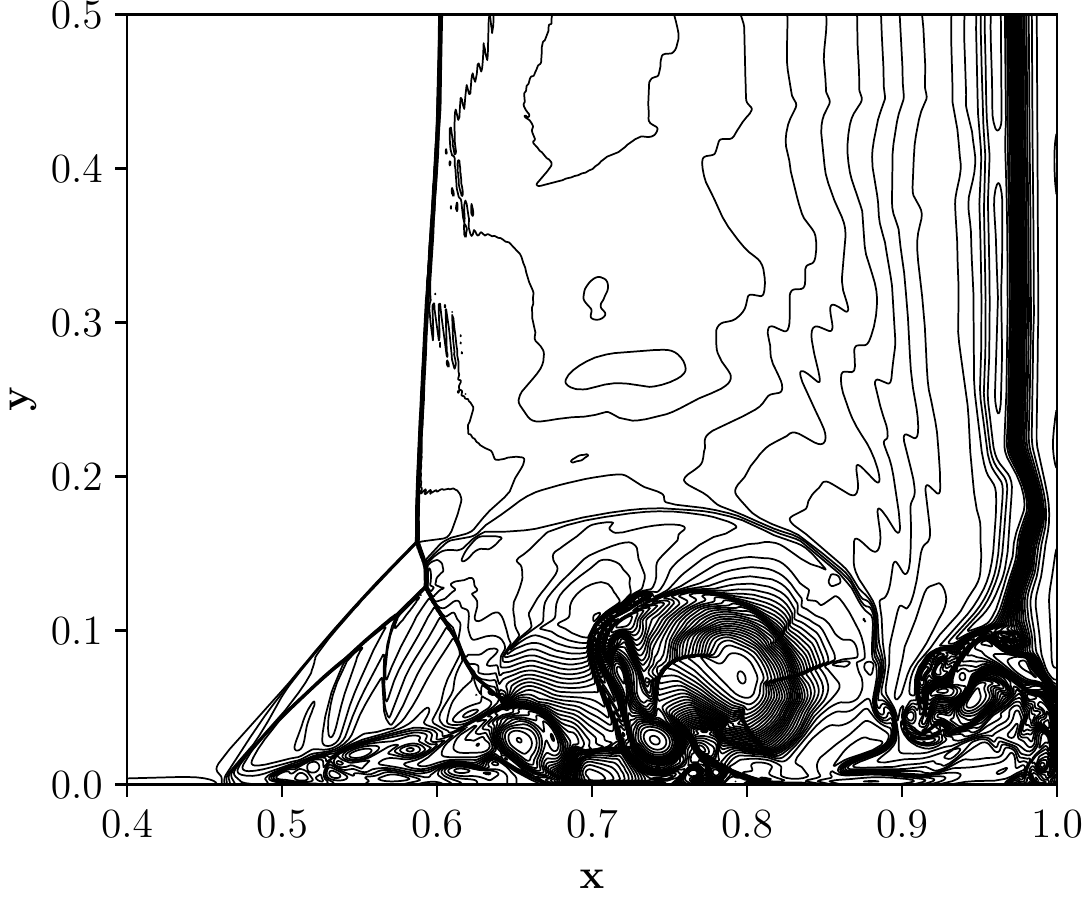}
\label{fig:VST_TENO-IG}}
\subfigure[\textcolor{black}{\textcolor{black}{MEG6, $\alpha$-EG}}]{\includegraphics[width=0.45\textwidth]{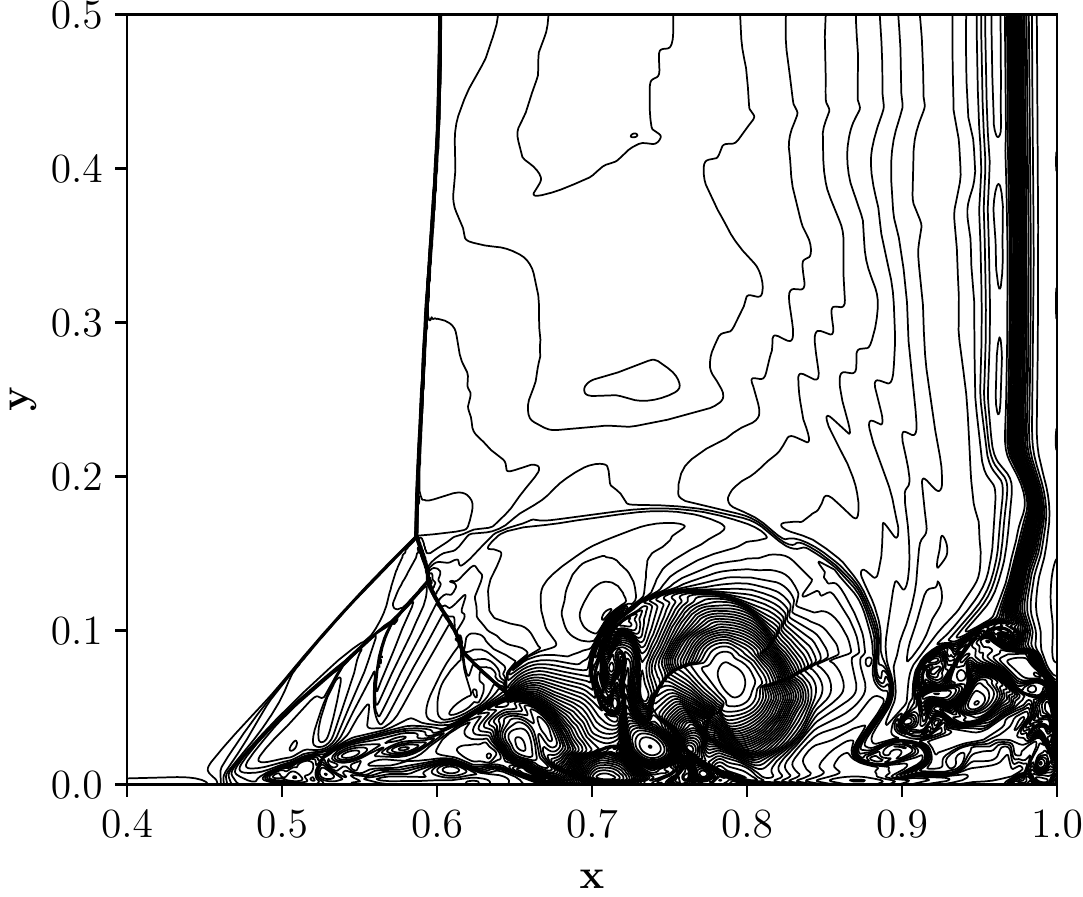}
\label{fig:VST_MP5-no-damp}}
\subfigure[\textcolor{black}{MIG4, $\alpha$-IG}]{\includegraphics[width=0.45\textwidth]{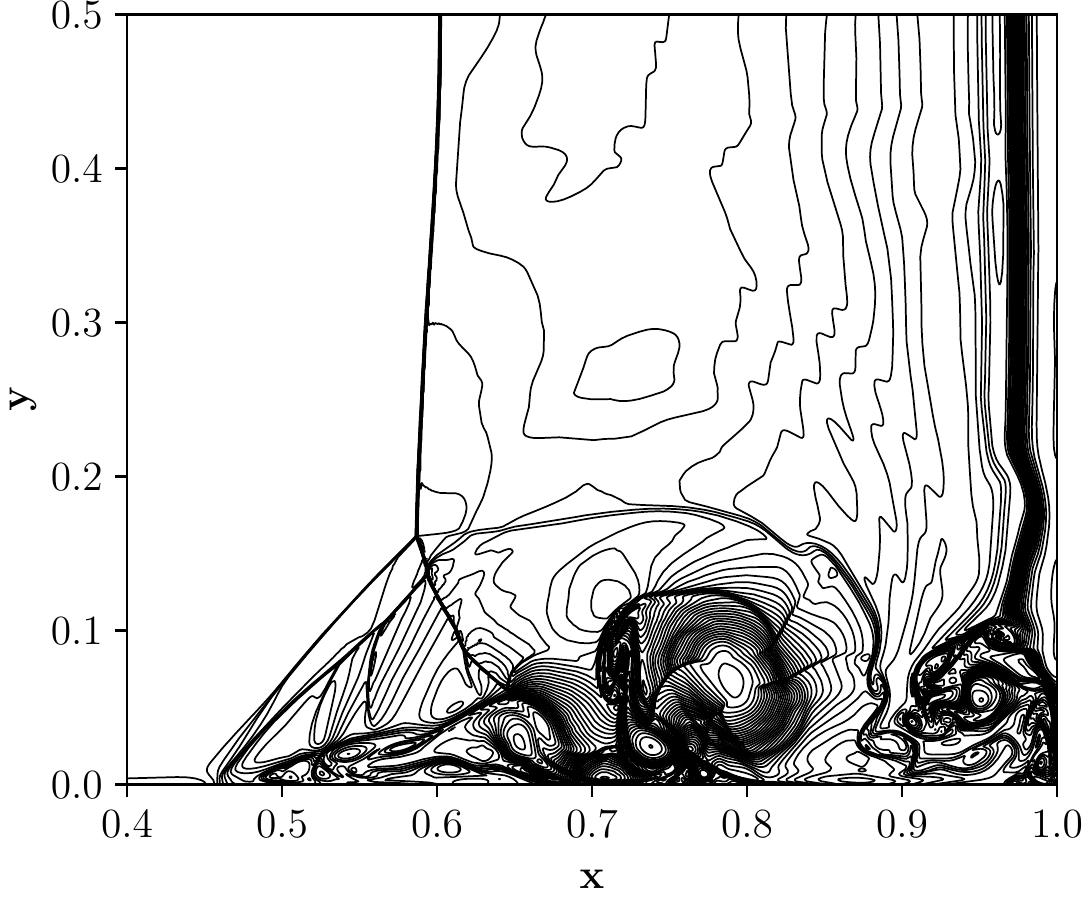}
\label{fig:VST_MIGE-no-damp_4e}}
\caption{Density contours for different schemes for the Example \ref{ex:vs} and $Re=1000$ using various schemes on a grid size of 1280 $\times$ 640, respectively, at $t$=1.0.}
\label{fig_damp-3}
\end{onehalfspacing}
\end{figure}

In this first test \textcolor{black}{case,} we show the effect of viscous flux discretizations proposed in Section \ref{sec-3.2}. For this \textcolor{black}{purpose,} the viscous shock-tube problem in a square shock tube with insulated walls on a domain of unit length and half unit height is considered \cite{daru2009numerical}. In this problem,  A diaphragm separating two different states is initially located in the middle of the tube (x = 0.5). The initial conditions are:
\begin{equation}\label{vst}
\begin{aligned}
\left( {\rho , u,v, p} \right) = \left\{ \begin{array}{l}
\left( {120, 0 ,0,120/\gamma } \right),  \quad 0 < x < 0.5,\\
\left( {1.2 , 0 ,0, 1.2/\gamma } \right),  \quad 0.5 \le x < 1.
\end{array} \right.
\end{aligned}
\end{equation}
The flow is simulated for time $t$ = 1, keeping the Mach number of the shock wave at 2.37. Constant dynamic viscosity $\mu$ = 1/1000 is assumed, and the gas constant, R, is taken as 1. By choosing the reference values as the initial speed of sound, unit density, and unit length, the Reynolds number, Re, is 1000. The ratio of specific heats of $\gamma = 7/5$. The domain for this test case is  $x \in [0,1], y \in [0,0.5]$.  The problem is solved on a grid size of 1280 $\times$ 640 with various schemes. Wall boundary conditions are used for all the boundaries except for the top boundary, where a symmetry boundary condition is used. Grid converged results for this test case can be found in Ref.  \cite{chamarthi2022} (their Fig. 2b), which are carried out on a grid resolution of 4000 $\times$ 2000. The following observations are made from the simulations:

\begin{itemize}
\item Fig. \ref{fig_damp-3} shows the results obtained by various schemes at $t$ = 1. There are no differences between the results obtained when the TENO5 scheme is used for inviscid fluxes and $\alpha$-EG and $\alpha$-IG, Fig. \ref{fig:VST_TENO-EG} and \ref{fig:VST_TENO-EG}, respectively, are used viscous fluxes. The results obtained with the MIG4 scheme, Fig. \ref{fig:VST_MIGE-no-damp_4e}, are closer to the fine grid simulations, specifically the main vortex (eyeball norm). Table \ref{tab:comp_eff_vst} shows the computational time taken by the various schemes. MEG6 scheme is efficient as the gradients are computed explicitly and are shared between the viscous and inviscid fluxes. TENO5 scheme with $\alpha$-EG viscous fluxes takes nearly the same time as the MIG4 scheme despite computing the gradients using a tridiagonal solver. TENO5 scheme with $\alpha$-IG viscous fluxes is the most expensive. These results show the computational efficiency of the current approach.
\item Fig. \ref{fig:1000_teno_useless} shows the results obtained with the sixth order finite-difference approach of Shen \cite{shen2010large} at $t$=1 and the primary vortex is distorted compared with the results obtained using $\alpha$-damping approach. The results are consistent with the findings in \cite{chamarthi2022}. TENO5 scheme would also benefit from using the $\alpha$-damping approach. 
\item Figs. \ref{fig:1000_alpha_eg_teno} - \ref{fig:1000_mig4_50} shows results at an intermediate time of $t$=0.5 obtained with the $\alpha$-damping approach using TENO5, MEG6 and MIG4 schemes for the inviscid fluxes. As expected, all the results showed no evidence of odd-even decoupling. Fig. \ref{fig:1000_shen_nodamp_50} shows the results obtained using Shen's scheme and TENO5, and the saw-tooth like oscillations due to the lack of high-frequency damping are obvious. Similar oscillations are also with the fourth-order scheme of Shen et al.\cite{shen2009high} (not shown here).
\end{itemize}

\begin{table}[H]
  \centering
  \caption{Computational time for the Example \ref{ex:vs} using TENO5, MEG6 and MIG4 schemes using different $\alpha$-damping schemes.}
    \begin{tabular}{||c|c|c|c|c||}
        \hline
    \hline
    Inviscid scheme & \multicolumn{2}{c|}{TENO5} & MEG6  & MIG4 \\
        \hline
            \hline
    Viscous Scheme & $\alpha$-IG & $\alpha$-EG & $\alpha$-EG & $\alpha$-IG \\
    \hline
    \hline
    Computational time & 58680s & 50880s & 48506s & 51146s \\
    \hline
    Ratio & 1.147 & 0.994 & 0.948 & 1 \\
    \hline
    \hline
    \end{tabular}%
  \label{tab:comp_eff_vst}%
\end{table}%

\begin{example}\label{ex:ssl}{Shock-shear layer} (Shows the advantage of implicit gradients)
\end{example}

\begin{figure}[H]
\centering\offinterlineskip
\subfigure[TENO5, $\alpha$-IG, $320 \times 80$]{\includegraphics[width=0.8\textwidth]{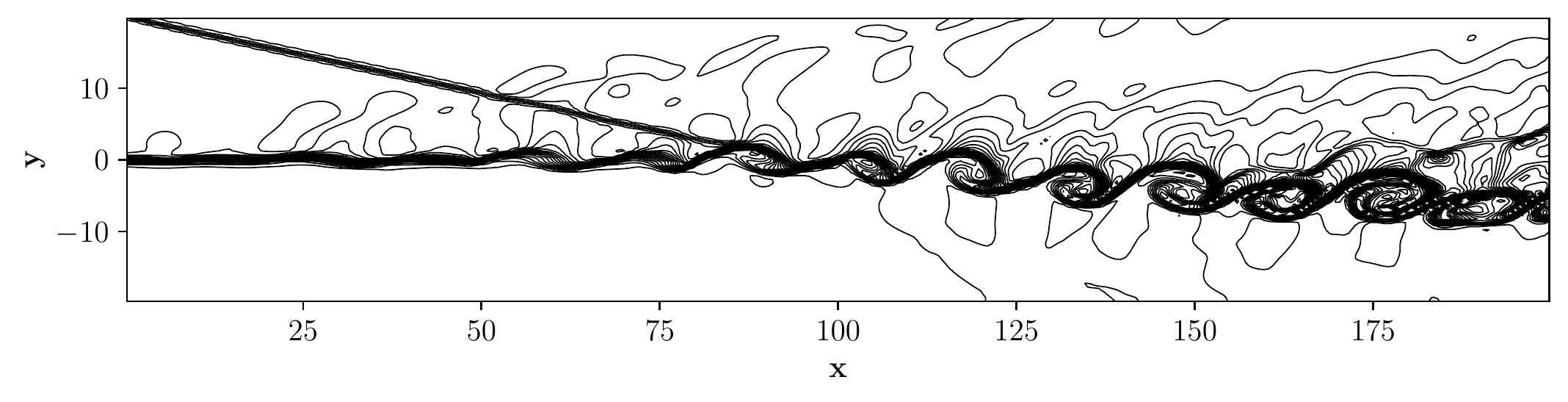}
\label{fig:TENO_SSL_fine}}
\subfigure[MEG6, $\alpha$-IG, $320 \times 80$]{\includegraphics[width=0.8\textwidth]{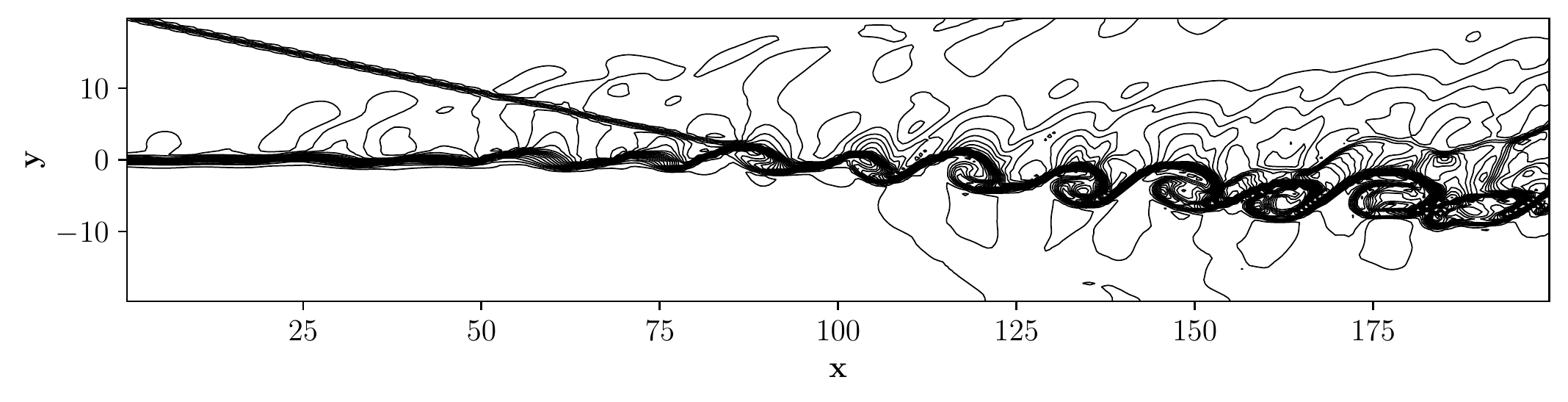}
\label{fig:MEG_SSL_fine}}
\subfigure[MIG4, $\alpha$-IG, $320 \times 80$]{\includegraphics[width=0.8\textwidth]{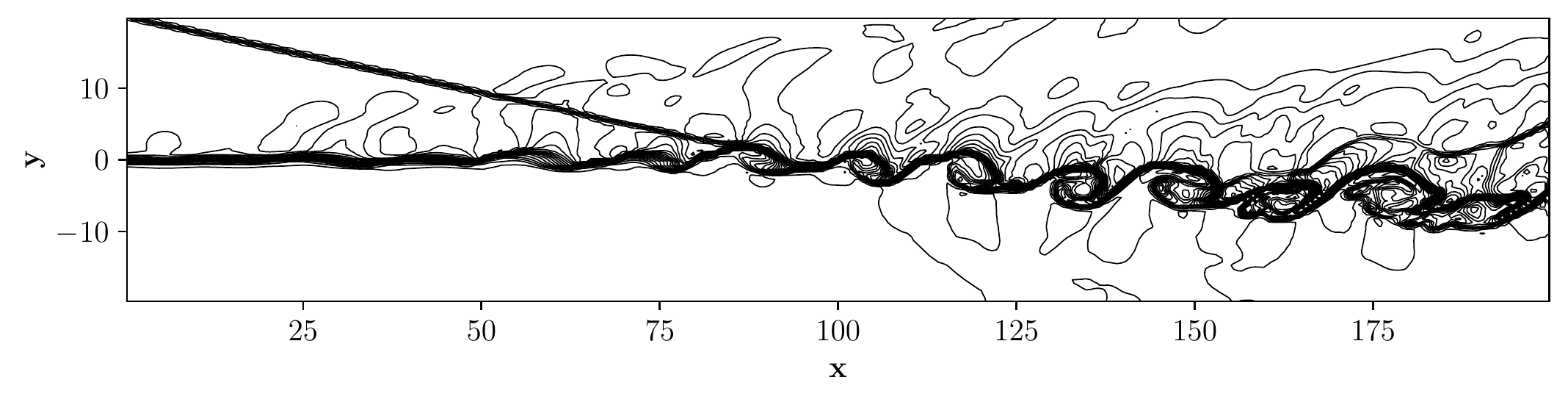}
\label{fig:MIG_SSLc}}
\caption{Computed density contours of the shock-shear layer interaction described in Example \ref{ex:ssl} on grid size of $320 \times 80$ for the proposed schemes and TENO5.}
\label{fig_SSL}
\end{figure}

The second two-dimensional viscous test case we considered was the shock-shear layer of Yee et al. \cite{yee1999low}. In this case, a 12-degree oblique shock propagates from the left boundary and impinges on a spatially evolving mixing layer. An expansion fan forms above the shear layer and reflects from the bottom boundary. The Reynolds number is, $Re = 1000$ and the initial convective Mach number of the mixing layer is $Ma = 0.6$. The Prandtl number is set as $Pr = 0.72$. The computational domain is $[x,y] = [0,200] \times [-20,20]$, using a uniform grid of $320 \times 80$. The initial stream-wise velocity is set as a hyperbolic tangent profile:

\begin{equation}
    u = 2.5 + 0.5 \tanh(2y)
\end{equation}
\noindent Fluctuations are added to the wall-normal velocity at the inlet:

\begin{equation}
    v' = \sum^{2}_{k=1} \alpha_k \cos(2 \pi k t/T + \phi_k) \exp(-y^2/b)
\end{equation}

\noindent where the period, $T = \lambda/u_c$, the wavelength, $\lambda = 30$, the convective velocity $u_c = 2.68$, and $b = 10$. When $k = 1$, $\alpha_1 = 0.05$ and $\phi = 0$. When $k = 2$, $\alpha_2 = 0.05$ and $\phi = \pi/2$. The bottom boundary is set as a slip \textcolor{black}{wall,} and the outflow boundary is specified as a supersonic outlet. A complete description of the boundary conditions can be found in Yee et al. \cite{yee1999low}. The case is run until a final time, $t = 120$. Fig. \ref{fig_SSL} displays the density contours of the shock-shear layer \textcolor{black}{case,} and all schemes resolve the oblique shock without any oscillations. 

For this test case, all the simulations are carried out with the $\alpha$-IG scheme for viscous fluxes for a fair comparison of the inviscid schemes. By comparing the Figs. \ref{fig:TENO_SSL_fine}, \ref{fig:MEG_SSL_fine} and \ref{fig:MIG_SSLc} it can be seen that the shear-layer thickness is thinner for the MIG4 scheme compared to the MEG6 and TENO5 which indicates the low dissipation property of the MIG4 approach.

\begin{example}\label{ex:acc} {Order of accuracy}
\end{example}

Here we test the order of accuracy of the proposed schemes for two different test cases, a linear and nonlinear problem. First, the linear test case is considered where the initial profile is given by Equation \eqref{accu-euler} is convected on a computational domain of $x,y \in [-1, 1]$. The time-step is varied as a function of the grid size as $\Delta t = \text{CFL} \Delta x^{2}$ and the final solution is obtained at time $t=2$. 
\begin{align}\label{accu-euler}
(\rho,u,v,p)= (1+0.5 \sin (x+y),\ \ 1.0,\ \, 1.0 \ \ 1.0)
\end{align}
The $L_2$ norm of the error between the exact and the obtained solution is used to compute the order of accuracy and is presented in Table \ref{tab:accu}. \hl{Results indicate that all the gradient-based schemes, regardless of the gradients used, are fourth-order accurate in space.} \textcolor{black}{ This high-order accuracy is possible because it is a pointwise approximation, a high-order approximation to $\frac{d \mathbf F}{dx}_{i,j}$  constructed along the $x$-direction and similarly applied to $\frac{d \mathbf G}{dy}_{i,j}$ in the $y$-direction as in Equation (\ref{eqn-differencing_residual})}. The TENO5 scheme is fifth-order accurate, but the absolute error of the gradient-based schemes on finer grids is almost the same as that of the TENO5.

\begin{table}[H]
  \centering
  \footnotesize
  \caption{Predicted order of accuracy for the considered schemes for linear test case.}
    \begin{tabular}{|| c | cccccc || }
    \hline
    \hline
    \multicolumn{7}{||c||}{Schemes} \\
     \hline
       N   & TENO5 & Order & MIG4  & Order & MEG6  & Order \\
          \hline
    $10^2$ & 6.79E-03 & -     & 1.19E-03 & -     & 1.15E-03 & - \\
    \hline
    $20^2$ & 2.24E-04 & 4.92  & 7.82E-05 & 3.89  & 6.11E-05 & 4.23 \\
    \hline
    $40^2$ & 7.06E-06 & 4.98  & 4.97E-06 & 3.91  & 3.77E-06 & 4.02 \\
    \hline
    $80^2$ & 2.21E-07 & 5.00  & 3.37E-07 & 3.97  & 2.35E-07 & 4.00 \\
    \hline
    \end{tabular}%
  \label{tab:accu}%
\end{table}%

Second, the nonlinear test case where an isentropic vortex is convected in an inviscid free stream \cite{yee1999low} is considered. The computations are done till a final time is $T$ = 10  with periodic boundary conditions on all four sides on a computational domain of [0, 10] $\times$ [0, 10]. At time T, the vortex returns to the initial position. To the mean flow, an isentropic vortex is added, and the initial flow field is initialized as follows:

\begin{equation}
\begin{doublespacing}
\begin{array}{l}
\rho=\left[1-\frac{(\gamma-1) \beta^{2}}{8 \gamma \pi^{2}} e^{\left(1-r^{2}\right)}\right]^{\frac{1}{p-1}}, \\
 	p=\rho^{\gamma},
 \quad r^{2}=\bar{x}^{2}+\bar{y}^{2} \\
 
(u, v)=(1,1)+\frac{\beta}{2 \pi} e^{\frac{1}{2}\left(1-r^{2}\right)}(-\bar{y}, \bar{x}), \\
 \bar{x}=x-x_{v}, \quad \bar{y}=y-y_{v} \\

\end{array}
\end{doublespacing}
\end{equation}
where $\beta$ = 5 and $(x_{v},y_{v})$ = (5, 5) are the coordinates of the center of the initial vortex. We can observe clearly from Table \ref{tab:accu_2_isen} that all the schemes are only second-order accurate for the nonlinear test \textcolor{black}{case,} including TENO5. Once \textcolor{black}{again,} the absolute error of the TENO5 scheme is nearly the same \textcolor{black}{as} the MEG6 and MIG4 schemes.

\begin{table}[H]
  \centering
  \footnotesize
 \caption{Predicted order of accuracy for the considered schemes for nonlinear test case.}
    \begin{tabular}{|| c | cccccc || }
    \hline
    \multicolumn{7}{c}{Schemes} \\
    \hline
    N     & \multicolumn{1}{c}{TENO5} & Order   & \multicolumn{1}{c}{MIG4} & Order   & \multicolumn{1}{c}{MEG6} & Order \\
    \hline
    \hline
    $25^2$ & \multicolumn{1}{c}{4.93E-03} & -     & 2.41E-03 & -     & 2.82E-03 & - \\
    \hline
    $50^2$ & \multicolumn{1}{c}{6.59E-04} & 2.90  & 6.47E-04 & 1.90  & 6.38E-04 & 2.14 \\
    \hline
    $100^2$ & 1.63E-04 & 2.01  & 1.63E-04 & 1.98  & 1.64E-04 & 1.96 \\
    \hline
    $200^2$ & 4.10E-05 & 1.99  & 4.10E-05 & 1.99  & 4.10E-05 & 2.00 \\
     \hline

    \end{tabular}%
  \label{tab:accu_2_isen}%
\end{table}%

{It is demonstrated by Zhang et al. in Ref. \cite{zhang2011order} that the finite-volume WENO scheme with mid-point rule will only be second-order accurate for non-linear systems, and for obtaining high-order accuracy, a Gaussian integral rule is necessary. However, they noted that for flows involving discontinuities, which are non-smooth, the resolution characteristics are often comparable despite the obvious difference in the formal order of accuracy. In recent papers, the BVD algorithm of Sun et al. \cite{sun2016boundary} is based on the finite-volume method with a mid-point rule and therefore is only second-order accurate. Deng et al. \cite{deng2019fifth} has used \textit{primitive variables} for the reconstruction procedure of the P4T2 scheme due to the restrictions of the BVD algorithm and is only second-order accurate for non-linear problems. Fu has proposed a variant of the TENO scheme in Ref. \cite{fu2019low} using conservative variables as the choice of reconstruction variables and is also second-order accurate. References \cite{sun2016boundary, deng2019fifth, fu2019low} did not formally compute the order of accuracy and mentioned that their schemes are second-order accurate by citing Zhang et al. work Ref. \cite{zhang2011order}.

\begin{example}\label{ex:dmr} {Double Mach reflection}
\end{example}

To further assess the proposed \textcolor{black}{schemes,} the Double Mach Reflection (DMR) case of Woodward and Collela \cite{woodward1984numerical} is considered. This problem consists of a Mach $10$ unsteady planar shock wave that impinges on a 30-degree inclined surface. To avoid modeling an inclined wall, we instead inclined the shock wave, as is common practice in the literature. The computational domain is $[x,y] = [0, 3] \times [0, 1]$, using a uniform grid of $768 \times 256$. The initial conditions for the test case are:

\begin{equation}
\begin{aligned}
(\rho,u,v,p)=
\begin{cases}
&(8,\ 8.25 \cos 30^\circ,\ -8.25 \sin 30^\circ,\
116.5),\quad x<1/6+\frac{y}{\tan 60^\circ},\\
&(1.4,\ 0,\ 0,\
1),\quad\quad\quad\quad\quad\quad\quad\quad\quad\quad\quad\ x>
1/6+\frac{y}{\tan 60^\circ}.
\end{cases}
\end{aligned}
\label{eu2D_mach}
\end{equation}

\noindent The case was run until a final time, $t = 0.2$. Post-shock flow conditions were set for the left boundary. The right boundary was set as zero-gradient. The bottom boundary was set with post-shock conditions for $0.0 \leq x \leq 0.1667$ and reflecting wall conditions for $0.1667 < x \leq 4.0$. The top boundary was set with the exact solution of the time-dependent oblique shock. 

Observing Figs. \ref{fig:TENO_DMR_prim}, \ref{fig:MP5_DBp} and \ref{fig:MIGE_DBp} the vortices generated along the slip line by the Kelvin Helmholtz instability and near-wall jet can be seen. The results obtained by the MIG4 scheme \textcolor{black}{are better than the MEG6 and TENO5 schemes,} which shows the superiority of the present scheme. Figs. \ref{fig:TENO_DMR_prim} shows the numerical results using the TENO5 scheme, and there are clear oscillations near the Mach stem, which are consistent with the observations available in the literature using the WENO methodology (see Fig. 7b in \cite{balsara2016efficient}, Fig. 4 of \cite{titarev2004finite} and Fig. 10 of \cite{fu2019low}).

\begin{figure}[H]
\centering\offinterlineskip
\subfigure[TENO5]{\includegraphics[width=0.32\textwidth]{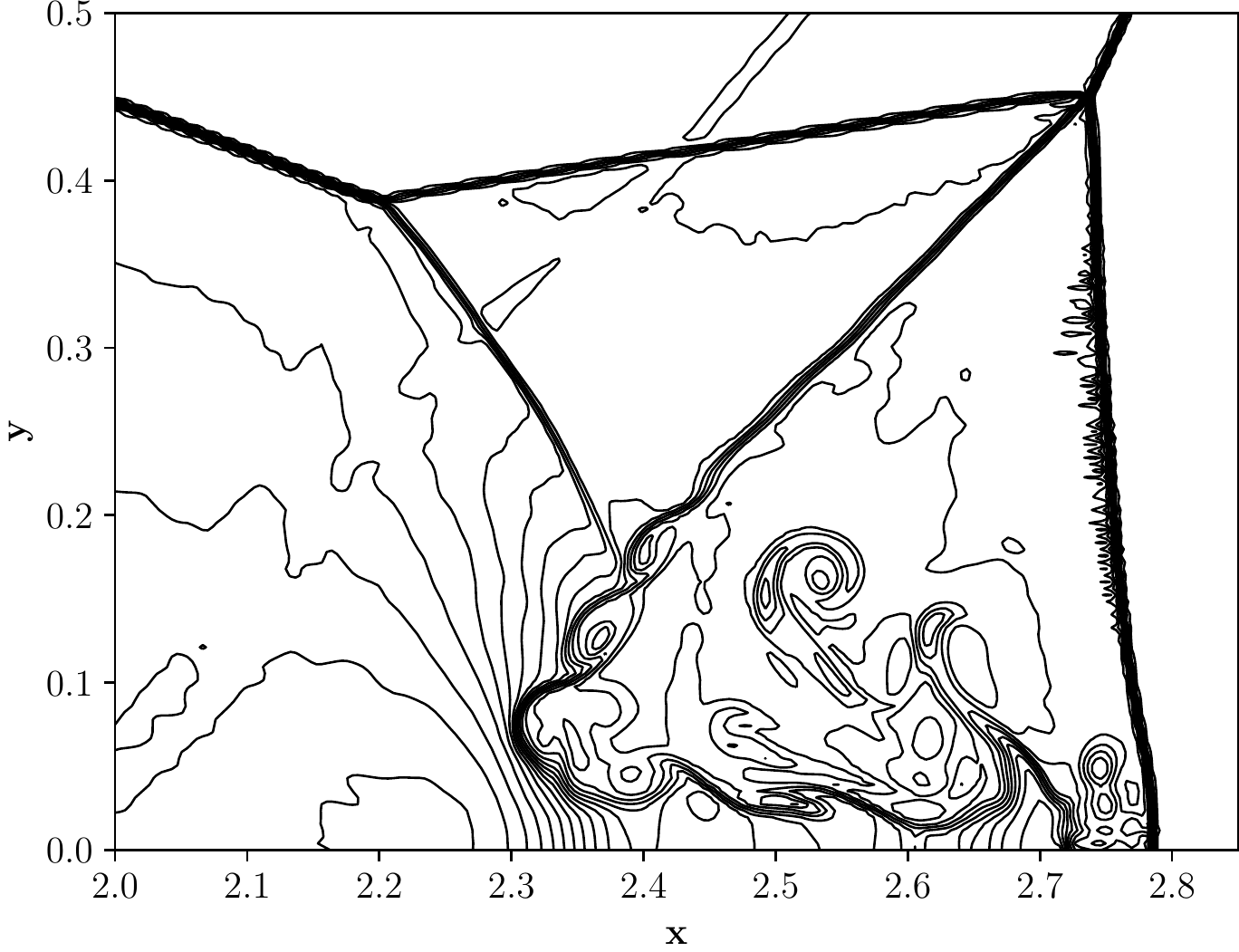}
\label{fig:TENO_DMR_prim}}
\subfigure[MEG6]{\includegraphics[width=0.32\textwidth]{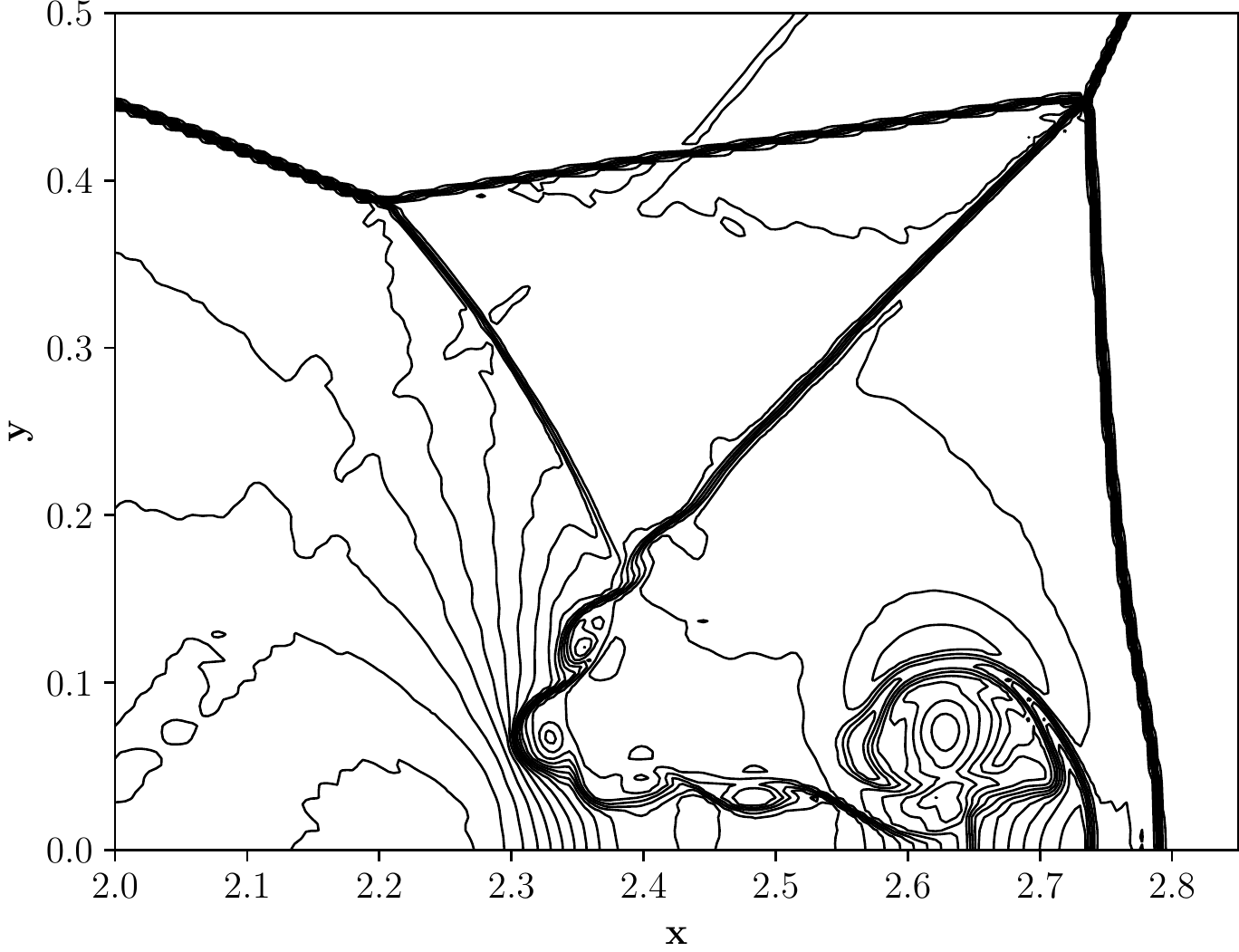}
\label{fig:MP5_DBp}}
\subfigure[MIG4]{\includegraphics[width=0.32\textwidth]{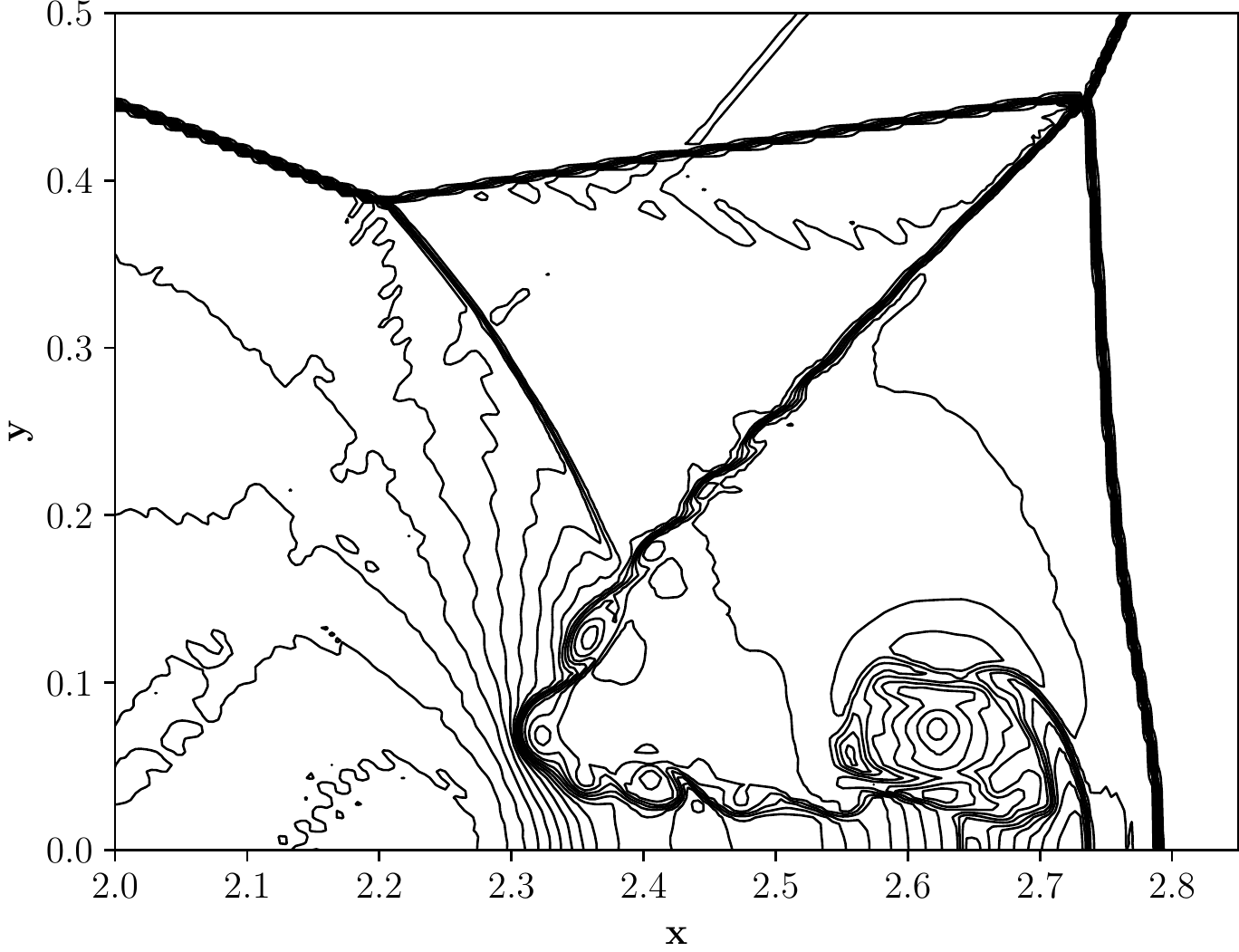}
\label{fig:MIGE_DBp}}

\caption{Computed density contours of the zoomed in Mach stem region of the Double Mach Reflection for the considered schemes. The figures are drawn with 38 contours.}
\label{fig_doublemach}
\end{figure}

\begin{example}\label{ex:TGV}{Taylor-Green Vortex} (Benefits of \textcolor{black}{gradient-based} reconstruction)
\end{example}

Next, the three-dimensional inviscid Taylor-Green vortex problem  with the following initial conditions is considered:  

\begin{equation}\label{itgv}
\begin{pmatrix}
\rho \\
u \\
v \\
w \\
p \\
\end{pmatrix}
=
\begin{pmatrix}
1 \\
\sin{x} \cos{y} \cos{z} \\
-\cos{x} \sin{y} \cos{z} \\
0 \\
100 + \frac{\left( \cos{(2z)} + 2 \right) \left( \cos{(2x)} + \cos{(2y)} \right) - 2}{16}
\end{pmatrix}.
\end{equation}

The simulations are carried out on a domain of size $x,y,z \in [0,2\pi)$ with periodic boundary conditions applied for all boundaries. The simulations are performed until time $t=10$ on a grid size of $64^3$ with the specific heat ratio as $\gamma=5/3$. The mean pressure is considerably large, so the flow problem can be considered incompressible. Therefore for this test case, only linear schemes, IG4H, EG6, and U5, are considered to evaluate the ability of different schemes to preserve kinetic energy and also the growth of enstrophy in time, i.e., the sum of vorticity of all the vortex structures, indicating the schemes ability to preserve as many structures as possible. The enstrophy (${\mathcal {E}}$) can be described as the integral of the square of the vorticity that can be computed as the integral of the magnitude of vorticity over the entire domain. 

\begin{equation}
{\mathcal {E}}(\mathbf {u} )\equiv \int _{S}|\nabla \times \mathbf {u} |^{2}\,dS.
\end{equation}

The kinetic energy evolution of all the numerical schemes is shown in Fig. \ref{fig:TGV_KE}. The unlimited linear schemes IGH4 and EG6 better preserved the kinetic energy than the U5 scheme. The velocity gradients necessary to compute the enstrophy are already available, another advantage of the proposed gradient-based reconstruction. The implicit gradient IG4H scheme has the highest enstrophy of all the considered schemes, whereas the U5 scheme is the lowest. Readers are referred to Appendix E of \cite{subramaniam2019high}\textcolor{black}{, where} the authors have shown the effect of the post-processing pipeline for velocity gradient statistics.
\begin{figure}[H]
\centering
\subfigure[Kinetic energy]{\includegraphics[width=0.45\textwidth]{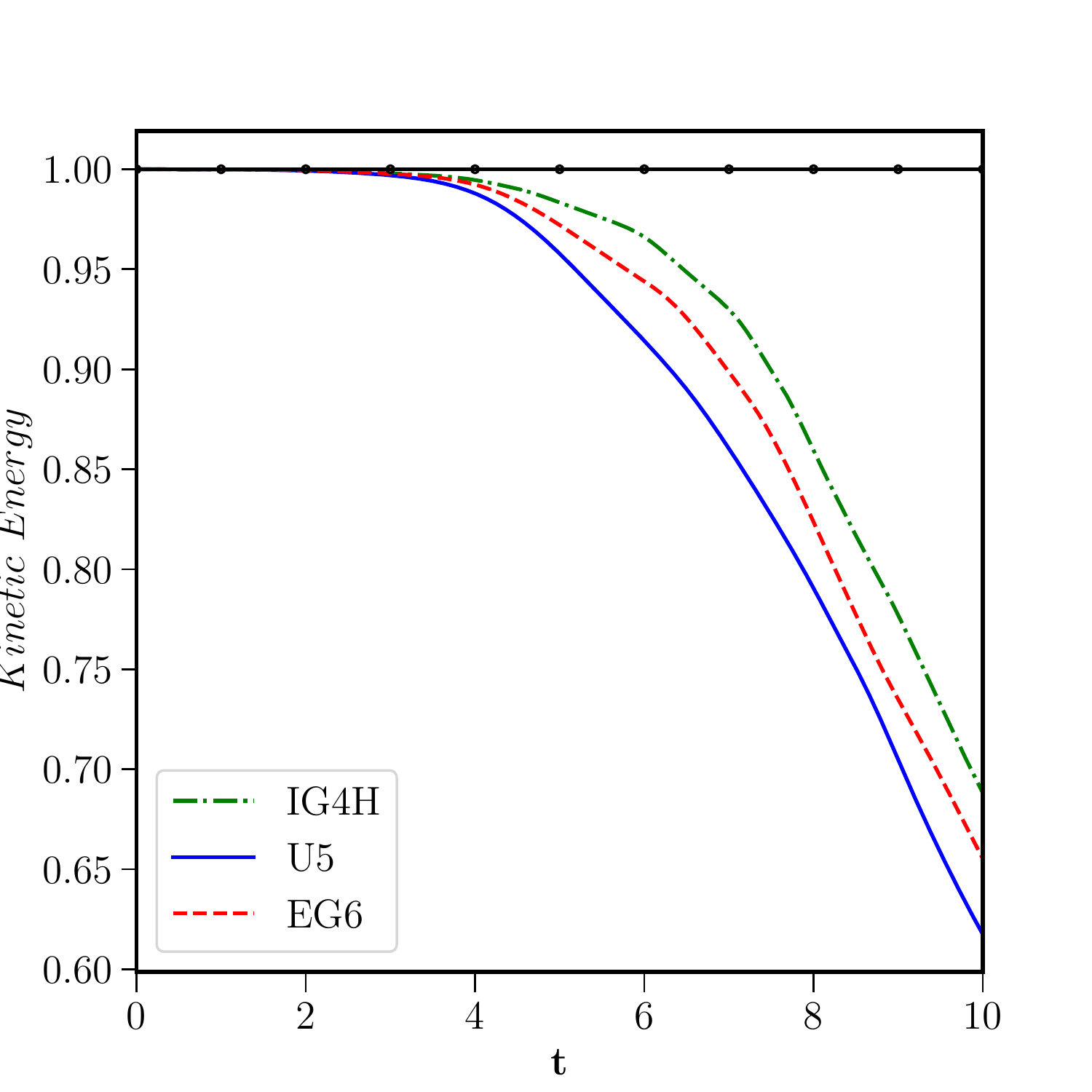}
\label{fig:TGV_KE}}
\subfigure[Enstrophy]{\includegraphics[width=0.45\textwidth]{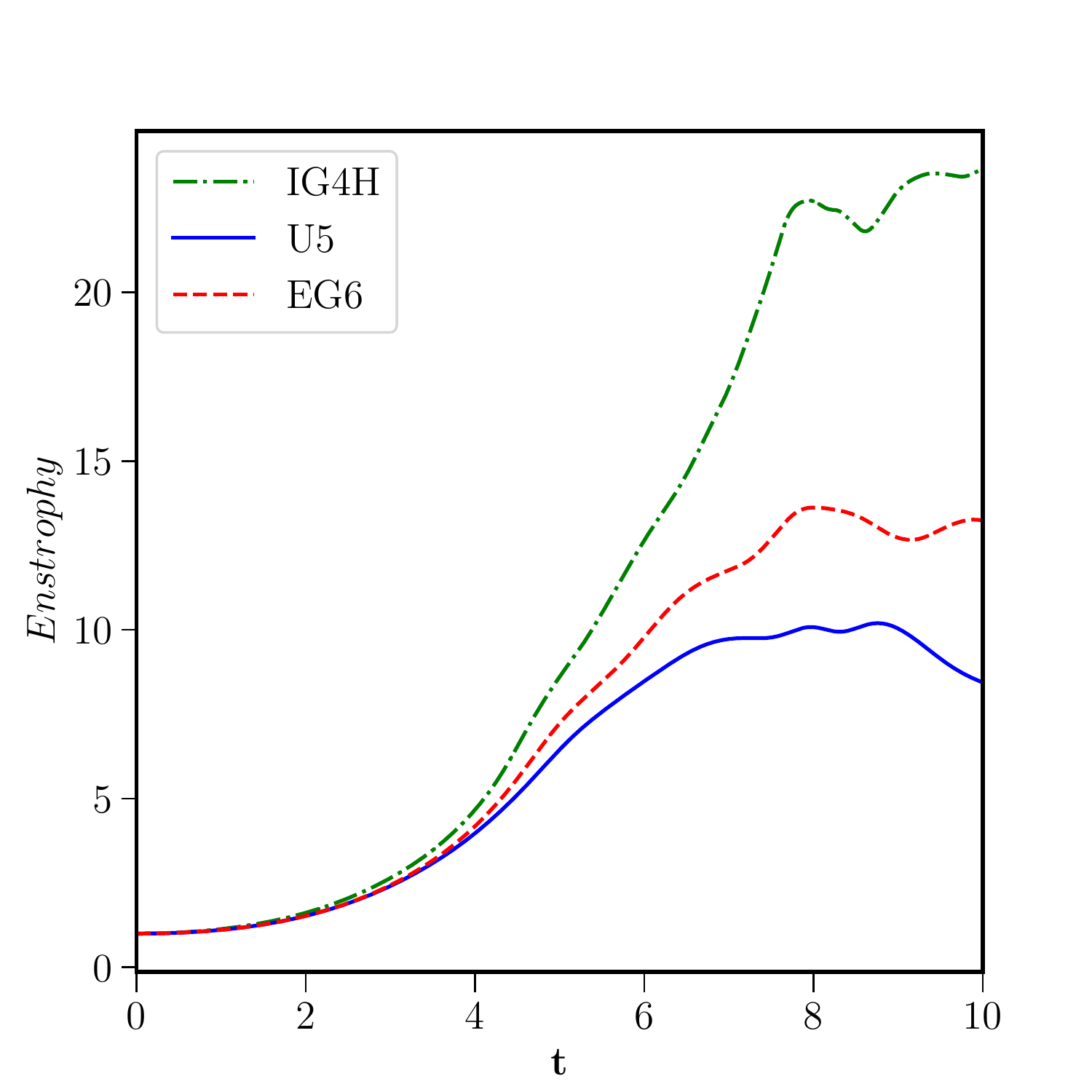}
\label{fig:TGV_ens}}
\caption{Normalised kinetic energy and enstrophy for different schemes presented in Example \ref{ex:TGV} on grid size of $64^3$. Solid line with circles: exact solution; solid blue line: U5; dash dotted green line: IG4H; dashed red line: EG6.}
\label{fig_TGV}
\end{figure}

\begin{example}\label{shock-entropy}{2D shock-entropy wave test} (Advantage of implicit gradients)
\end{example}
The two-dimensional shock-entropy wave interaction problem proposed in \cite{deng2019fifth} with the following initial conditions
\begin{align}\label{shock_entropy}
(\rho,u,p)=
\begin{cases}
&(3.857143, \ \ 2.629369,\ \ 10.3333),\quad x<-4,\\
&(1+0.2\sin(10x \cos\theta+10y\sin\theta),\ \ 0,\ \ 1),\quad otherwise,
\end{cases}
\end{align}
over a domain of $[-5,5]\times [-1,1]$ is considered in this test case. The value of $\theta$ = $\pi/6$. Simulations are carried out on a mesh size of $400\times 80$, which corresponds to $\Delta x= \Delta y =1/40$. The ``exact'' solution in Fig. \ref{fig:SSE-Compare} is computed on a very fine mesh of $1600 \times 320$ using the MP5 scheme.

 Numerical results in Fig. \ref{fig_shock_entropy} indicate that the resolution of the flow structures is significantly improved by the MIG4 scheme, Fig. \ref{fig:SE-BVD}, in comparison with the \textcolor{black}{MEG6}, and TENO5, shown in  Figs. \ref{fig:SE-HOCUS5} and \ref{fig:SE-MP5}, respectively. The local density profile in the high-frequency region along $y=0$, shown in Fig. \ref{fig:SSE-Compare}, indicates that the MIG4 approach predicts the density amplitudes with lower numerical dissipation compared to the other methods. The improved dispersion properties of the MIG4 scheme by improving the MP limiting approach can be seen in Fig. \ref{fig:SSE-Compare}. The density peaks, shown in green circles, are better captured by the MIG4 scheme over the base scheme. For this test case, the TENO5 scheme showed oscillations in the solution, shown in \ref{fig:SE-MP5}, which may indicate insufficient dissipation or deficiencies in capturing the shocks robustly.

\begin{figure}[H]
\centering\offinterlineskip
\subfigure[TENO5]{\includegraphics[width=0.48\textwidth]{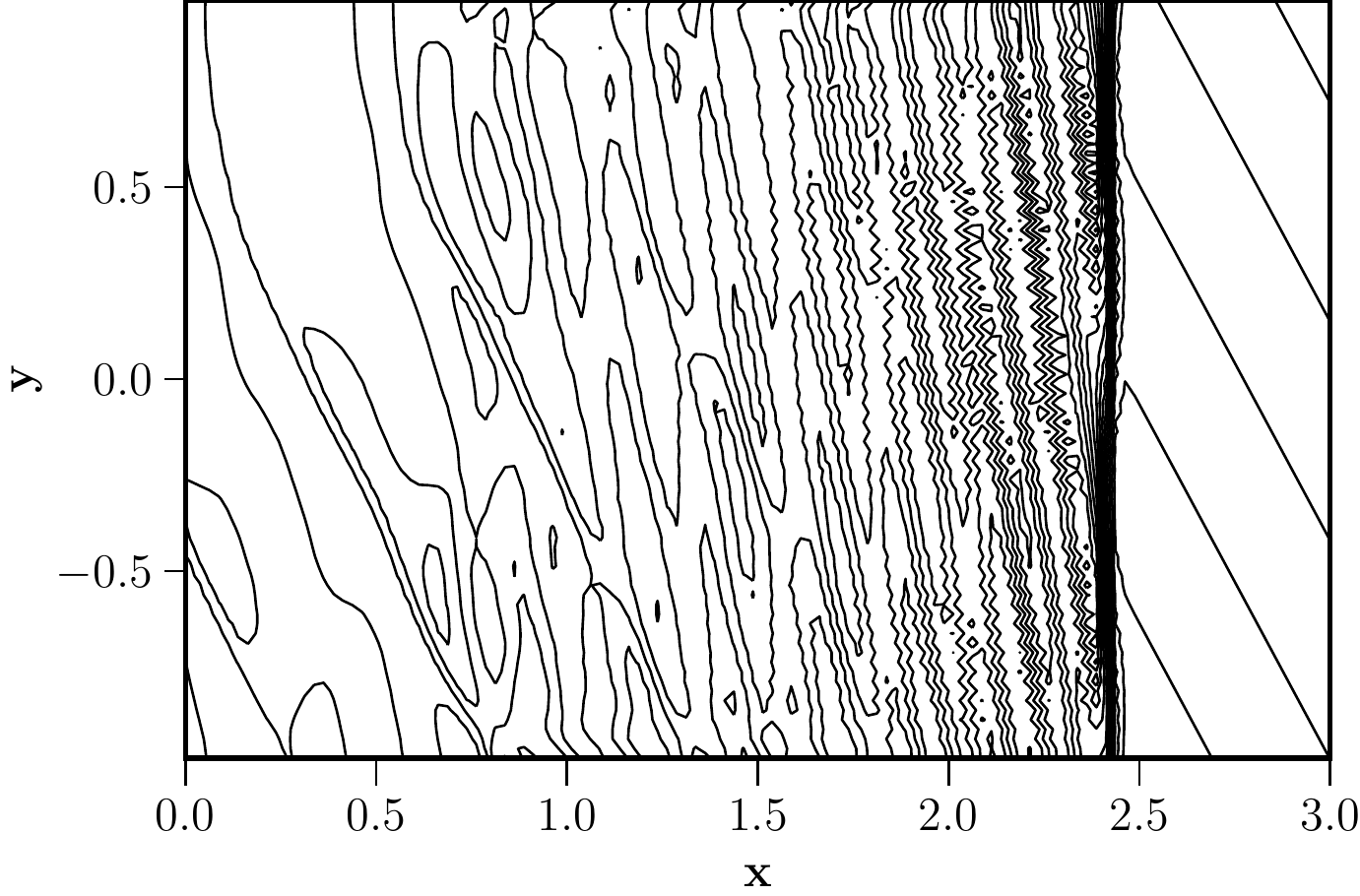}
\label{fig:SE-MP5}}
\subfigure[\textcolor{black}{MEG6 }]{\includegraphics[width=0.48\textwidth]{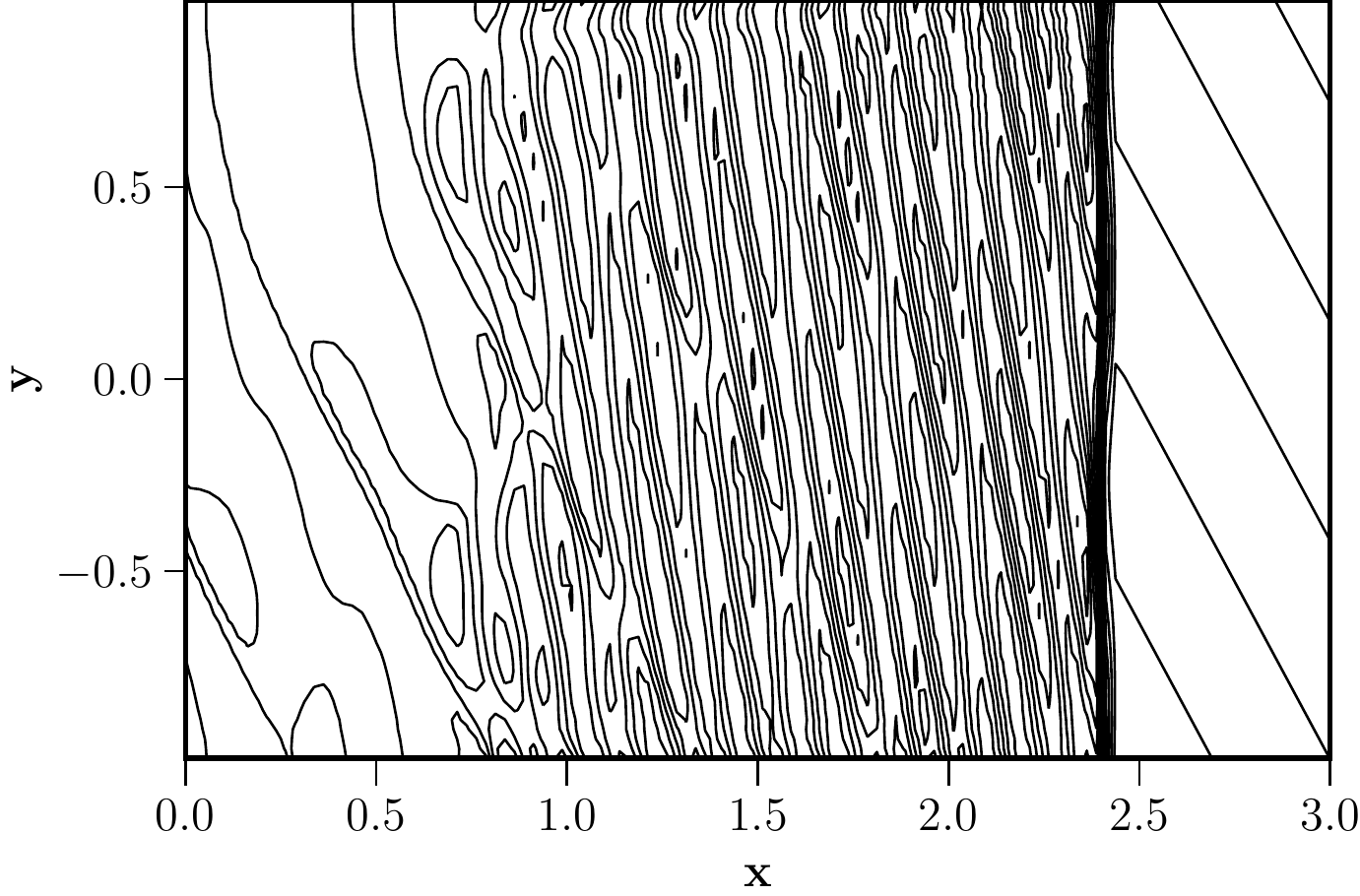}
\label{fig:SE-HOCUS5}}
\subfigure[MIG4]{\includegraphics[width=0.48\textwidth]{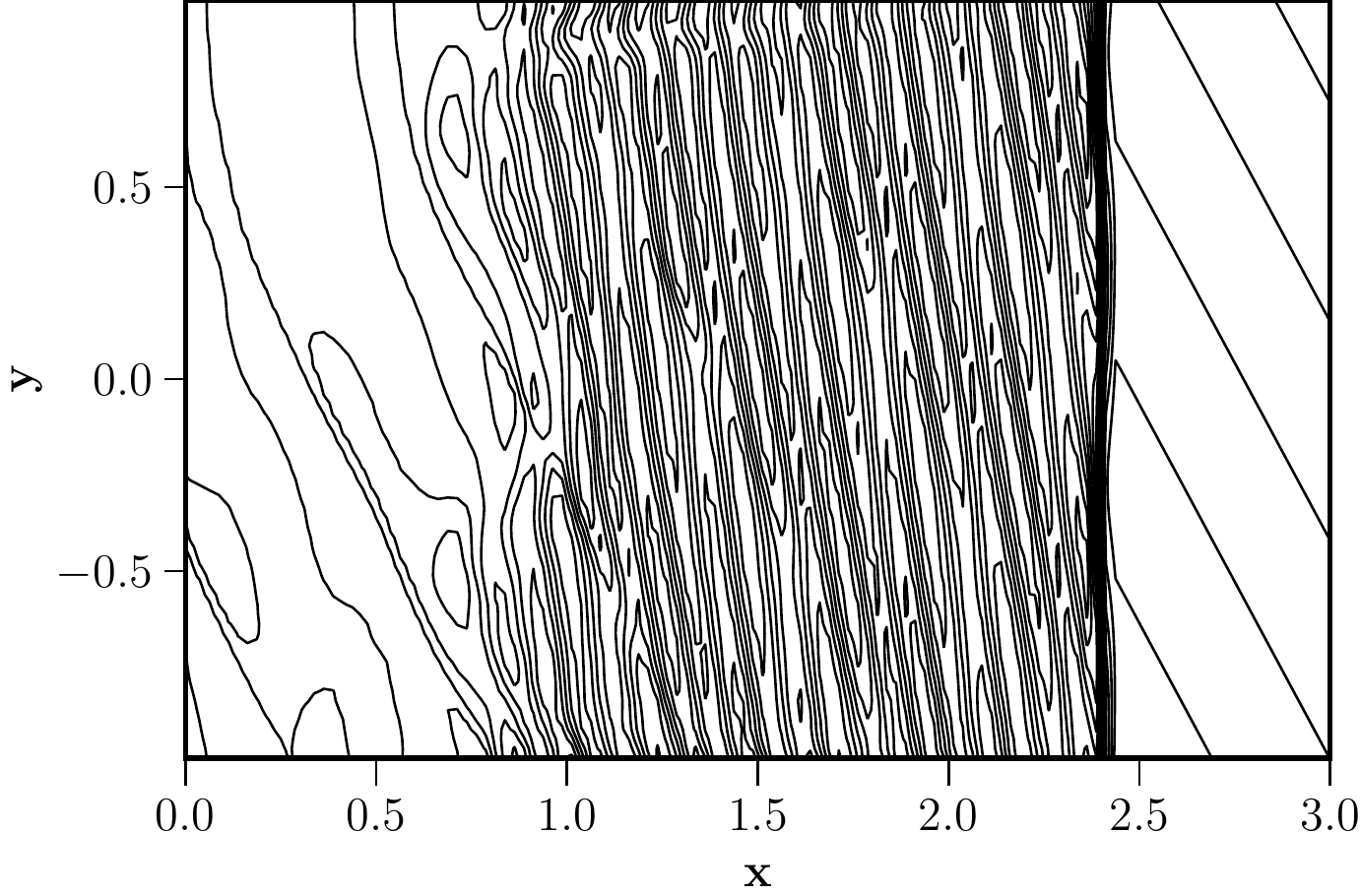}
\label{fig:SE-BVD}}
\subfigure[Local profile]{\includegraphics[width=0.495\textwidth]{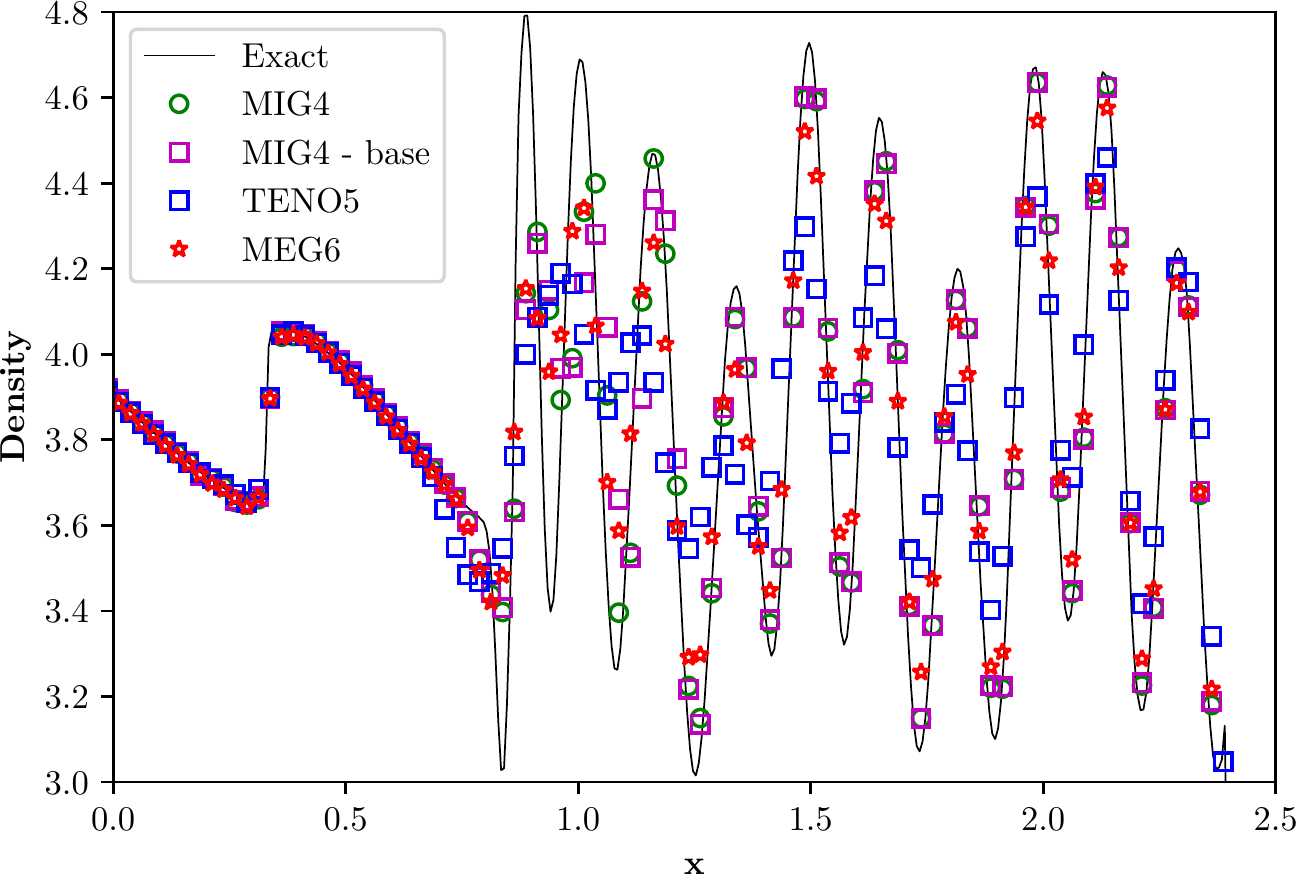}
\label{fig:SSE-Compare}}
\caption{\textcolor{black}{15 density contours for the 2D shock-entropy wave test at $t=1.8$, Example \ref{shock-entropy}, for various schemes are shown in Figs. (a), (b) and (c). Fig. (d) shows the local density in the region with high-frequency waves for all the schemes where solid  line: reference solution; green circles: MIG4; magenta squares: MIG4-base; blue squares: TENO5; red stars: MEG6.}}
\label{fig_shock_entropy}
\end{figure}

\begin{example}\label{ex:rp}{Riemann Problem} (Improvement of MP shock-capturing)
\end{example}

In this example, the two-dimensional Riemann problem taken from Schulz-Rinne et al. \cite{schulz1993numerical} is considered for testing the proposed schemes. Small-scale structures are formed for this test case due to the Kelvin-Helmholtz instability along the slip lines. The resolution of these structures commonly serves as a benchmark for numerical dissipation of a given scheme. At all four boundaries, non-reflective boundary conditions are used for these test cases. The initial conditions for the configuration considered are as follows:

\begin{equation}\label{ex:rp1}
        \left( \rho,u,v,p \right) =
        \begin{cases}
            (1.5,0,0,1.5), & \text{if } x > 0.8, y > 0.8, \\
            (33/62,4/\sqrt{11},0,0.3), & \text{if } x \leq 0.8, y > 0.8, \\
            (77/558,4/\sqrt{11},4/\sqrt{11},9/310), & \text{if } x \leq 0.8, y \leq 0.8, \\
            (33/62,0,4/\sqrt{11},0.3), & \text{if } x > 0.8, y \leq 0.8.
        \end{cases}
\end{equation}

\noindent This initial condition produces four shocks at the interfaces of the four quadrants. The case is run until a final time, $t_f=0.8$. The computational domain for this test case is $[x,y] = [0,1]\times [0,1]$ and simulations are carried out using a uniform grid of $400 \times 400$. In Figs. \ref{fig_riemann}, the density contours of the considered schemes are presented. Observing the figures, while TENO5 alone was capable of resolving small-scale structures, the present schemes, MEG6 and MIG4, can capture more structures. By comparing the Figs. \ref{fig:MIGE_mann} and \ref{fig:MP5_mann}, \textcolor{black}{one can observe that the MIG4 scheme better resolves the vortical structures and the jet-like structure compared to the MEG6}. Also, vortices are better resolved by the MIG4 scheme compared to the MIG4-base scheme, shown in Fig. \ref{fig:MIGE_mann_base}, which shows the improvements made to the MP scheme by reusing the gradients to estimate the curvature terms.

\begin{figure}[H]
\begin{onehalfspacing}
\centering\offinterlineskip
\subfigure[TENO5]{\includegraphics[width=0.4\textwidth]{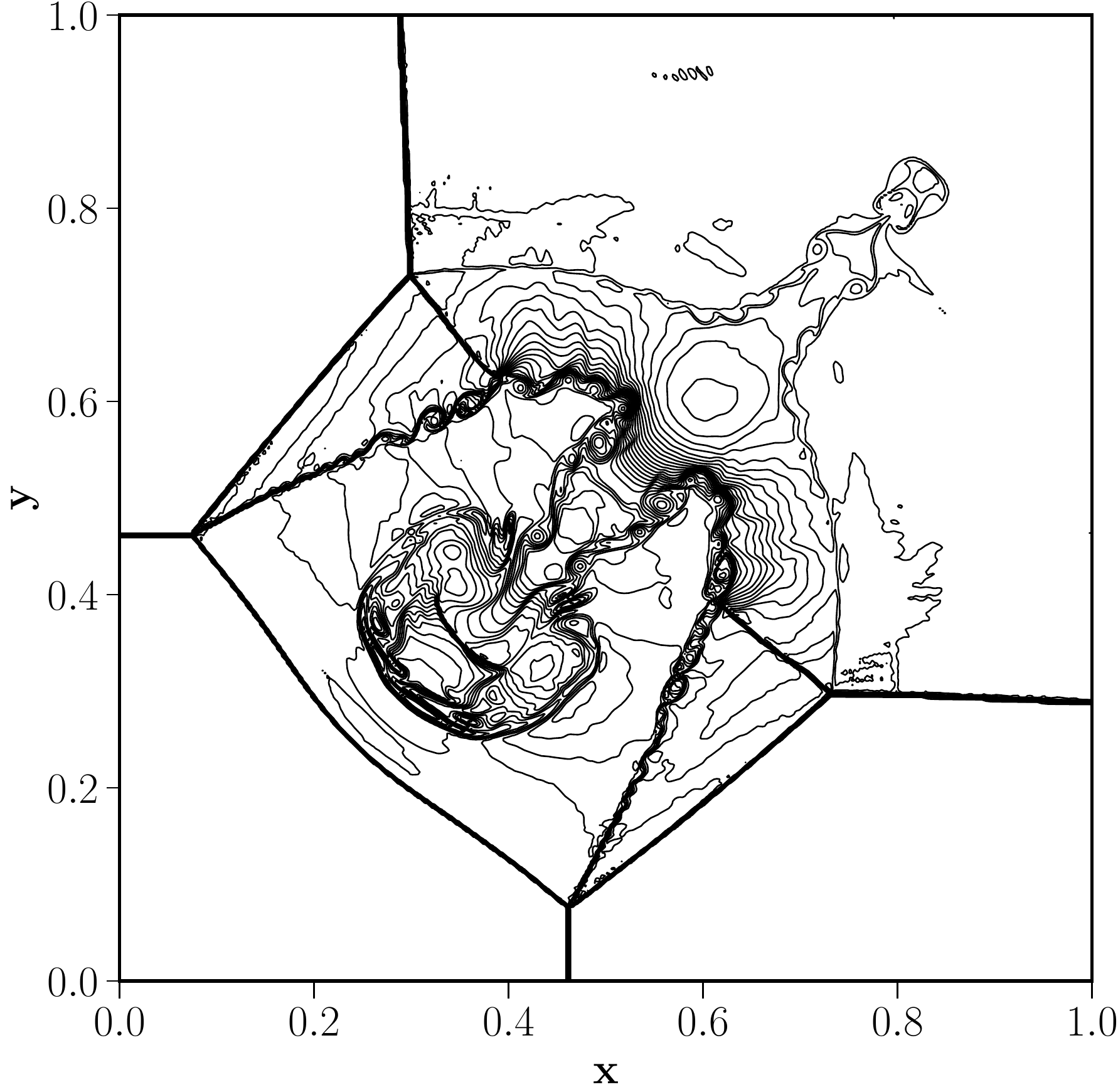}
\label{fig:TENO_prim12}}
\subfigure[MEG6]{\includegraphics[width=0.4\textwidth]{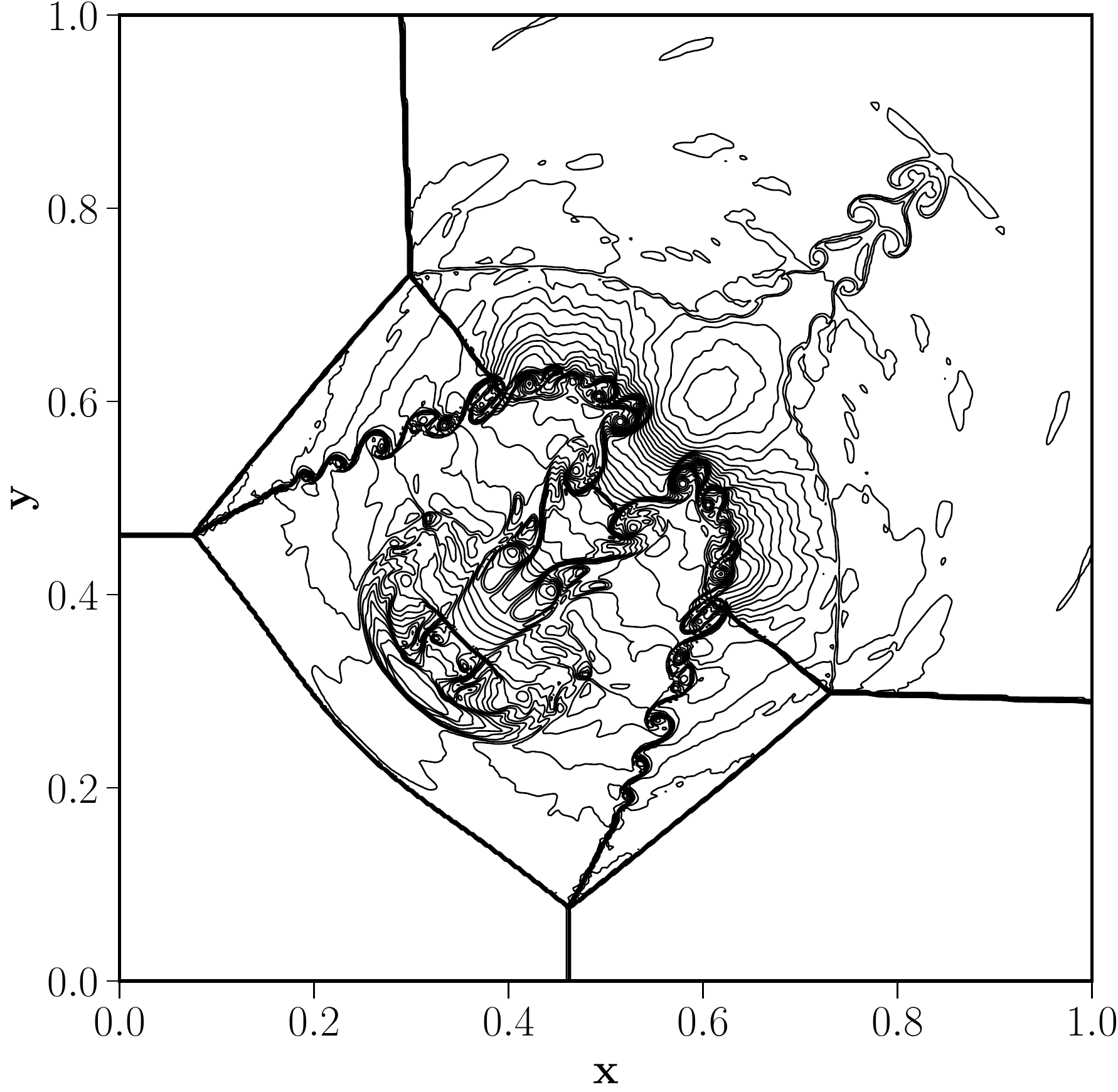}
\label{fig:MP5_mann}}
\subfigure[MIG4-base]{\includegraphics[width=0.4\textwidth]{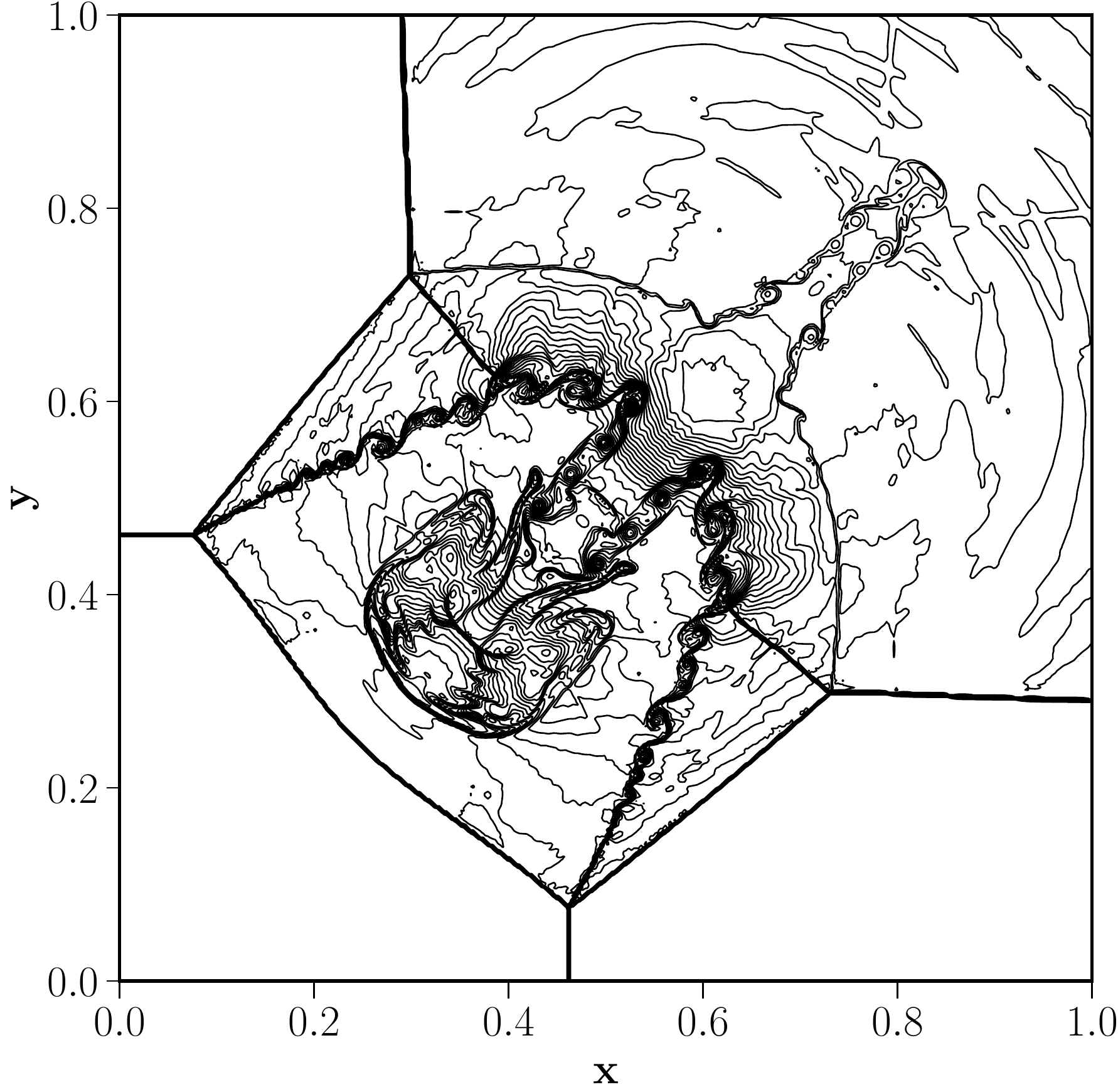}
\label{fig:MIGE_mann_base}}
\subfigure[MIG4]{\includegraphics[width=0.4\textwidth]{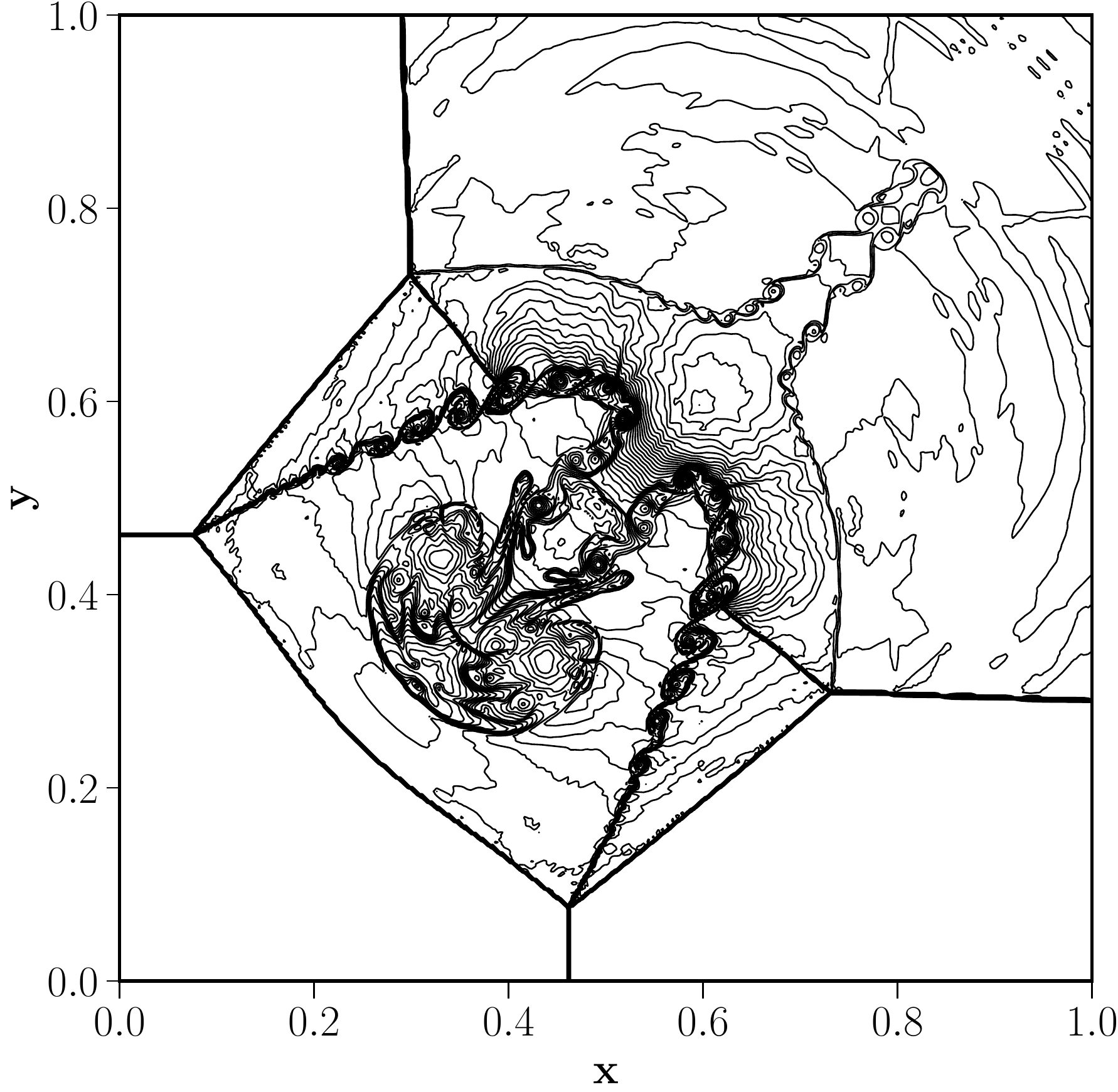}
\label{fig:MIGE_mann}}
\caption{Computed density contours of the Riemann problem described in Example \ref{ex:rp} for the considered schemes using primitive variables. The figures are drawn with 40 contours.}
\label{fig_riemann}
\end{onehalfspacing}
\end{figure}

\begin{example}\label{ex:rm} {Richtmeyer - Meshkov instability}
\end{example}

In this example, two-dimensional single-mode \textcolor{black}{Richtmeyer--Meshkov} instability problem \cite{terashima2009front} with the following initial conditions is considered:
\begin{equation*}
	\left( \rho, u, v, p \right)
 =
 \begin{cases}
 \left(5.04, 0, 0, 1 \right) , &\mbox{if x $<$ 2.9 - 0.1 sin(2$\pi$(y+0.25), perturbed interface}, \\
 	\left(1, 0, 0, 1 \right), &\mbox{if x $<$ 3.2}, \\
    \left(1.4112, -665/1556, 0, 1.628 \right), &\mbox{for $\mathrm{otherwise}$}. \\
 \end{cases}
\end{equation*}
Richtmeyer - Meshkov instability occurs when an incident shock accelerates an interface between two fluids of different densities. This test case is usually considered with multi-species. However, to assess the resolution capability of the various schemes, both the fluids are assumed to have the same specific heat ratio of $\gamma$ =1.4 as is considered in \cite{chamarthi2021high}. The computational domain has size of $\left[0, 4 \right] \times \left[ 0, 1 \right]$ and simulation is conducted until $t=9$, on the uniform mesh size of $320 \times 80$. The left and right boundary values were fixed to initial conditions, and the upper and lower boundaries were treated as periodic boundary conditions. 
\begin{figure}[H]
\centering\offinterlineskip
\subfigure[TENO5]{\includegraphics[width=0.23\textheight]{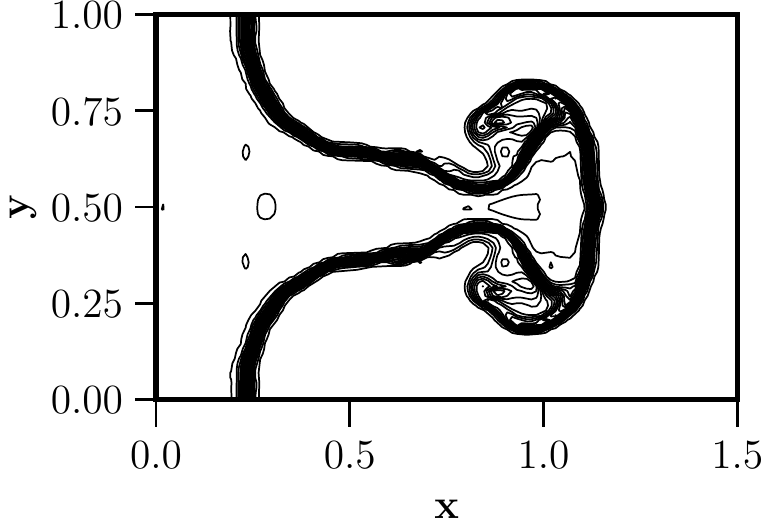}
\label{fig:TENO_RMp}}
\subfigure[MEG6]{\includegraphics[width=0.23\textheight]{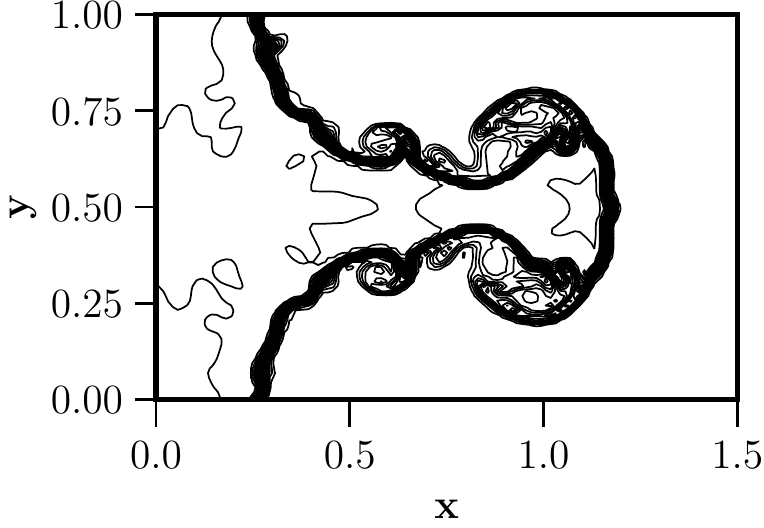}
\label{fig:MIGE_RMp-MP}}
\subfigure[MIG4]{\includegraphics[width=0.23\textheight]{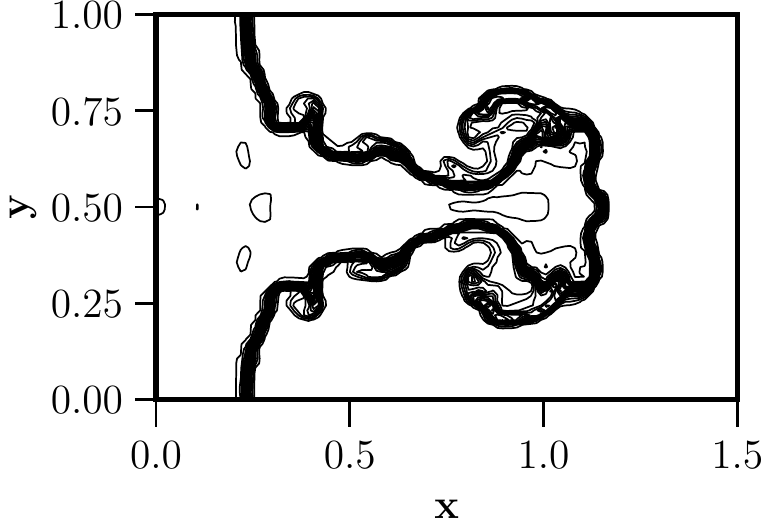}
\label{fig:MIGE_RMp}}
\caption{Numerical results of Richtmeyer - Meshkov instability described in Example \ref{ex:rm} for the considered schemes.}
\label{fig_RM}
\end{figure} 

The density distribution contours computed using different schemes are shown in Fig. \ref{fig_RM}. One can observe that the MIG4 scheme has lower numerical dissipation and produces \textcolor{black}{small-scale} roll-up vortices due to the instability compared to the MEG6 scheme, which is, in turn, better than the TENO5 scheme. Surprisingly TENO5 scheme showed no roll-up vortices at all. The material interface is also thinner for the MIG4 scheme, which is also observed in the multi-component flow test case and will be shown later in Example \ref{ex:RM-viscous}.

\subsection{One-dimensional inviscid single species test cases}\label{1D-euler}
\begin{example}\label{sod}{Shock tube problems}
\end{example}

The first of the one-dimensional inviscid cases is the Sod shock tube problem \cite{sod1978survey}, which is a one-dimensional Riemann problem with initial conditions:
\begin{equation}
        \left( \rho,u,p \right) =
        \begin{cases}
            (1,0,1), & \text{if } 0 \leq x < 0.5, \\
            (0.125,0,0.1), & \text{if } 0.5 \leq x \leq 1.
        \end{cases}
\end{equation}

\noindent The case was computed on a computational domain $x = [0,1]$ using $N = 200$ uniformly distributed grid points until a final time, $t = 0.2$. Observing Fig. \ref{fig_sod}, all employed schemes  clearly capture the rarefaction wave, contact discontinuity, and the shock wave.
\begin{figure}[H]
\centering
\subfigure[Computed density profiles.]{\includegraphics[width=0.4\textwidth]{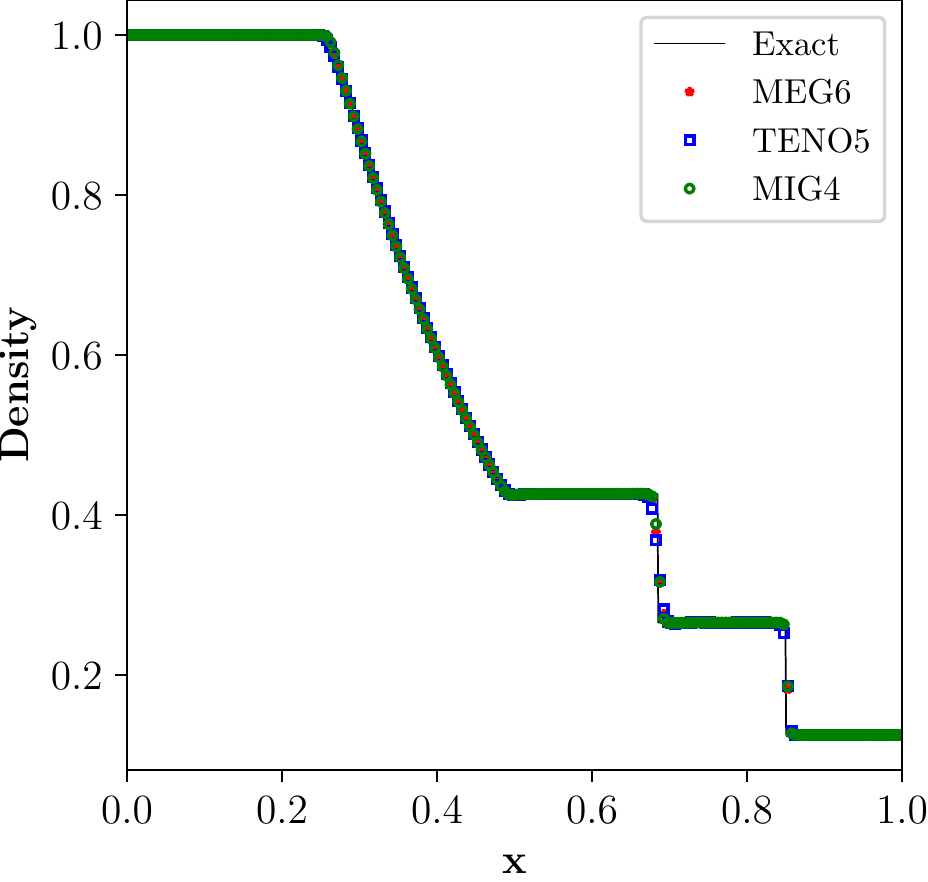}
\label{fig:sod-den}}
\subfigure[Computed velocity profiles.]{\includegraphics[width=0.4\textwidth]{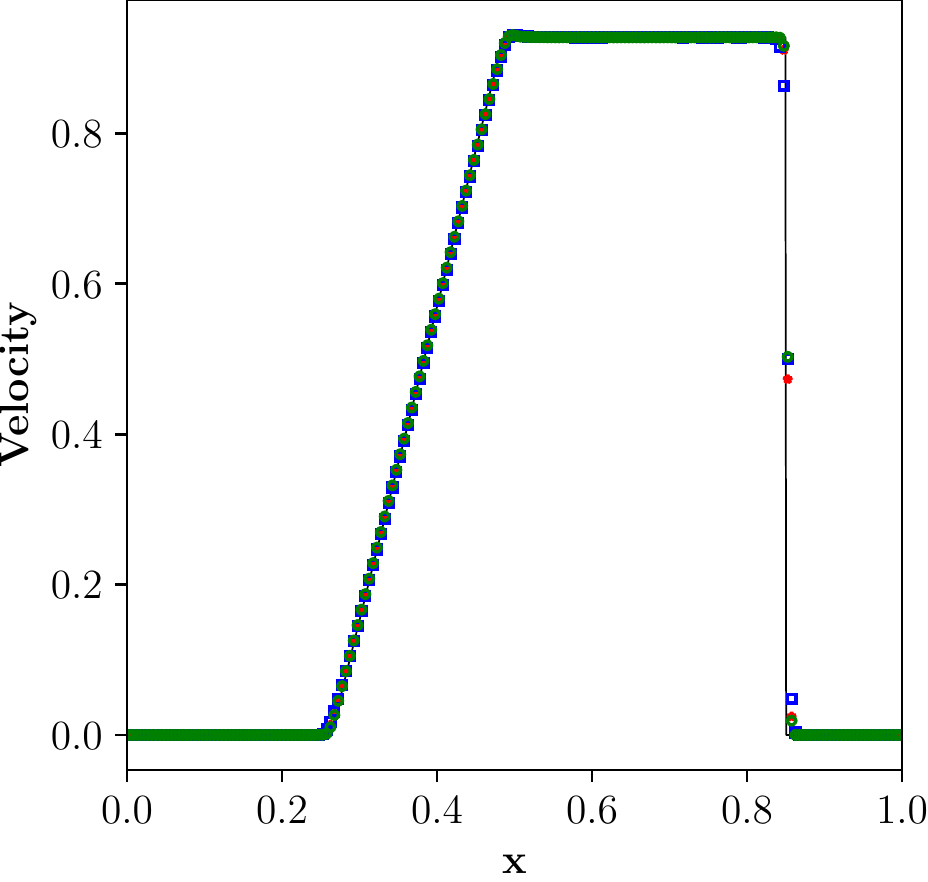}
\label{fig:sod-vel}}
\caption{Numerical solution for Sod shock tube problem using $N = 200$ grid points at $t = 0.2$, for Sod test case of Example \ref{sod}.}
\label{fig_sod}
\end{figure}

The second one-dimensional inviscid case is the Lax problem \cite{lax1954weak}, which has initial conditions:
\begin{equation}\label{lax}
        \left( \rho,u,p \right) =
        \begin{cases}
            (0.445,0.698,3.528), & \text{if } 0 \leq x < 0.5, \\
            (0.5,0,0.571), & \text{if } 0.5 \leq x \leq 1. 
        \end{cases}
\end{equation}
\noindent The case was computed on a computational domain $x = [0,1]$ using $N = 200$ uniformly distributed grid points until a final time, $t = 0.14$. The results are shown in Fig. \ref{fig_lax}, with all schemes performing well and producing no numerical oscillations near discontinuities. For both the test cases, the discontinuities are captured in fewer cells by the MIG4 scheme.
\begin{figure}[H]
\centering
\subfigure[Computed density profiles. ]{\includegraphics[width=0.45\textwidth]{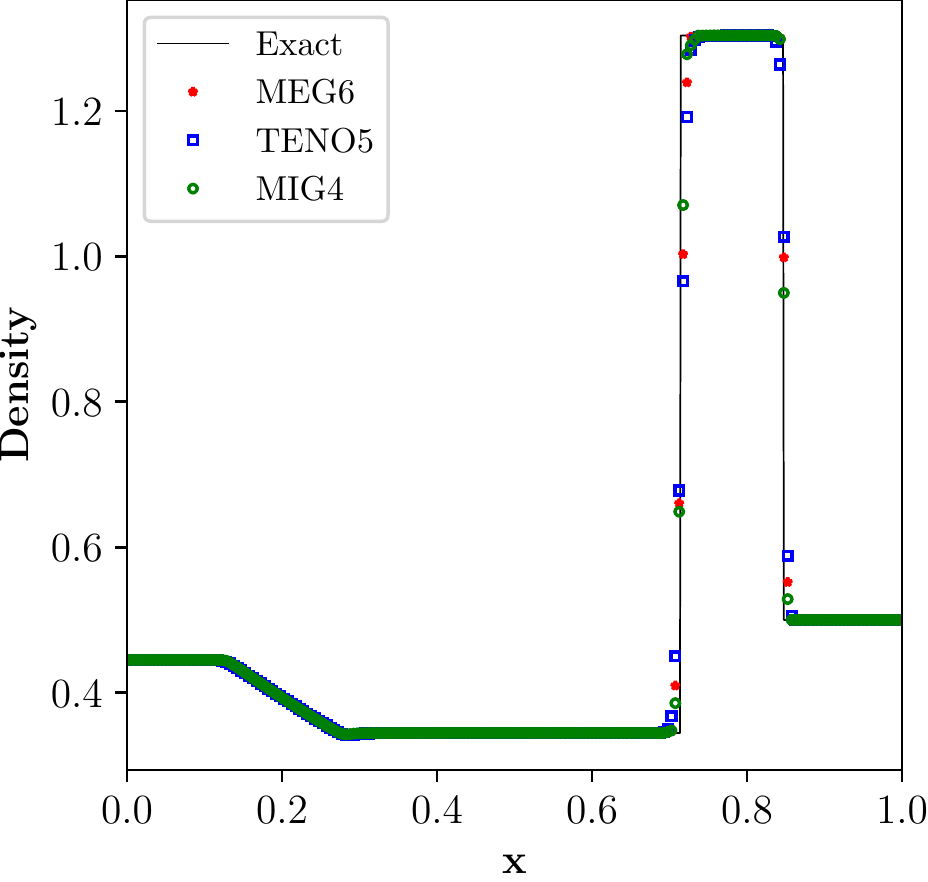}
\label{fig:lax-den}}
\subfigure[Computed velocity profiles.]{\includegraphics[width=0.45\textwidth]{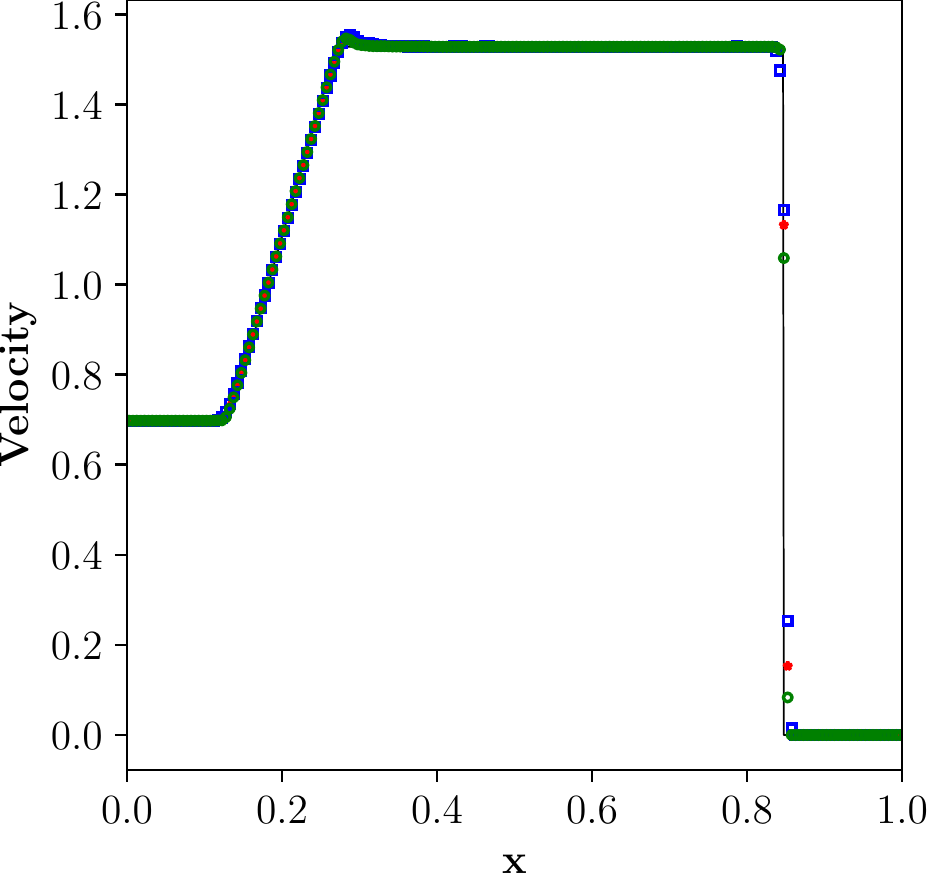}
\label{fig:lax-vel}}
\caption{Numerical solution for Lax problem using $N = 200$ grid points at $t = 0.14$, for Lax test case of Example \ref{sod}.}
\label{fig_lax}
\end{figure}

\begin{example}\label{leblanc}{Le Blanc problem}
\end{example}
\textcolor{black}{The Le Blanc problem \cite{loubere2005subcell}, an extreme shock-tube problem, with the following initial conditions, is considered in this example:}

\begin{align}\label{blanc_prob}
(\rho,u,p)=
\begin{cases}
&(1.0\ \ ,\ \ 0,\ \ \frac{2}{3}\times10^{-1}\ \ ),\quad 0<x<3.0,\\
&(10^{-3},\ \ 0,\ \ \frac{2}{3}\times10^{-10}),\quad 3.0<x<9,
\end{cases}
\end{align}

The exact solution is computed by an exact Riemann solver \cite{toro2009riemann} and the specific heat ratio for this test case is $\frac{5}{3}$. Numerical results for density and velocity obtained for MEG6, \textcolor{black}{MIG4,} and TENO5 schemes on a $N=900$ grid for the final time $t=6$ are shown in Fig. \ref{fig_blanc}. Density profiles obtained by the TENO5 scheme show clear oscillations with larger amplitude, whereas the solutions obtained by the MEG6 and MIG4 schemes are smoother in comparison. Oscillations in the TENO5 scheme are consistent with the results in the literature, see Fig. 19 of \cite{li2021low}, where the MP scheme is used to filter the oscillations of the TENO8A scheme. The results for velocity profile, zoom region of Fig. \ref{fig:blanc-vel}, are far more striking as the MIG4 scheme has minimal oscillations compared to the TENO5 and MEG6, respectively. Moreover, the discontinuity at the location $x$ = 8 is captured within a cell by the MIG4 scheme, unlike the TENO5 scheme.

These results indicate that the proposed scheme, MIG4, is capable of resolving small scale features as shown in earlier cases and is also robust compared to the TENO5 scheme.

\begin{figure}[H]
\centering
\subfigure[Density]{\includegraphics[width=0.43\textwidth]{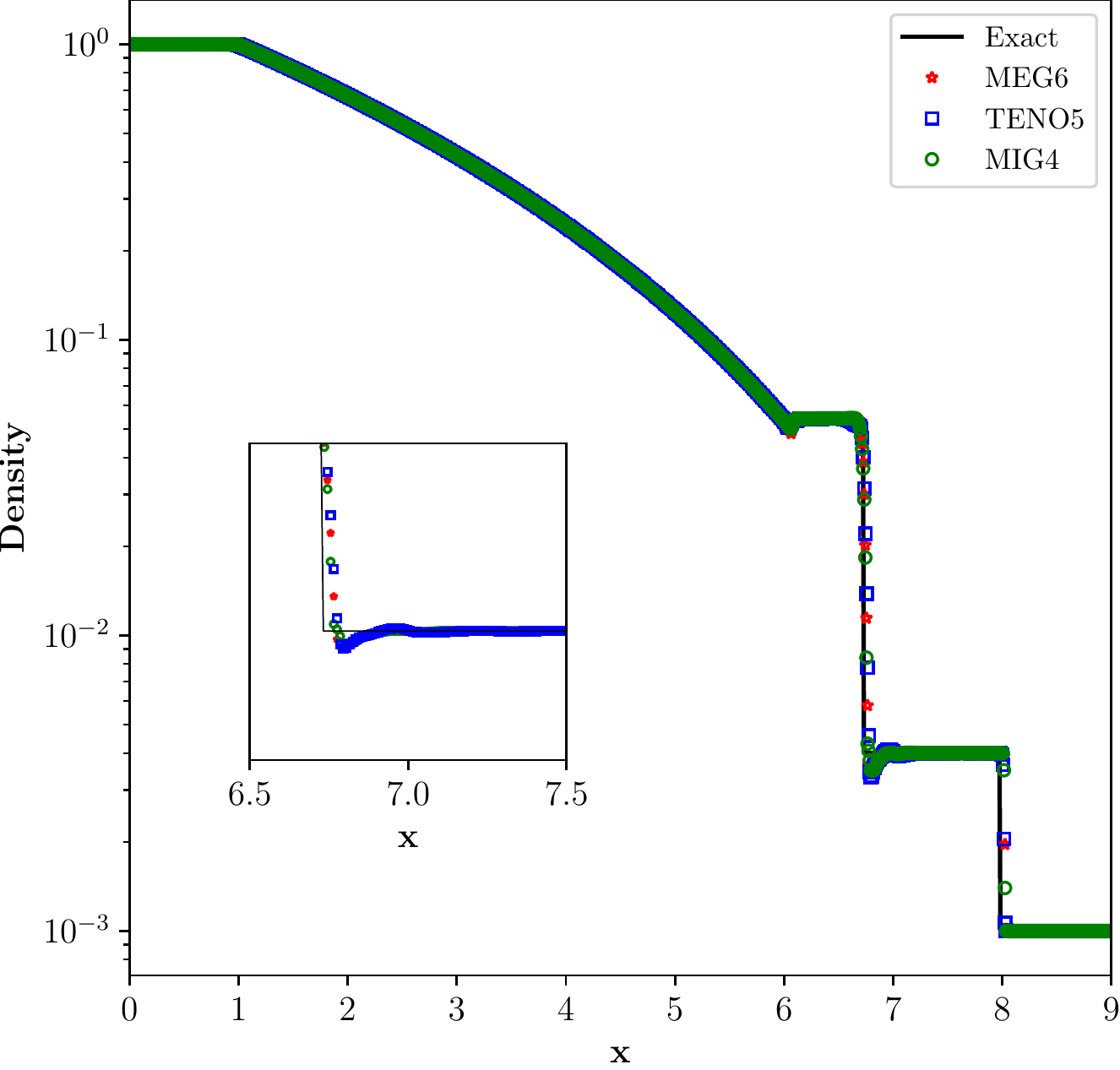}
\label{fig:blanc-den}}
\subfigure[Velocity]{\includegraphics[width=0.42\textwidth]{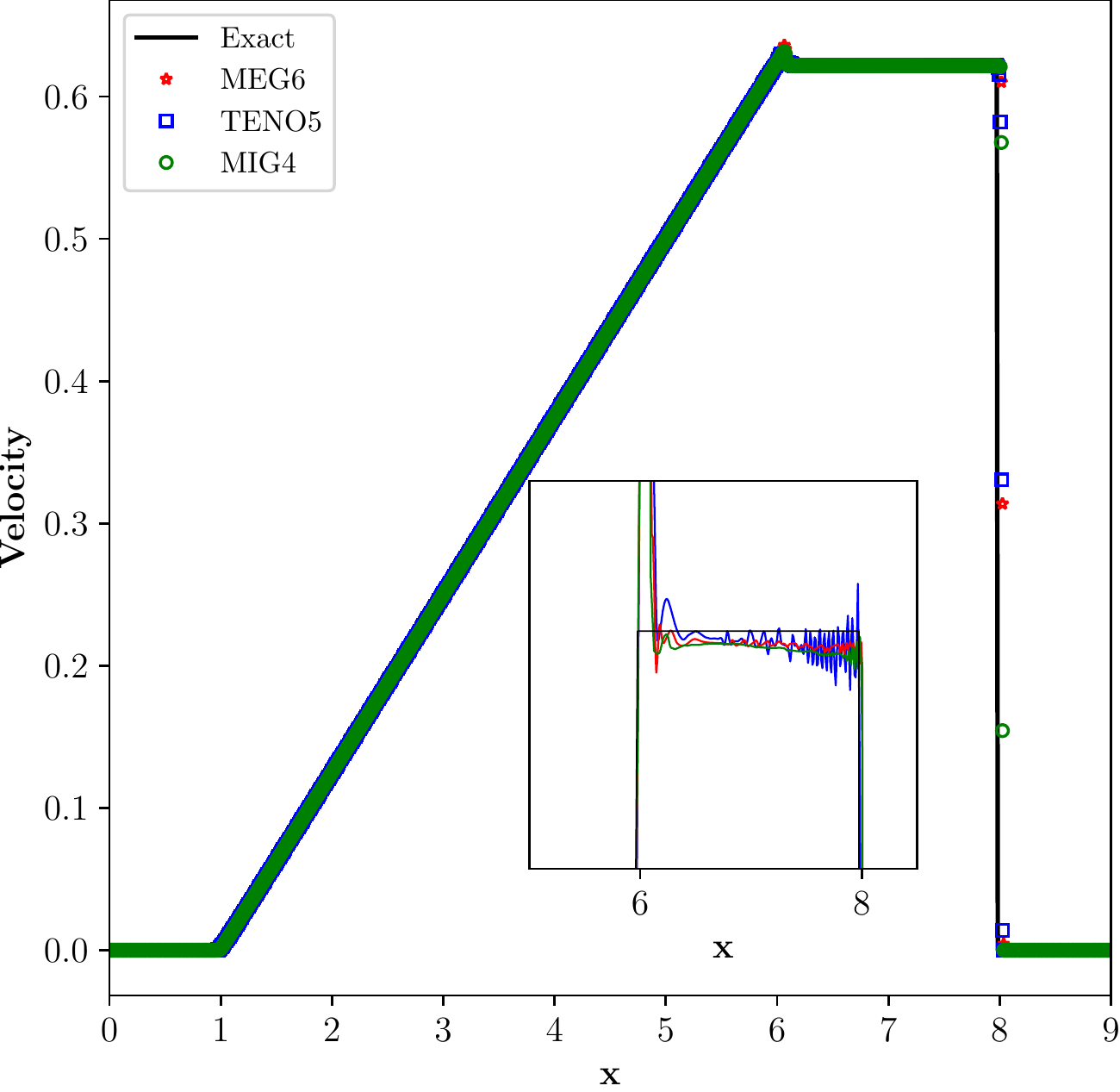}
\label{fig:blanc-vel}}
\caption{\textcolor{black}{Numerical solution for Le Blanc problem with the initial conditions given by Equation\ref{blanc_prob}}}
\label{fig_blanc}
\end{figure}

\begin{example}\label{Shu-Osher}{Shu-Osher problem}
\end{example}

The third one-dimensional inviscid test case considered was the Shu-Osher problem \cite{Shu1988}, with initial conditions:

\begin{equation}
        \left( \rho,u,p \right) =
        \begin{cases}
            (3.857,2.629,10.333), & \text{if } -5 \leq x < -4, \\
            (1+0.2\sin(5(x-5)),0,1), & \text{if } -4 \leq x \leq 5.
        \end{cases}
\end{equation}

\noindent The case was solved on a computational domain $x = [-5,5]$ with $N = 300$ uniformly distributed grid points until a final time, $t = 1.8$. The Shu-Osher problem is a common case to evaluate a \textcolor{black}{scheme's} one-dimensional shock and disturbance capturing capabilities. As shown in Fig. \ref{fig:shu-global}, the presented schemes match the reference solution well. Observing Fig. \ref{fig:shu-local}, MEG6 and MIG4 perform slightly better than TENO5 at the density extrema.

\begin{figure}[H]
\centering 
\subfigure[Global density profiles.]{\includegraphics[width=0.41\textwidth]{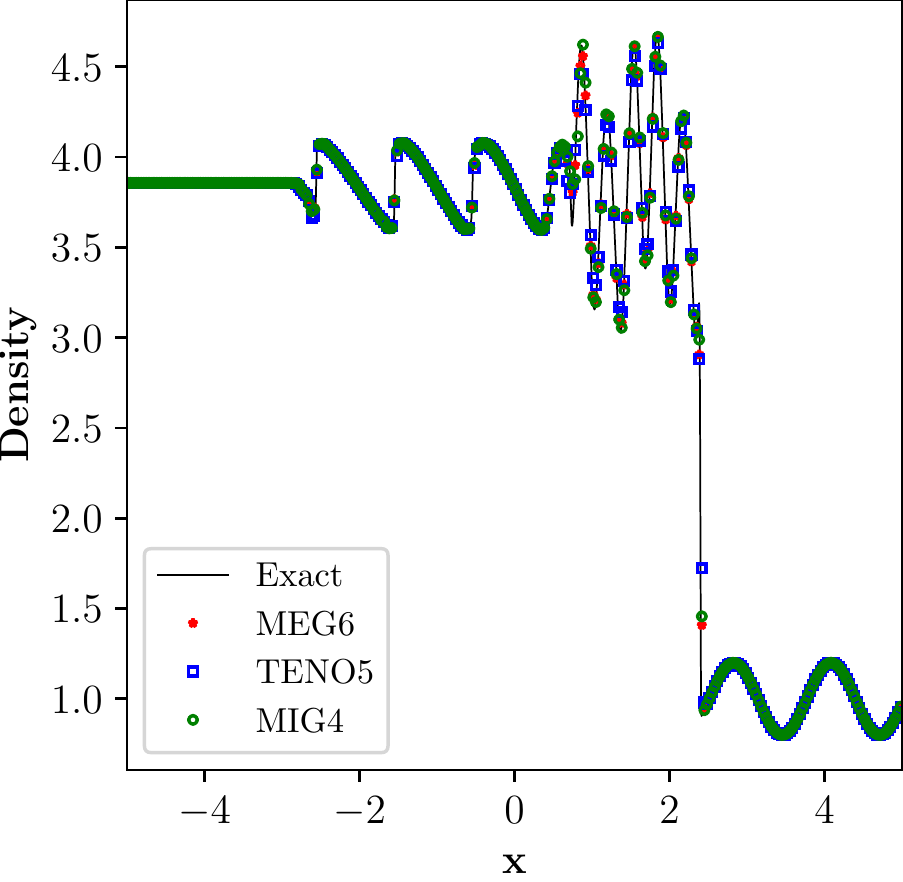}
\label{fig:shu-global}}
\subfigure[Local density profiles.]{\includegraphics[width=0.42\textwidth]{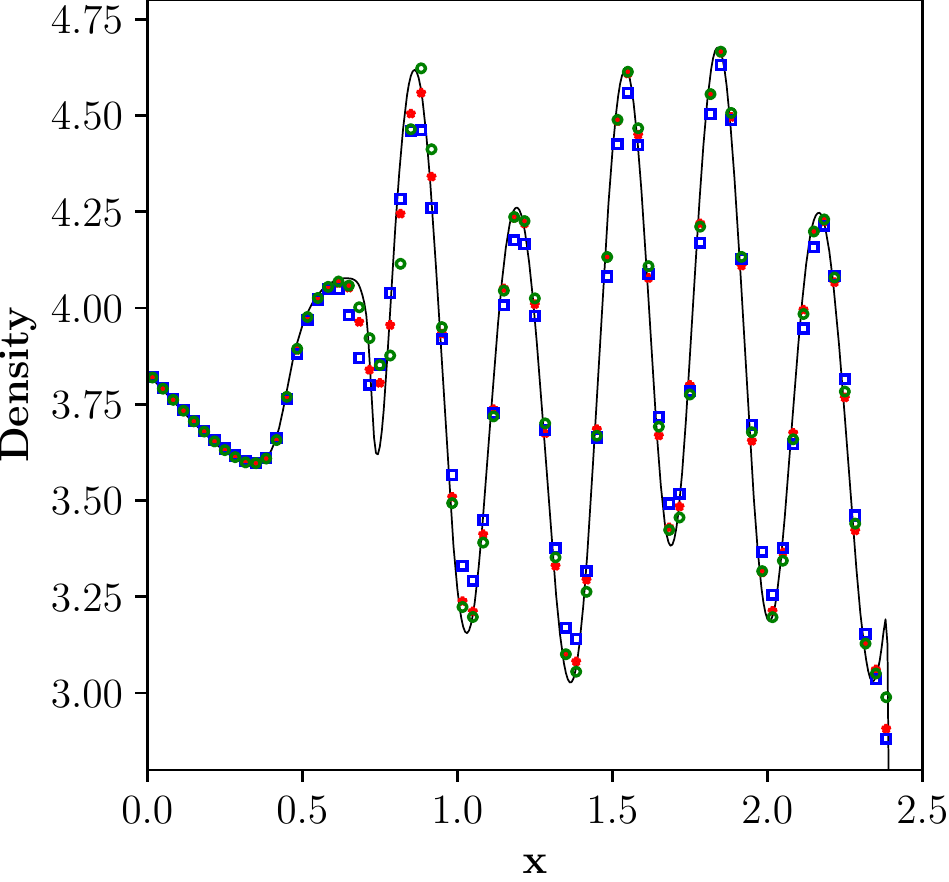}
\label{fig:shu-local}}
\caption{Numerical solution for Shu-Osher problem using $N = 300$ grid points at $t = 1.8$, Example \ref{Shu-Osher}.}
\label{fig:1d-SO}
\end{figure}

\newpage
\begin{example}\label{Titarev-Toro}{Titarev-Toro problem}
\end{example}

The fourth one-dimensional inviscid test case considered is the Titarev-Toro problem \cite{titarev2004finite}, with initial conditions:
\begin{equation}
        \left( \rho,u,p \right) =
        \begin{cases}
            (1.515695,0.523346,1.805), & \text{if } 0 \leq x < 0.5, \\
            (1+0.1\sin(20\pi(x-5)),0,1), & \text{if } 0.5 \leq x \leq 10.
        \end{cases}
\end{equation}

\noindent The case was solved on a computational domain $x = [0,10]$ with $N = 1000$ uniformly distributed grid points until a final time, $t = 5$.  In this problem, a shock interacts with a highly oscillatory wave, with extrema that are very difficult for a scheme to capture. Observing Fig. \ref{fig:tita2}, it is clear that TENO5 severely under-resolves the fluctuations compared to MEG6 and MIG4. Both \textcolor{black}{gradient-based} reconstructions resolved the fluctuations well, with MIG4 capturing the extrema slightly better than MEG6.

\begin{figure}[H]
\centering
\subfigure[Global density profiles.]{\includegraphics[width=0.48\textwidth]{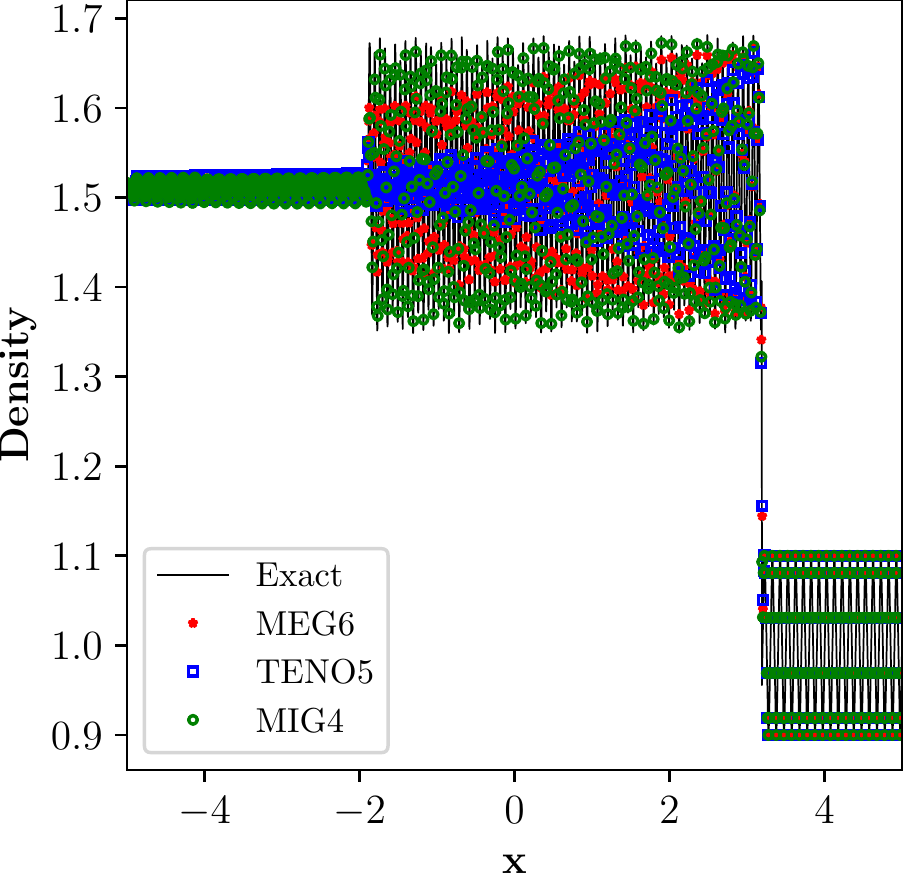}
\label{fig:tita1}}
\subfigure[Local density profiles.]{\includegraphics[width=0.48\textwidth]{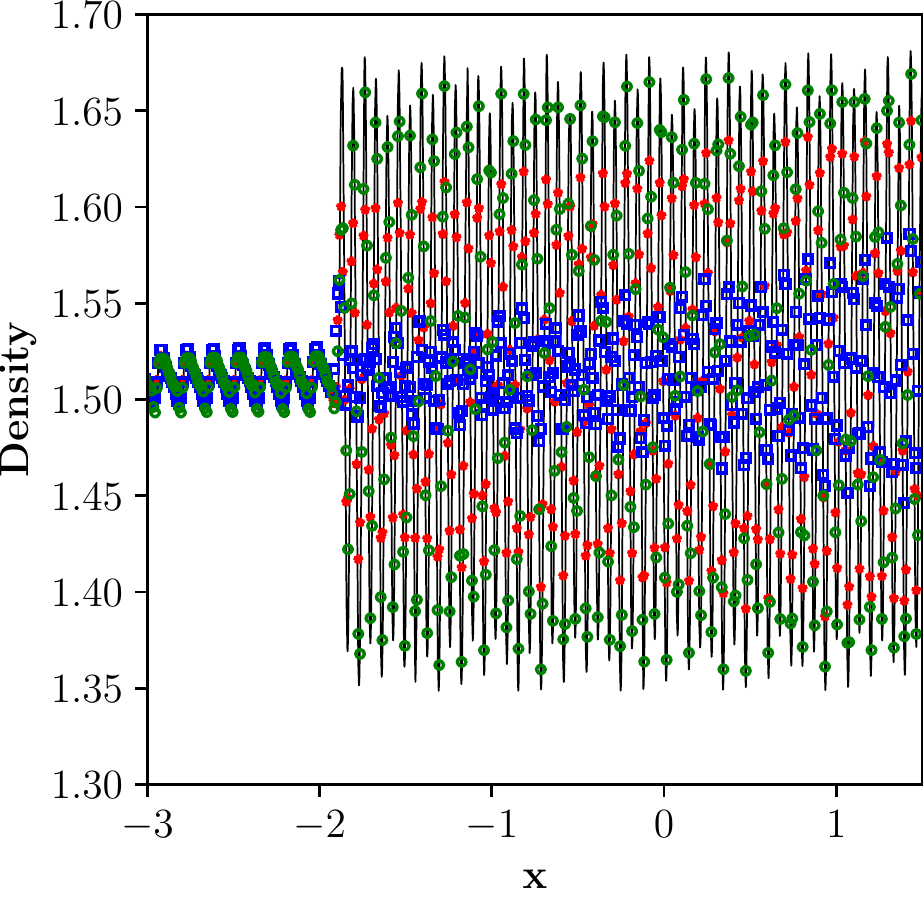}
\label{fig:tita2}}
\caption{Numerical solution for Titarev-Toro problem using $N = 1000$ grid points at $t = 5$, Example \ref{Titarev-Toro}. }
\label{fig_tita}
\end{figure}

\begin{example}\label{blast}{Blast wave problem}
\end{example}

The fifth one-dimensional inviscid test case considered was the blast wave problem \cite{woodward1984numerical}, with initial conditions:

\begin{equation}
        \left( \rho,u,p \right) =
        \begin{cases}
            (1,0,1000), & \text{if } 0.0 \leq x < 0.1, \\
            (1,0,0.01), & \text{if } 0.1 \leq x < 0.8, \\
            (1,0,100), & \text{if } 0.8 \leq x \leq 1.0.
        \end{cases}
\end{equation}

\noindent The case was solved on a computational domain $x = [0,1]$ with $N = 600$ uniformly distributed grid points until a final time, $t = 0.038$. Observing Fig. \ref{fig:blast-400}, all considered schemes perform well globally. When looking at the local density profile in Fig. \ref{fig:blast-local}, the density extrema is best predicted by MIG4 and MEG6, with both giving virtually identical results. However, both capture the extrema more accurately than TENO5 alone. The discontinuity in velocity, at $x$$\approx$0.87, is captured within a cell by the MIG4 scheme, whereas the TENO5 and MEG6 schemes require more cells, indicating the low dissipation property of the implicit gradient approach.
\begin{figure}[H]
\centering
\subfigure[Global density profiles.]{\includegraphics[width=0.4\textwidth]{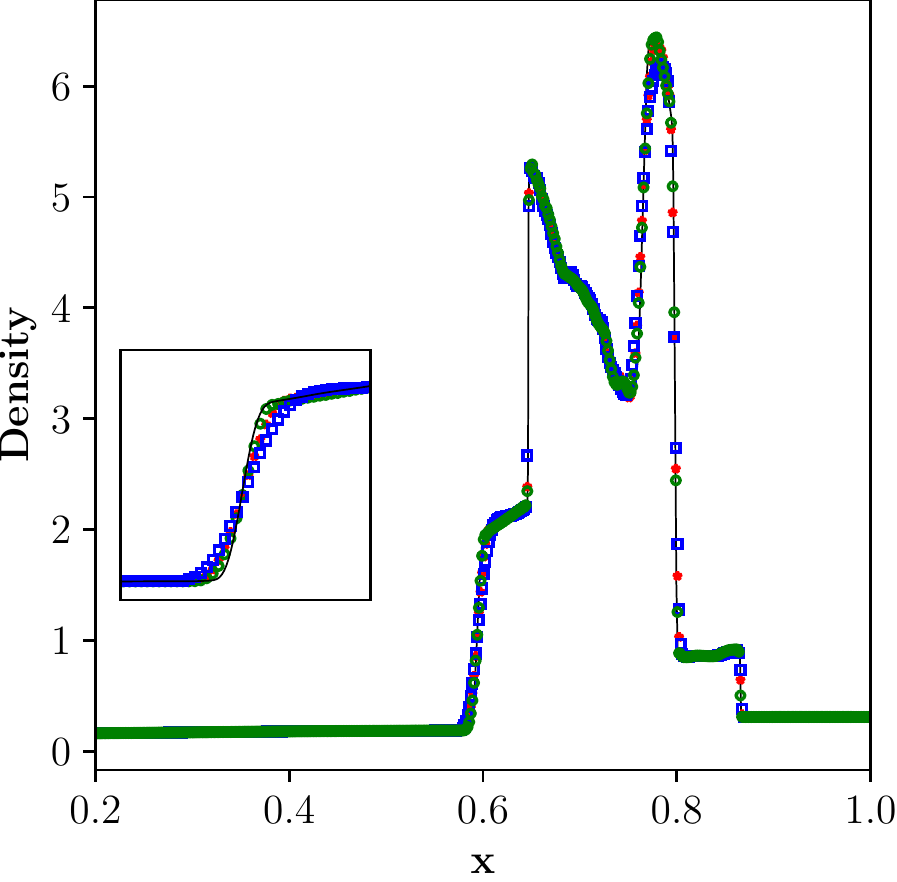}
\label{fig:blast-400}}
\subfigure[Global velocity profiles.]{\includegraphics[width=0.4\textwidth]{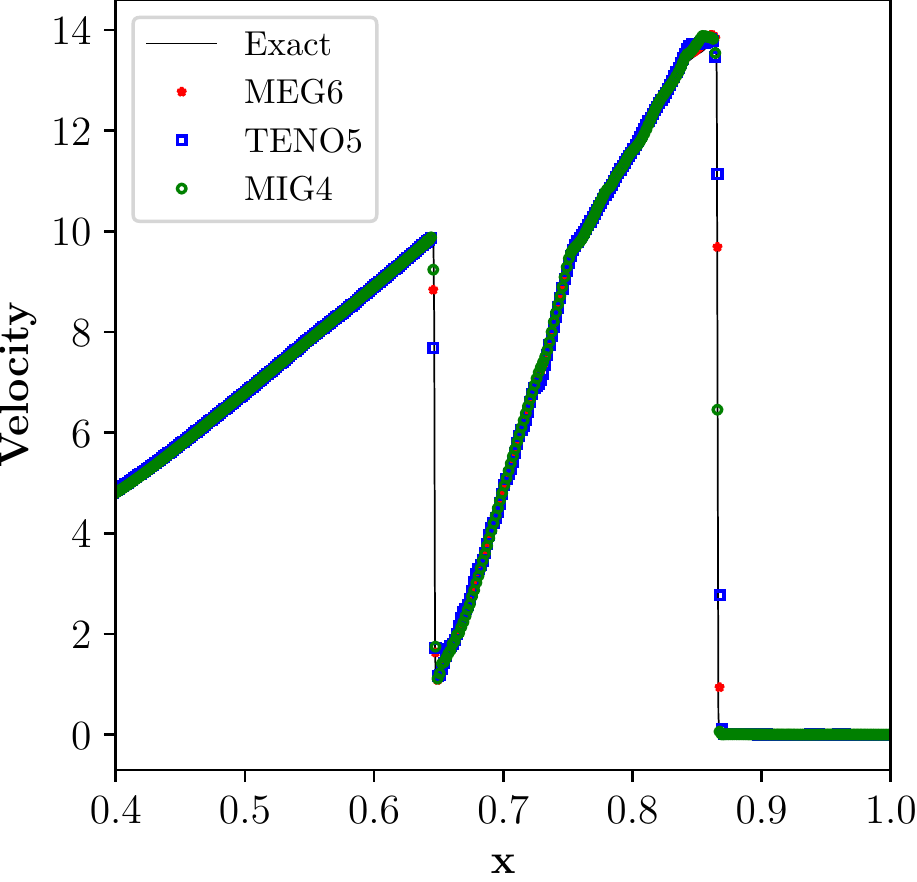}
\label{fig:blast-local}}
\caption{Numerical solution for blast wave problem using $N = 600$ grid points at $t = 0.038$, Example \ref{blast}.}
\label{fig_blast}
\end{figure}

\subsection{Inviscid one-dimensional multi-species test cases}\label{multi-species-tests}

\begin{example}\label{ex:multi-species}{Multi-species shock tube}
\end{example}

The first one-dimensional test case is the two-fluid modified shock tube of Abgrall and Karni \cite{abgrall2001computations}. The initial conditions of the test case are as follows:
\begin{equation}
\left(\alpha_{1} \rho_{1}, \alpha_{2} \rho_{2}, u, p, \alpha_{1}, \gamma \right)=\left\{\begin{array}{ll}
\left(1 , 0, 0, 1, 1, 1.4 \right) & \text { for } x<0 \\
\left(0,0.125,0,0.1, 0, 1.6 \right) & \text { for }  x \geq 0 .
\end{array}\right.
\end{equation}
Simulations are carried out on a uniformly spaced grid with $200$ cells on the spatial domain $-0.5 \leq x \leq 0.5$ with a constant CFL of $0.1$ and the final time is $t=0.2$. 

\begin{figure}[H]
\centering
\subfigure[Density]{\includegraphics[width=0.4\textwidth]{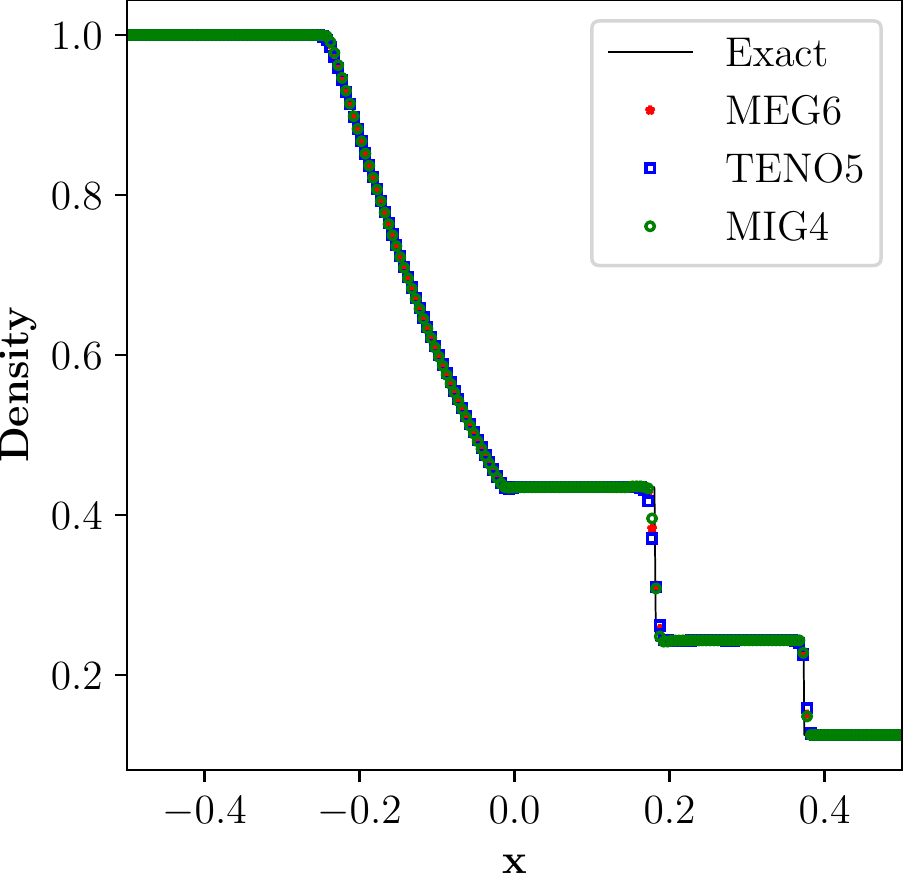}
\label{fig:multi_sod-den}}
\subfigure[Volume fraction]{\includegraphics[width=0.4\textwidth]{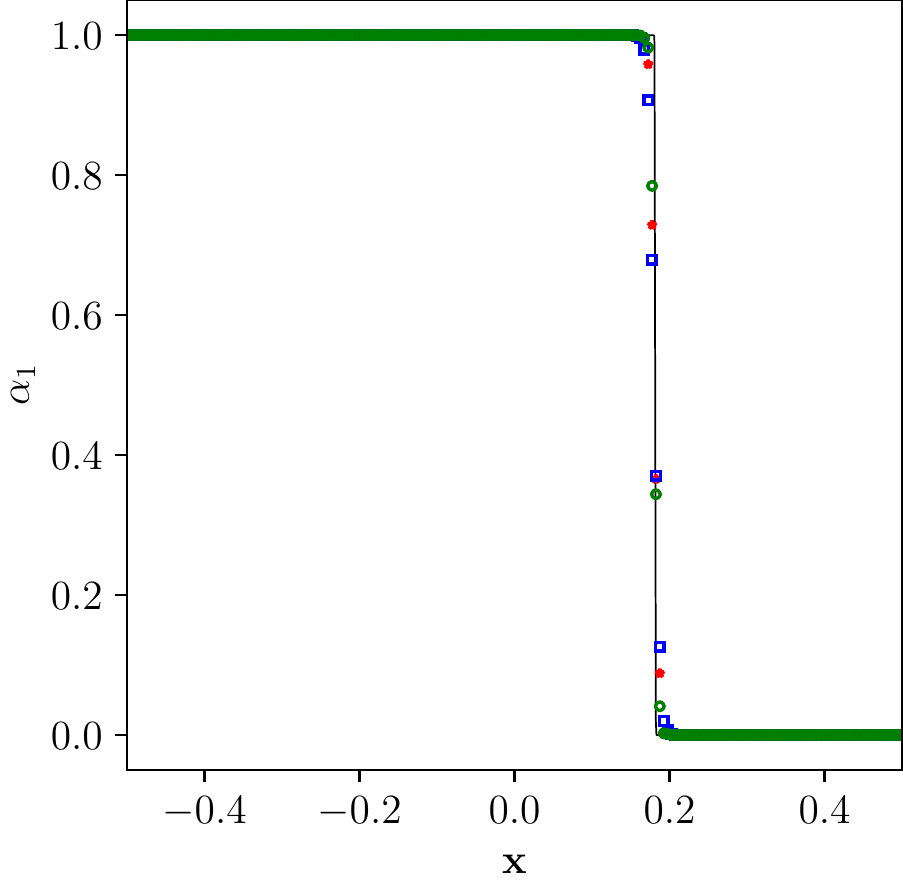}
\label{fig:multi_sod-volf}}
\caption{Numerical solution for multi-species shock tube problem in Example \ref{ex:multi-species} on a grid size of $N=200$. Dashed line: reference solution; green circles: MIG4; blue squares: TENO5; red stars: MEG6.}
\label{fig_multisod}
\end{figure}

\textcolor{black}{Fig. \ref{fig_multisod} shows various schemes' density and volume fraction profiles.} Proposed schemes capture the shock wave and the material interface without any oscillations. The solution profile of the volume fraction and the contact discontinuity in the density profile are captured \hl{within a few cells} by the MIG4 scheme over MEG6 in Figs. \ref{fig:multi_sod-volf}, and \ref{fig:multi_sod-den}, respectively.

\begin{example}\label{ex:curtain}{Inviscid shock-curtain interaction}
\end{example}
  The last one-dimensional inviscid shock-curtain interaction problem introduced by Abgrall \cite{abgrall1996prevent} is considered in this final one-dimensional test case. The initial conditions are as follow:
\begin{equation}
\left(\alpha_{1} \rho_{1}, \alpha_{2} \rho_{2}, u, p, \alpha_{1}, \gamma \right)=\left\{\begin{array}{ll}
\left(1.3765 , 0.0,0.3948,1.57, 1, 1.4\right), & 0 \leq x<0.25 \\
\left(1.0 , 0.0,0.0,1.0, 1, 1.4\right), & 0.25 \leq x<0.4 \text { or } 0.6 \leq x<1 \\
\left(0.0 , 0.138,0,1.0, 0, 1.67\right), & 0.4 \leq x<0.6
\end{array}\right.
\end{equation}
The shock wave travels in the air and moves to the right to interact with a helium bubble/curtain in the region $0.4 \leq x \leq 0.6$. Simulations are carried out on a grid size of $200$ cells until $t = 0.3$. Figs. \ref{fig:multi_curtain-den} and \ref{fig:multi_curtain-pres} show the density and pressure profiles obtained for all the schemes. Both the \textcolor{black}{gradient-based} reconstruction schemes capture the shock wave and the material interface without any oscillations or numerical instabilities. {Once again, the volume fraction profile shows that the material interface is captured within \textcolor{black} {a few cells} by the MIG4 schemes as shown in Fig.} \ref{fig:multi_curtain-alpha}.

\begin{figure}[H]
\centering
\subfigure[Density]{\includegraphics[width=0.4\textwidth]{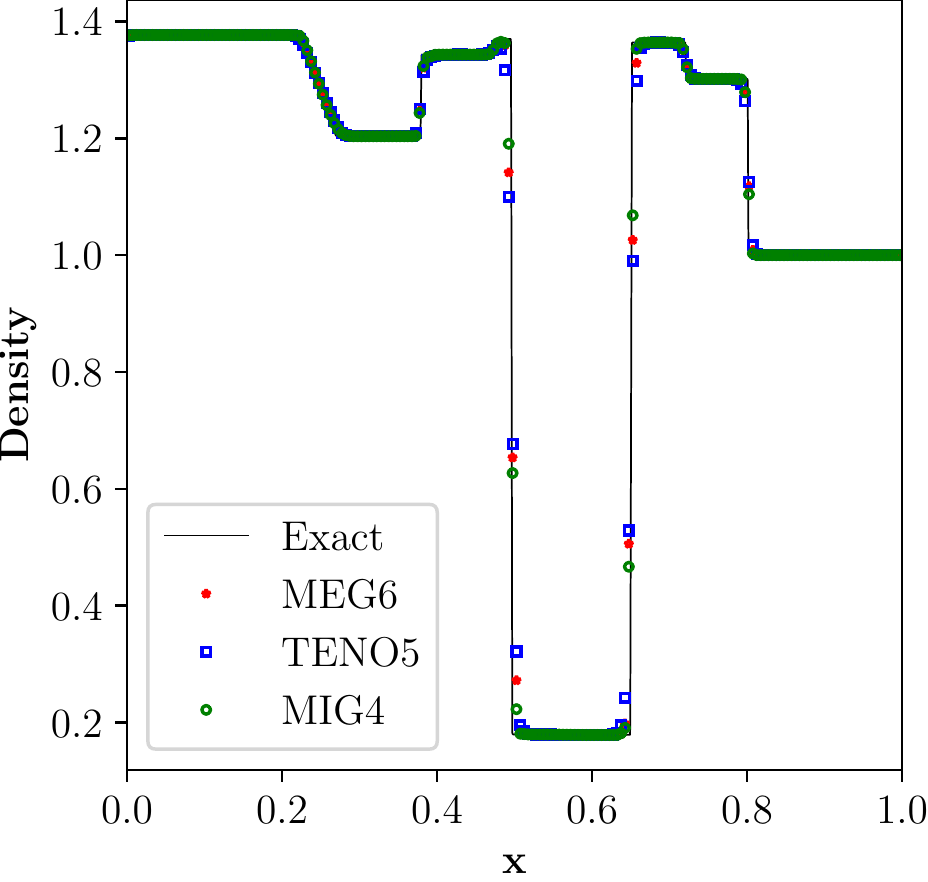}
\label{fig:multi_curtain-den}}
\subfigure[Velocity]{\includegraphics[width=0.4\textwidth]{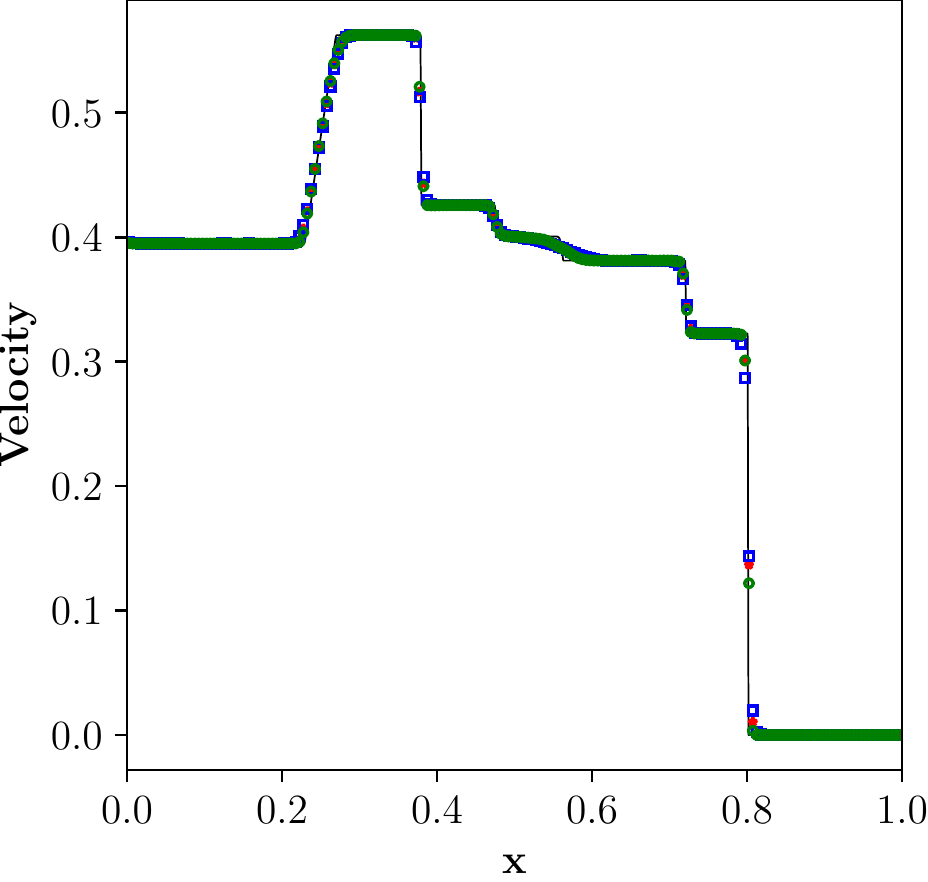}
\label{fig:multi_curtain-vel}}
\subfigure[Pressure]{\includegraphics[width=0.4\textwidth]{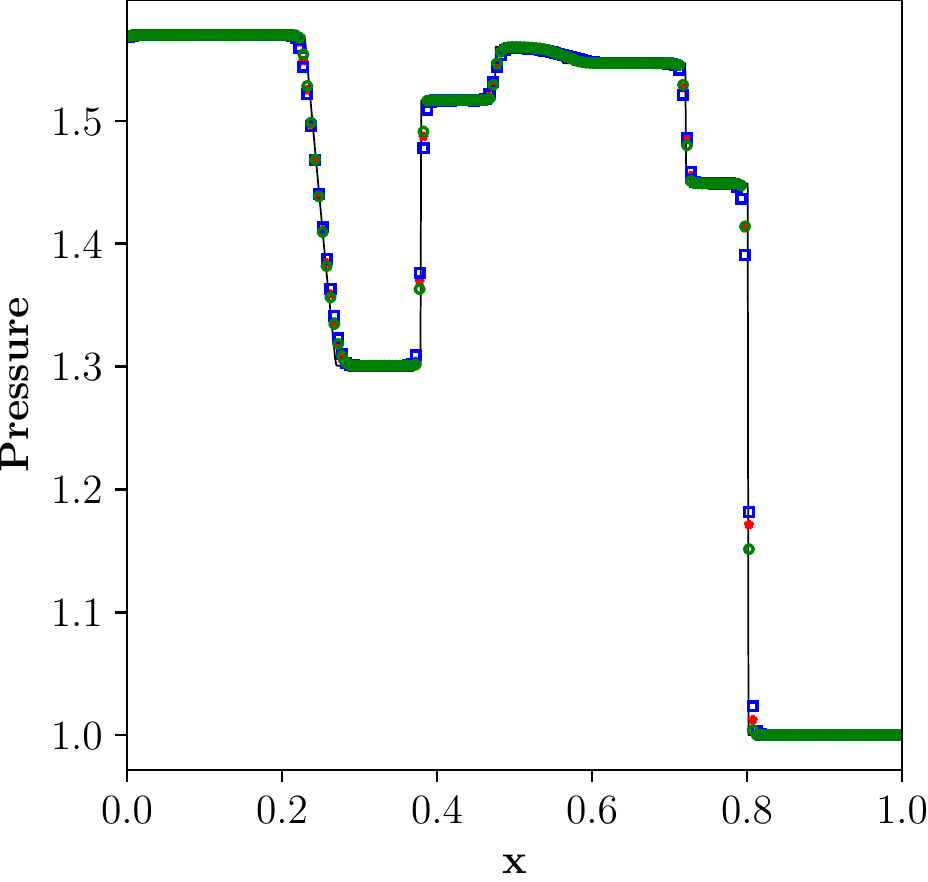}
\label{fig:multi_curtain-pres}}
\subfigure[Volume fraction, $\alpha_1$]{\includegraphics[width=0.4\textwidth]{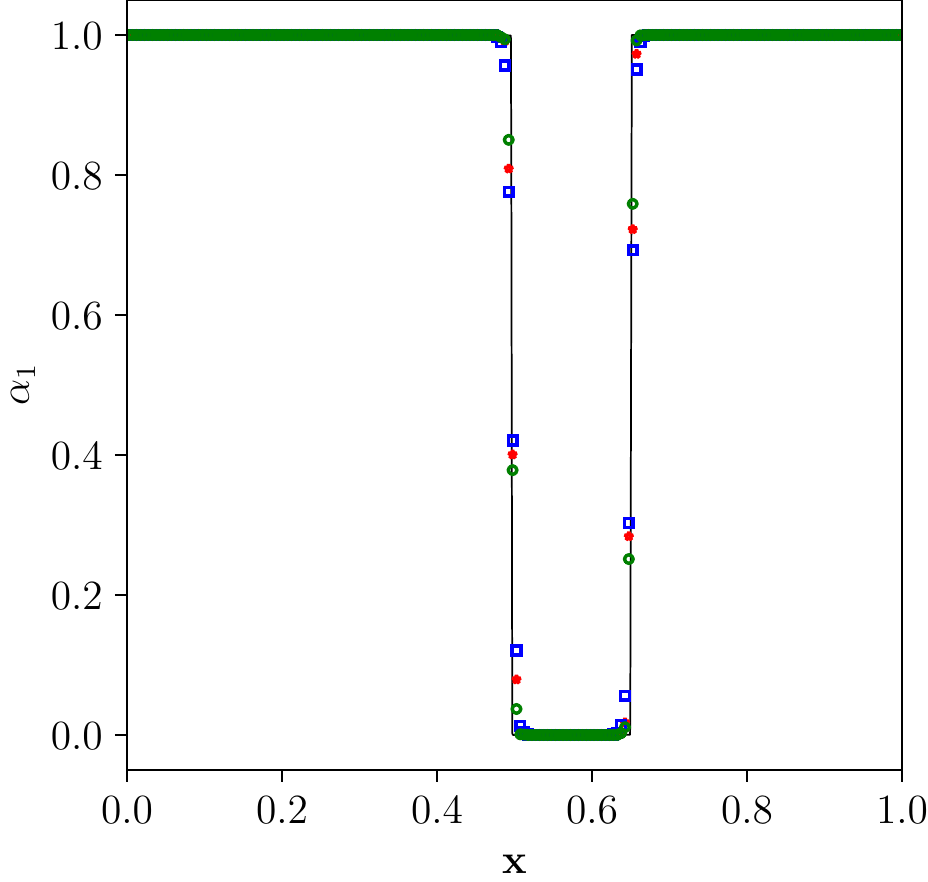}
\label{fig:multi_curtain-alpha}}
\caption{Numerical solution for multi-species shock-curtain interaction problem in Example \ref{ex:curtain}  on a grid size of $N=200$. Dashed line: reference solution; green stars: MIG4; blue squares: TENO5; red circles: MEG6.}
\label{fig_multi_curtain}
\end{figure}



\subsection{Two-dimensional inviscid and viscous multi-species  test cases}\label{2D-multi}

\begin{example}\label{ex:He-bubble}{Multi species Shock-Cylinder interaction}
\end{example}

In this two-dimensional test case, the shock-cylinder interaction where a Helium bubble with a density lighter than the air surrounding it interacts with a planar shock wave. Hass and Sturtevant  \cite{haas1987interaction}   studied this problem experimentally, and many researchers carried out computational studies for the same \cite{kawai2011high, Nonomura2012, Wong2017, wang2020consistent}. After the shock wave's impact on the helium cylinder, the \textcolor{black}{small-scale} vortices generated due to the baroclinic effects along the He-air interface can demonstrate the capabilities of the numerical scheme. The initial conditions of the test case are similar to that of Refs. \cite{kawai2011high, Wong2017} and are given by:

\begin{equation}
(\rho, u, v, p, \gamma)=\left\{\begin{array}{ll}
(1.3764,-0.3336,0,0,1.5698 / 1.4,1.4), & \text { for post-shock air, } \\
(0.1819,0,0,1 / 1.4,1.648), & \text { for helium cylinder, }\\
(1,0,0,1 / 1.4,1.4), & \text { for pre-shock air }.
\end{array}\right.
\end{equation}

The computational domain extends from $0.0 \leq x \leq 6.5D$ and $0.0 \leq y \leq 1.78D$. Initially, the Helium bubble is located at $[3.5$D$, 0.89$D$]$, where $D$ denotes the bubble's diameter and is taken as $D$ = 1\textcolor{black}{, and} a left moving normal shock of M 1.22 is placed at $x$=4.5$D$. All simulations are carried out with a CFL of 0.1, and slip wall boundary condition is imposed on top and bottom boundaries. The right boundary condition is extrapolated from the inside, and the left boundary is imposed as outflow. Simulation is carried out on two grid resolutions of 1300 $\times$ 356 and 2600 $\times$ 712. The results of the normalized density gradient magnitude $\phi = $exp$(|\nabla \rho|/|\nabla \rho|_{max} )$ obtained by various schemes are shown in Fig. \ref{fig_bubble_He_fine} for the grid resolution 1300 $\times$ 356.  To the best of the authors' knowledge, \textcolor{black}{the} implementation of the TENO scheme for the multi-component flows, specifically for the 5-equation model, is not presented in the open literature. 

It can be seen from these figures that there are no noticeable spurious oscillations in both the \textcolor{black}{gradient-based} schemes, MEG6 and MIG4. \hl{The material interface is thinner with the MIG4 scheme, in comparison with the MEG6 and TENO5}. Fig. \ref{fig_bubble_He_finer} shows the results of the normalized density gradient magnitude on a finer grid size of 2600 $\times$ 712 at various time instances. It can be seen that more secondary instabilities can be seen in the later stages of the simulations, and the interface thickness is thinner for the MIG4 scheme\textcolor{black}{, which} indicates less numerical dissipation.

\begin{figure}[H]
\centering
 \includegraphics[width=1.0\textwidth]{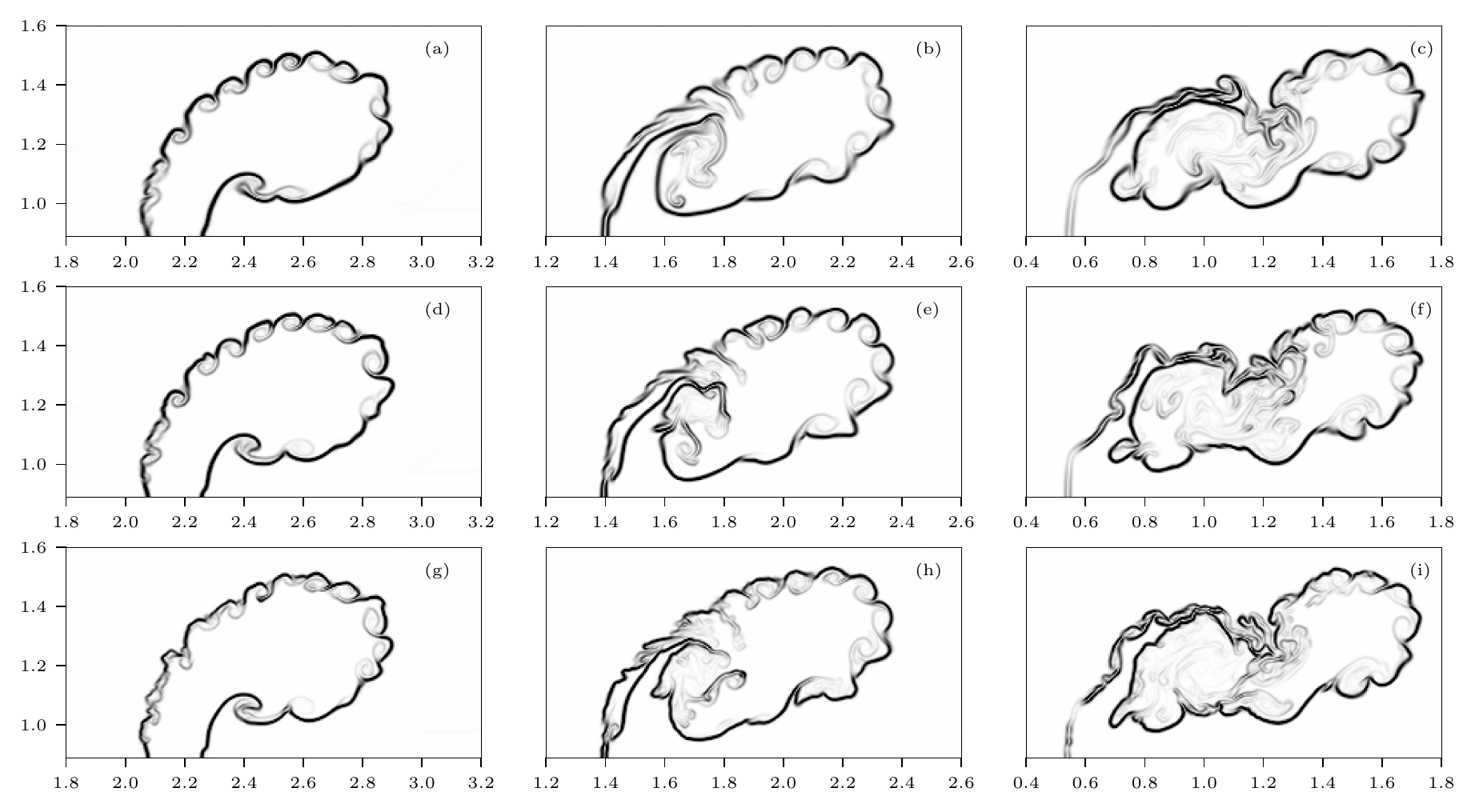}
\caption{Comparison of normalized density gradient magnitude, $\phi$, contours for Example \ref{ex:He-bubble} on a grid resolution of 1300 $\times$ 356. Contours are from 1 to 1.7 at different times $t$ using different schemes. Top row: TENO5; middle row: MEG6; and bottom row: MIG4.}
\label{fig_bubble_He_fine}
\end{figure}

\begin{figure}[H]
\centering
 \includegraphics[width=1.0\textwidth]{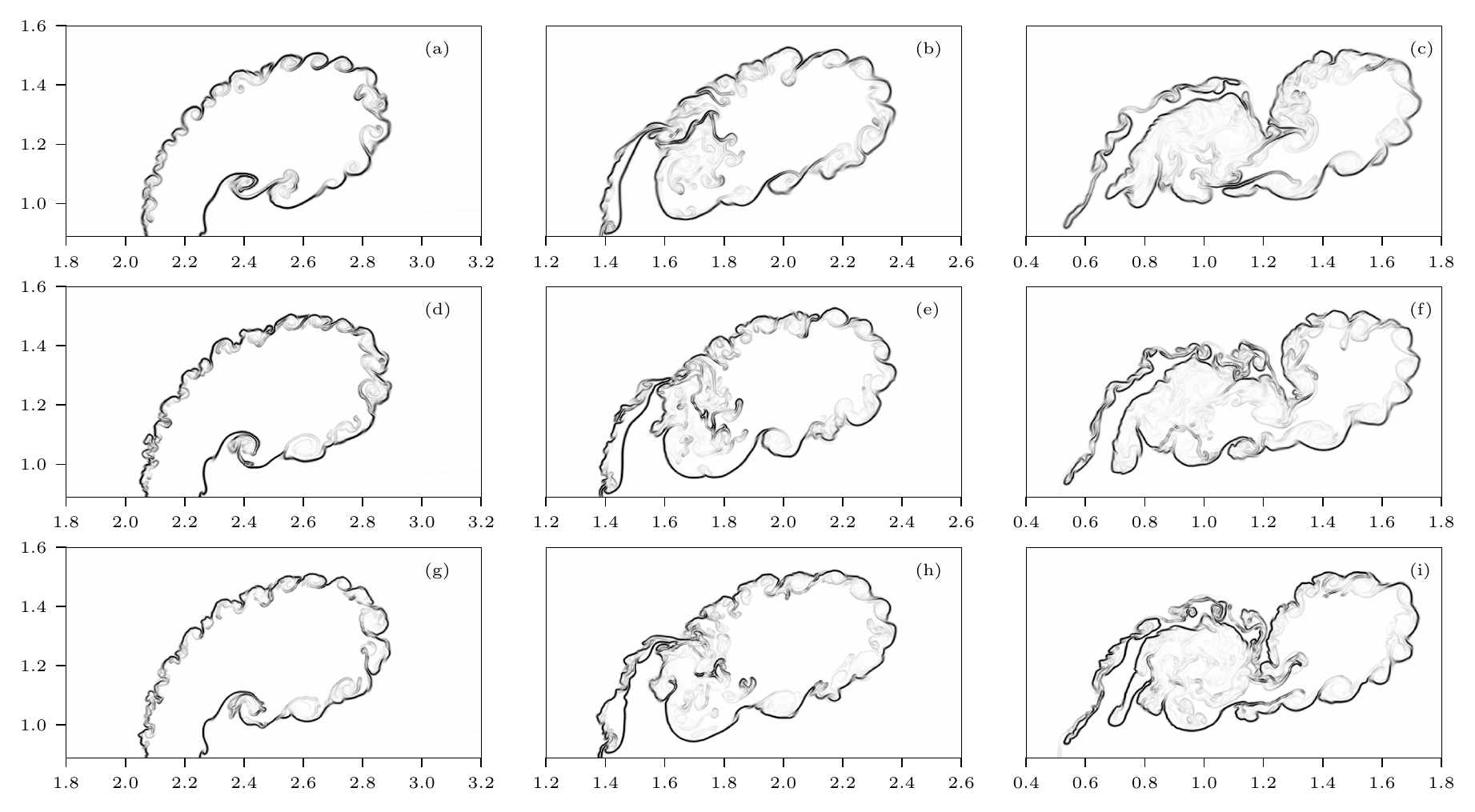}
\caption{\textcolor{black}{Normalized density gradient magnitude, $\phi$, contours for Example \ref{ex:He-bubble} on a grid resolution of 2600 $\times$ 712. Contours are from 1 to 1.7 at different times $t$ using different schemes. Top row: TENO5; middle row: MEG6; and bottom row: MIG4.}.}
\label{fig_bubble_He_finer}
\end{figure}

\begin{example}\label{ex:triple}{Compressible triple point problem} 
\end{example}

In the final inviscid test case, the multi-material compressible triple point problem, a three-state, two-dimensional Riemann problem with different materials, is considered. The computational domain is [0, 7] $\times$ [0, 3], and the initial conditions are the same as that of Pan et al. \cite{pan2018conservative}  which are as follows:  
\begin{equation}
(\rho, u, v, p, \gamma)=\left\{\begin{array}{ll}
(1.0~~~,0,0,1.0,1.5), & \text { \text{sub-domain} [0, 1]$\times$[0, 3], } \\
(1.0~~~,0,0,0.1,1.4), & \text { \text{sub-domain} [1, 1]$\times$[0, 1.5], }\\
(0.125,0,0,0.1,1.5), & \text { \text{sub-domain} [1, 7]$\times$[1.5, 3]}.
\end{array}\right.
\end{equation}
Simulation is carried out on a grid size of 1792 $\times$ 768, and the final time is 5.0. Reflective boundary conditions are imposed on all the boundaries. Fig. \ref{fig:TENO_trip_fine} shows the density gradient contours obtained by MEG6 and TENO5 scheme at $t$ = 5. It can be observed that the MEG6 scheme captures the material interface (shown with a red arrow in Fig. \ref{fig:TENO_trip_fine} ) within a few cells compared to the TENO5 scheme based on the thickness. Similarly, Fig. \ref{fig:MIG_trip} shows the density gradient contours obtained by MIG4 and TENO5 schemes and the material interface even thinner than the MEG6 scheme (shown with a green arrow in Fig. \ref{fig:MIG_trip} ) which indicates the low dissipation property of the scheme. The MIG4 scheme can reproduce the small vortices and complex flow structures with substantially improved resolution compared to the other schemes. Fig. \ref{fig:ivort} shows the vorticity contours of the results at $t$ = 5.0 \hl{computed by the MIG4 scheme, which is comparable in some regions and much more enhanced than that of} \cite{pan2018conservative}, see their Fig. 12. The results of Pan et al. \cite{pan2018conservative} is computed out on a finer mesh of 3584 $\times$ 1536. Numerical Schlieren images at various time instances $t$ = 0.2, 1.0, 3.0, 3.5, 4.0 and 5.0 are shown in Fig. \ref{fig:triple}, which depicts the development of the shock system and are consistent with Fig. 11 of \cite{pan2018conservative}.

$C1$ and $C2$ are the initial contact discontinuities generated by the initial Riemann problem, Fig. \ref{fig:triple} (a). Near the triple \textcolor{black}{point, a roll-up} region appears as the shockwave $S1$ travels faster than $S2$, Fig. \ref{fig:triple} (b). As the shockwave $S1$ travels ahead it interacts with the contact discontinuity $C3$, Fig. \ref{fig:triple} (c). At $t$ = 3.5 the shock wave $S1$ interacts with the right boundary and reflects into the domain, Fig. \ref{fig:triple} (d). As the shock wave $S1$ travels \textcolor{black}{inward} it generates transmitted shock waves $TS1$ and $TS2$, which interacts with the contact discontinuity $C1$, Fig. \ref{fig:triple} (e). These complex small-scale vortical structures generated along the contact discontinuities due to the Kelvin–Helmholtz instabilities are consistent with the inviscid flow physics.

\begin{figure}[H]
  \centering\offinterlineskip
  \subfigure[]{\includegraphics[width=0.48\textwidth]{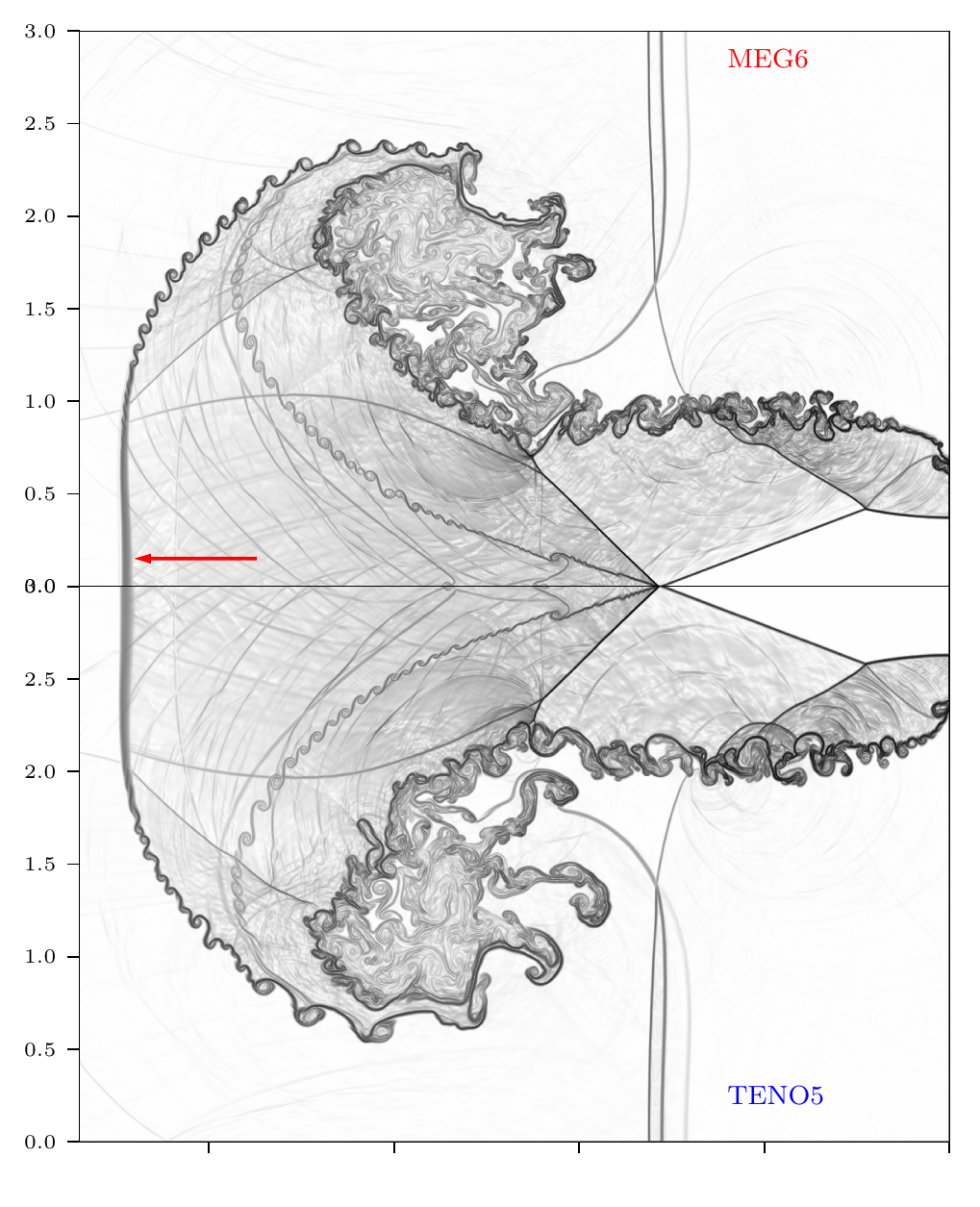}
  \label{fig:TENO_trip_fine}}
  \subfigure[]{\includegraphics[width=0.48\textwidth]{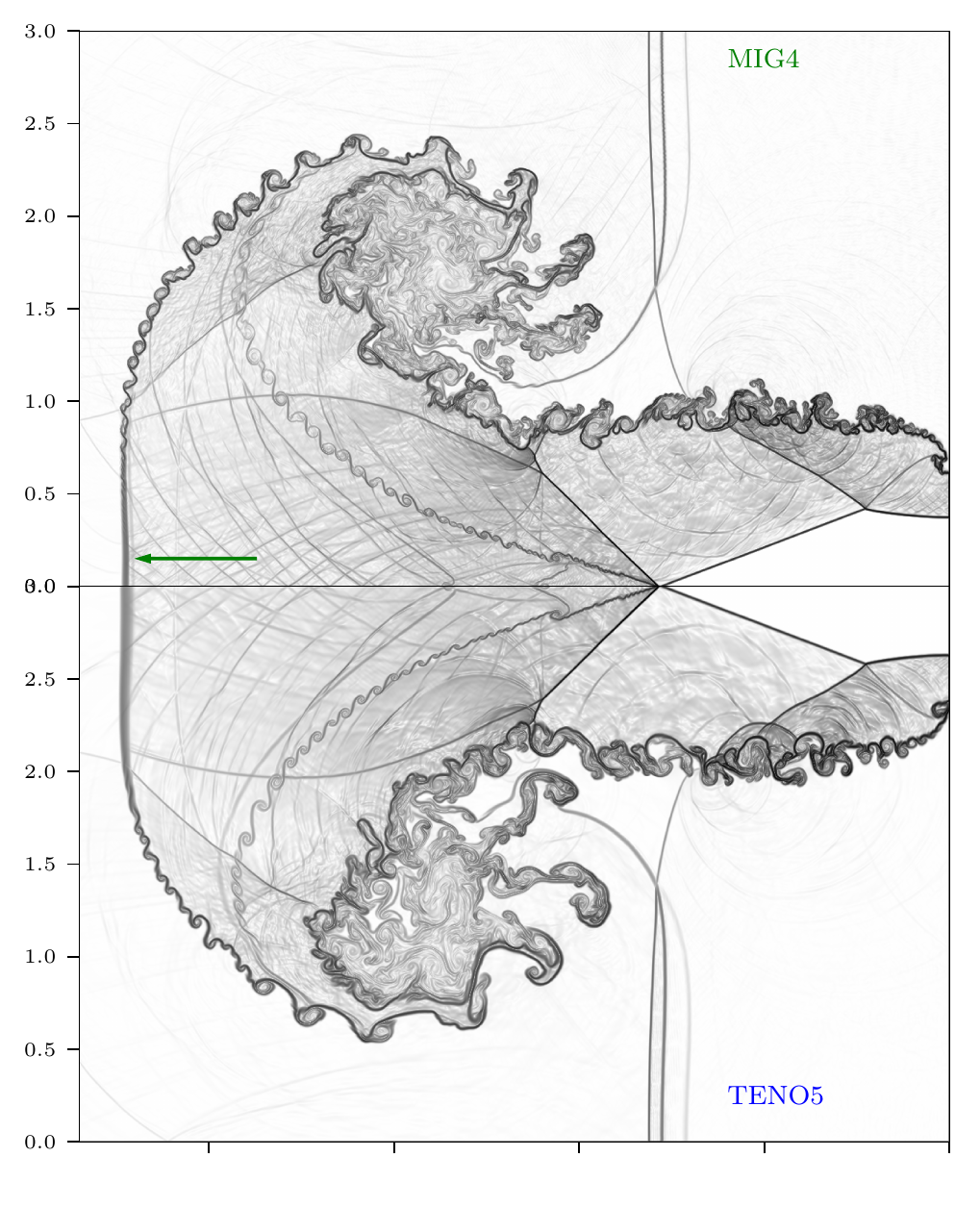}
  \label{fig:MIG_trip}}
  \caption{Density gradient contours computed by the proposed schemes and TENO5, Example \ref{ex:triple}. Fig. \ref{fig:TENO_trip_fine}, Top: MEG6 and bottom: TENO5. Fig. \ref{fig:MIG_trip}, Top: MIG4 and bottom: TENO5.}
  \label{fig_triple_den}
\end{figure}

\begin{figure}[H]
\centering
 \includegraphics[width=0.78\textwidth]{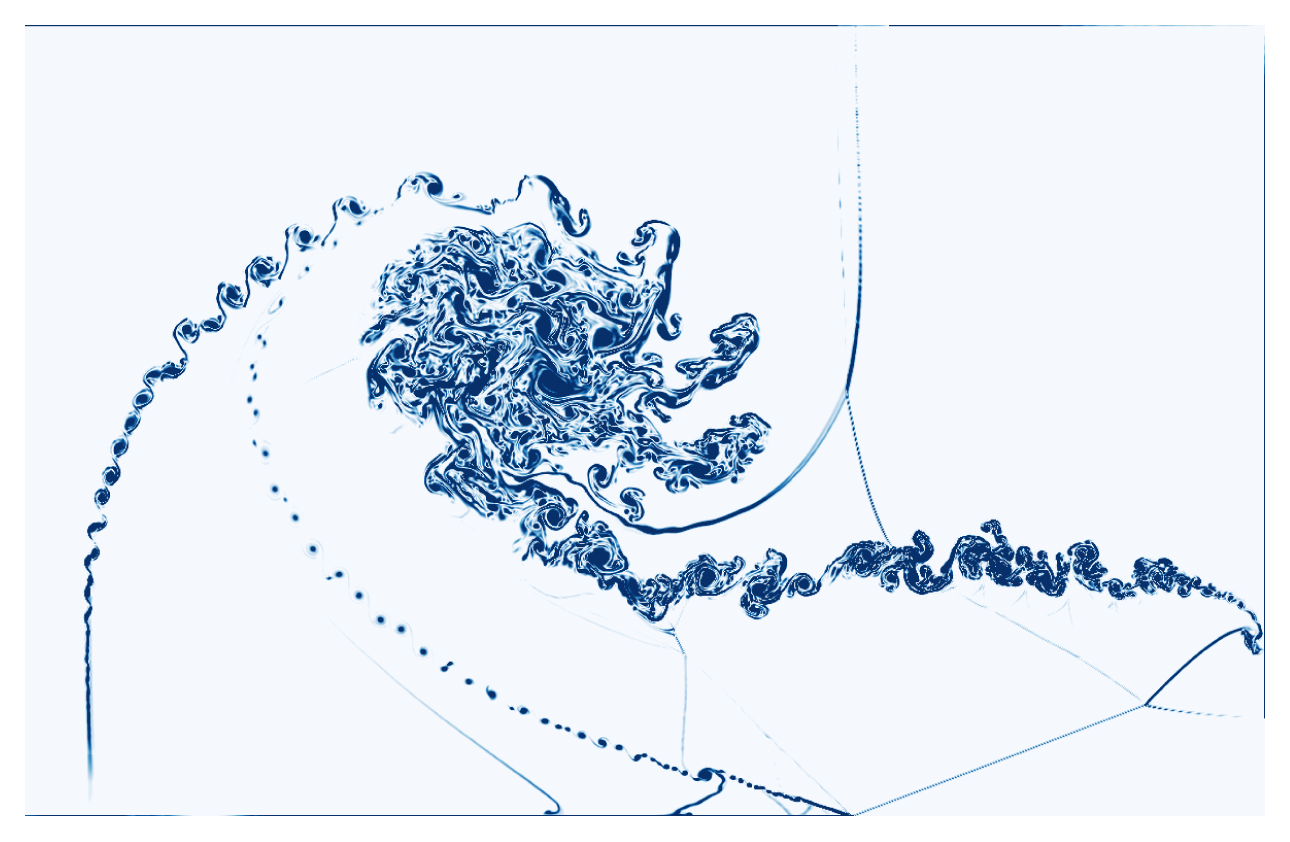}
 \caption{Vorticity contours at time instance $t=5$ using MIG4 scheme, Example \ref{ex:triple}, on a grid resolution of 1792 $\times$ 768.}
 \label{fig:ivort}
\end{figure}


\begin{figure}[H]
\centering
 \includegraphics[width=0.98\textwidth]{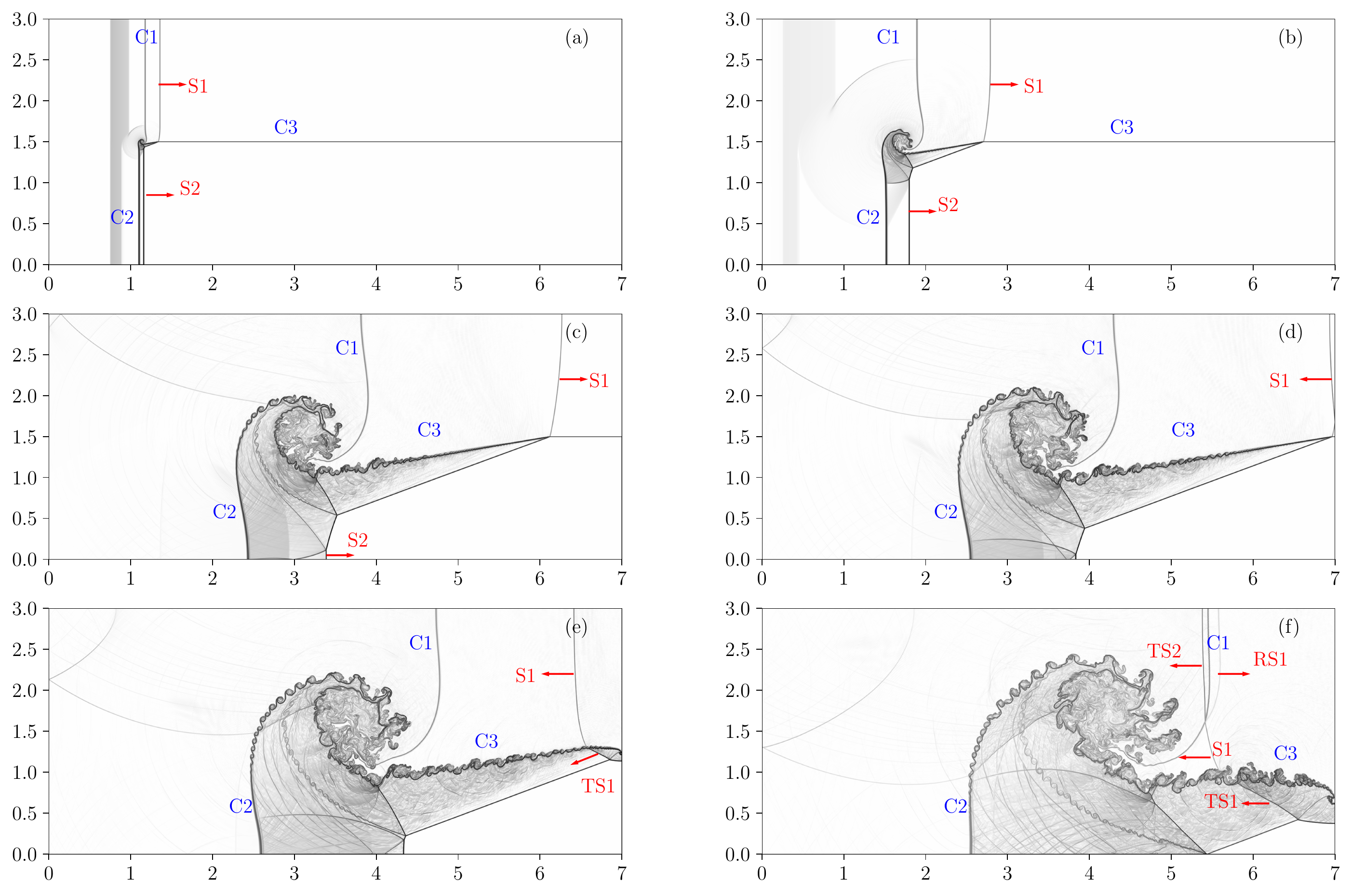}
 \caption{Numerical Schlieren images, ($\log(\text{abs}(|\nabla \rho|+1)$), at various time instances $t$ = 0.2, 1.0, 3.0, 3.5, 4.0 and 5.0 for the compressible triple point problem using MIG4 scheme, Example \ref{ex:triple}, on a grid resolution of 1792 $\times$ 768 .}
 \label{fig:triple}
\end{figure}

\begin{example}\label{ex:RM-viscous}{Two dimensional multi-species viscous Richtmeyer-Meshkov instability}
\end{example}
In this last test case, the two-dimensional viscous Richtmeyer-Meshkov (RMI) instability is computed. Numerical simulations of inviscid test cases are carried out by Nonomura et al.\cite{Nonomura2012}, and Kawai and Terashima \cite{kawai2011high} whereas Yee and B. Sjögreen \cite{yee2007simulation} have included physical viscosity. RMI occurs when an incident shock accelerates an interface between two fluids of different densities. 

The computational domain for this test case extends from $0.0 \leq x \leq 16.0\lambda$ and $0.0 \leq y \leq 1.0\lambda$ where $\lambda$ is the initial perturbation wavelength and the initial shape of the interface is given by 

\begin{equation}
\frac{x}{\lambda}=0.4-0.1 \sin \left(2 \pi\left(\frac{y}{\lambda}+0.25\right)\right),
\end{equation}
where, $\lambda$ =1 and the initial conditions are as follows:

\begin{equation}
(\rho, u, v, p, \gamma)=\left\{\begin{array}{ll}
(1,1.24,0,1 / 1.4,1.4), & \text { for pre-shock air } \\
(1.4112,0.8787,0,1.6272 / 1.4,1.4), & \text { for post-shock air } \\
(5.04,1.24,0,1 / 1.4,1.093), & \text { for } S F_{6}
\end{array}\right.
\end{equation}
In the current simulations a constant dynamic viscosity $\mu$ = 1.0 $\times$ $10^{-4}$ \cite{yee2007simulation} is considered. Periodic boundary conditions are imposed at the top and bottom boundaries of the domain, and the initial values are fixed at the left and right boundaries. Due to the cartesian grid considered in the present simulation, secondary instabilities will be generated at the material interface, which is not observed in experimental conditions \cite{jacobs2005experiments}. In order to avoid these secondary instabilities, the initial perturbation is smoothened by adding an artificial diffusion layer given by Ref. \cite{Wong2017}:

\begin{equation}
\begin{array}{c}
f_{sm}=\frac{1}{2}\left(1+\operatorname{erf}\left(\frac{\Delta D}{E_{i} \sqrt{\Delta x \Delta y}}\right)\right) \\
u=u_{L}\left(1-f_{sm}\right)+u_{R} f_{sm}
\end{array}
\end{equation}
where $u$ represents the primitive variables near the initial interface, the parameter $E_i$ introduces additional thickness to the initial material interface, $\Delta D$ is the distance from the initial perturbed material interface, and subscripts $L$ and $R$ denote the left and right interface conditions. Parameter $E_i$ is chosen as 5 in this test case.


\begin{figure}[H]
\centering\offinterlineskip

\subfigure[$t$  = 5.5]{\includegraphics[width=0.29\textwidth]{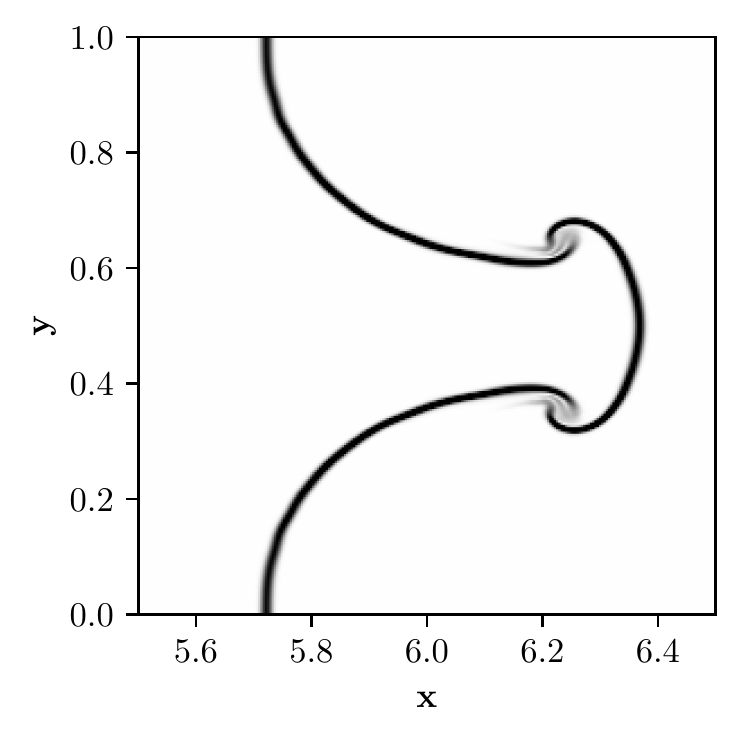}
\label{fig:MIGI_55_fine_RMI}}
\subfigure[$t$ = 8.25]{\includegraphics[width=0.29\textwidth]{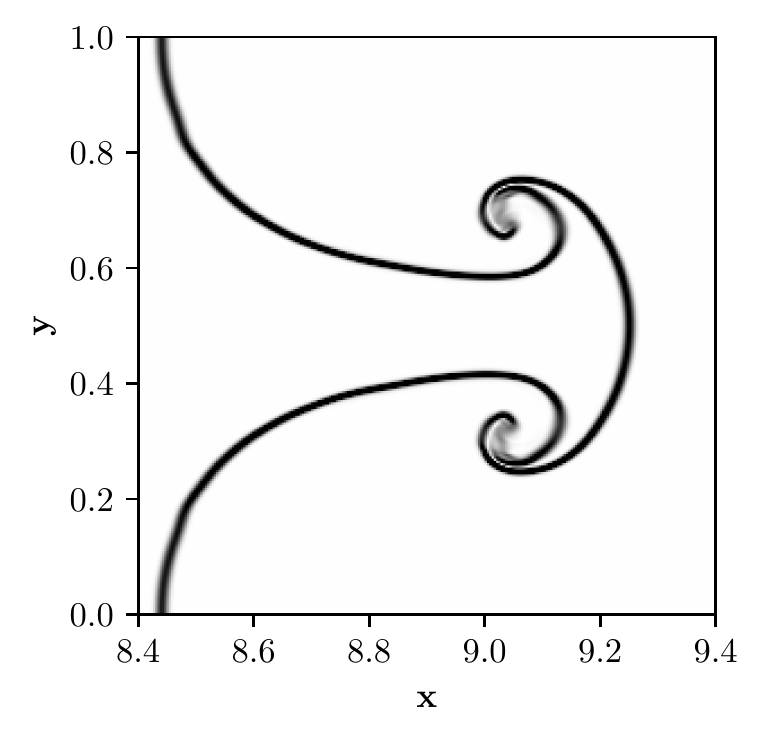}
\label{fig:MIGI_82_fine_RMI}}
\subfigure[$t$  = 11.0]{\includegraphics[width=0.35\textwidth]{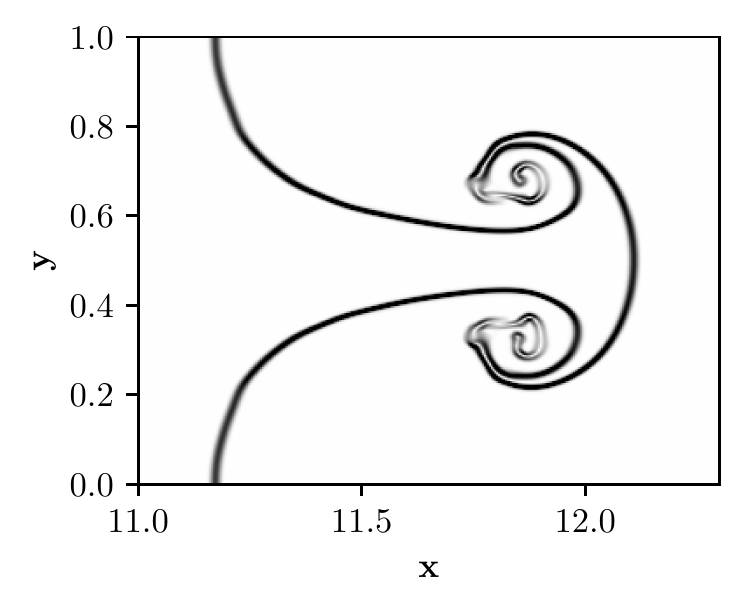}
\label{fig:MIGI_110_fine_RMI}}

\subfigure[$t$  = 5.5]{\includegraphics[width=0.29\textwidth]{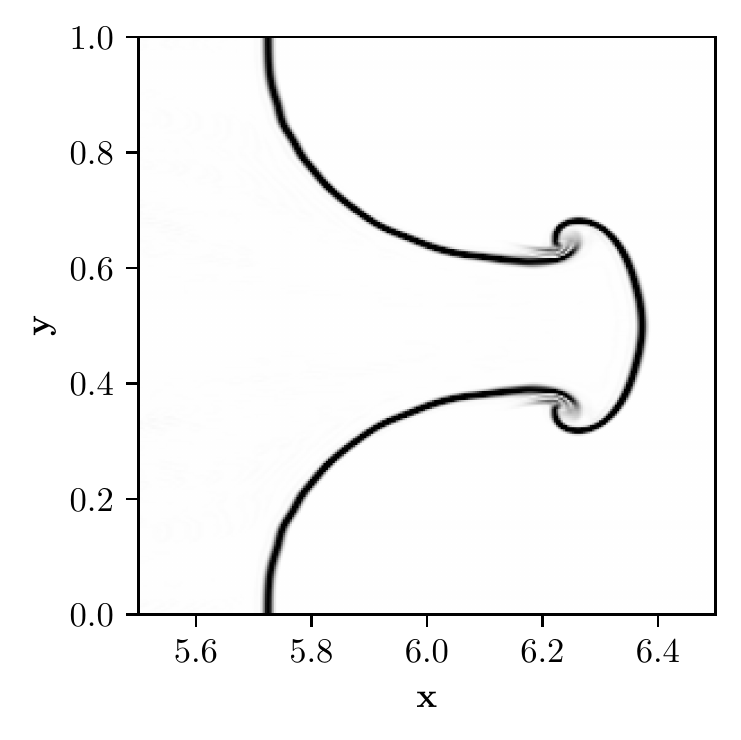}
\label{fig:MP5_55_fine_RMI}}
\subfigure[$t$  = 8.25]{\includegraphics[width=0.29\textwidth]{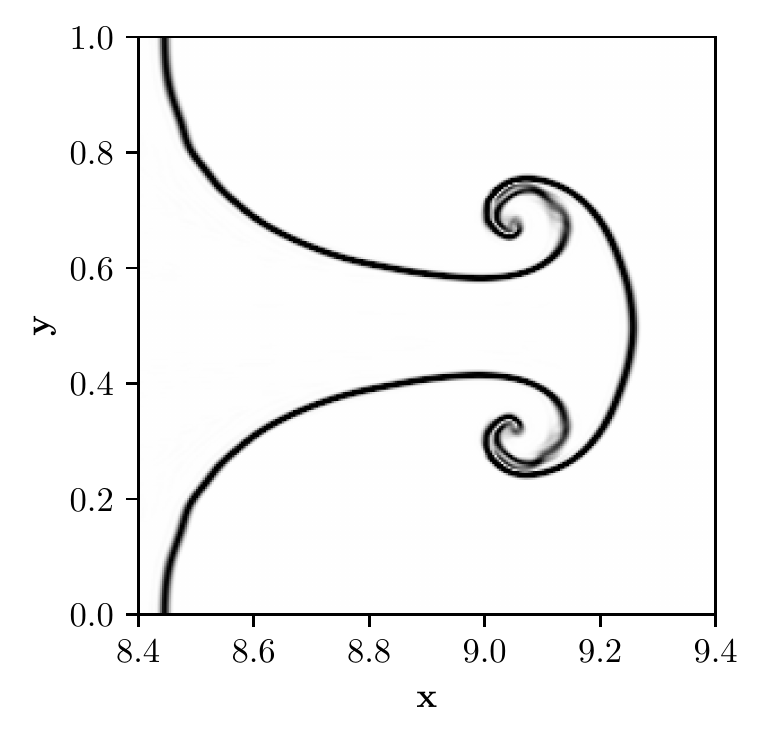}
\label{fig:MP5_82_fine_RMI}}
\subfigure[$t$  = 11.0]{\includegraphics[width=0.35\textwidth]{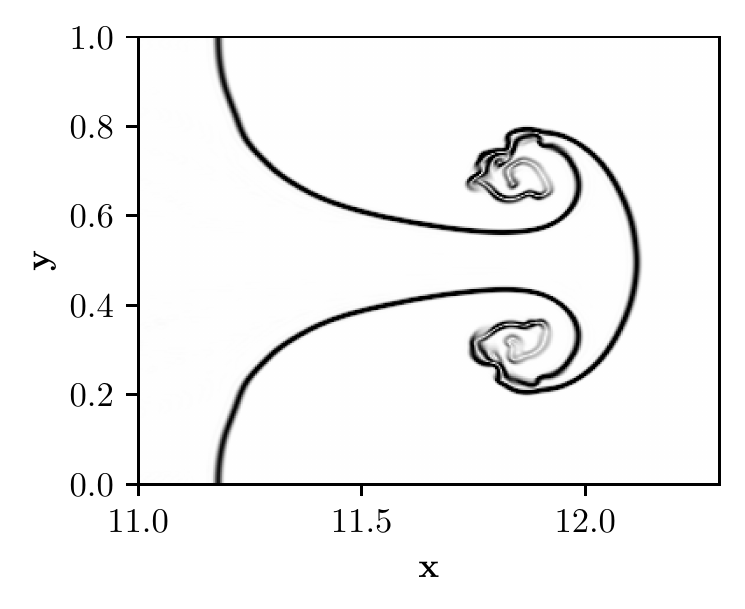}
\label{fig:MP5_110_fine_RMI}}

\subfigure[$t$  = 5.5]{\includegraphics[width=0.29\textwidth]{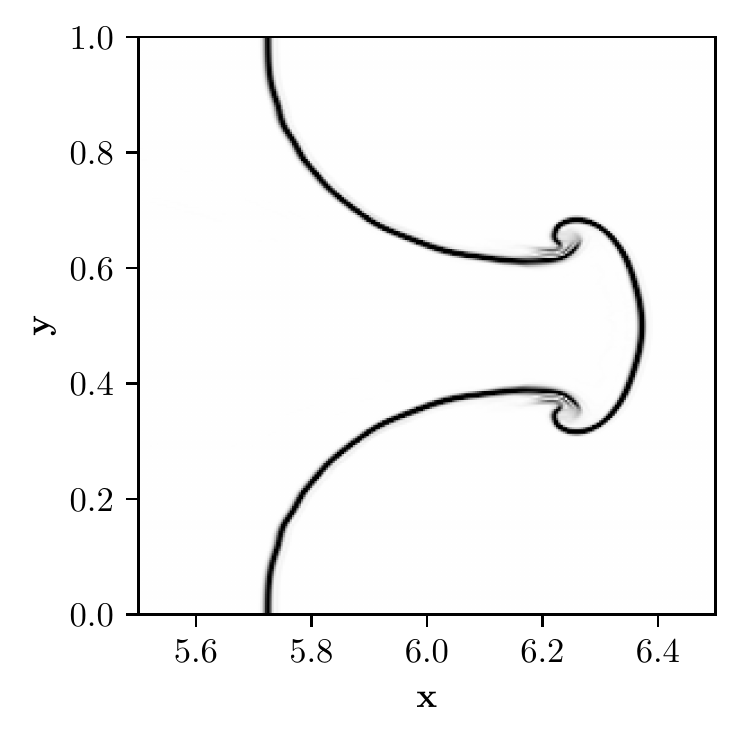}
\label{fig:MIGE_55_fine_RMI}}
\subfigure[$t$  = 8.25]{\includegraphics[width=0.29\textwidth]{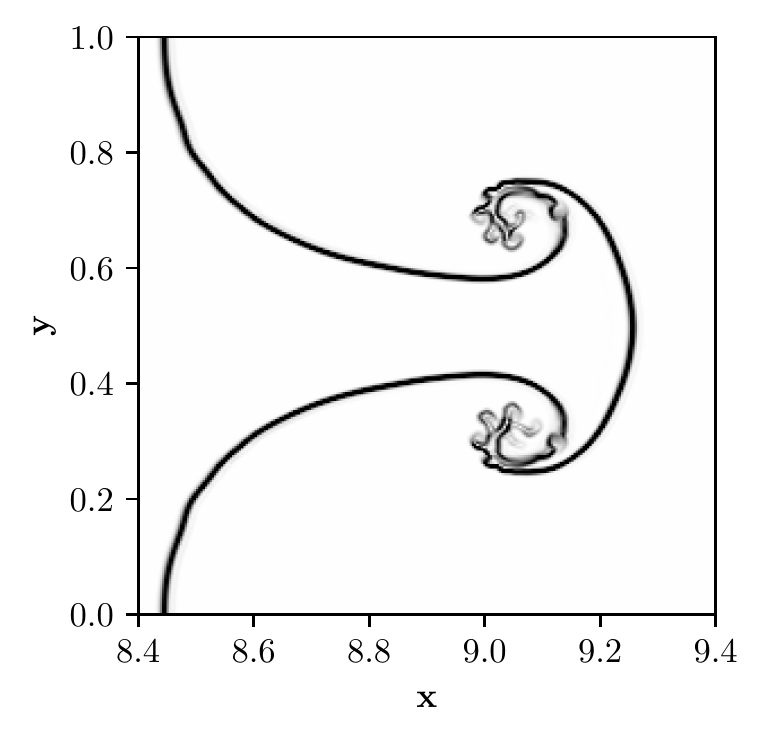}
\label{fig:MIGE_82_fine_RMI}}
\subfigure[$t$  = 11.0]{\includegraphics[width=0.35\textwidth]{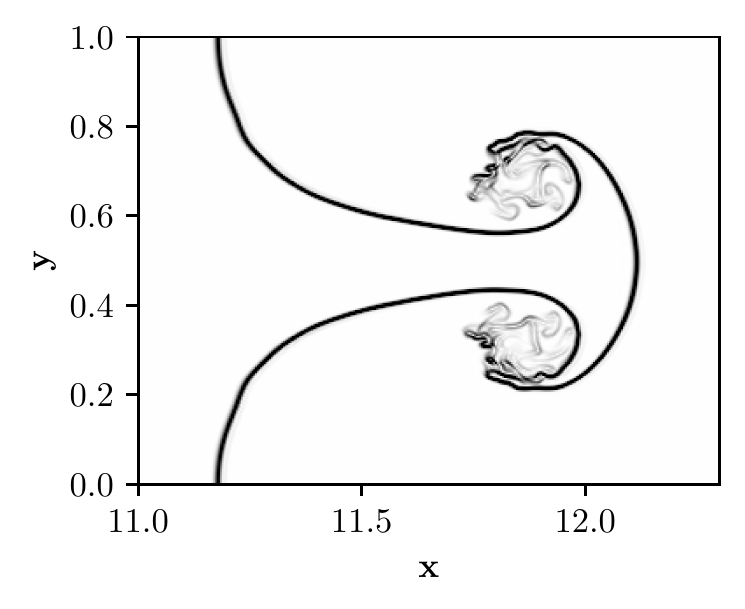}
\label{fig:MIGE_110_fine_RMI}}

\caption{\textcolor{black}{Comparison of normalized density gradient magnitude, $\phi$, contours for two-dimensional viscous Richtmeyer-Meshkov instability problem in Example \ref{ex:RM-viscous}  on a grid size of 4096 $\times$ 256. Contours are from 1 to 1.7 at time t$=$5.5, 8.25 and 11.0 using the proposed schemes. Top row: TENO5; middle row: MEG6; and bottom row: MIG4. All the simulations are carried out with $\alpha$-IG scheme for viscous fluxes for fair comparison.}}
\label{fig_RM_viscous}
\end{figure}
Simulations are conducted on a grid size of 4096 $\times$ 256 cells, with a constant CFL of 0.1. Computational results of normalized density gradient magnitude $\phi = $exp$(|\nabla \rho|/|\nabla \rho|_{max} )$ obtained at $t$ = 5.5, 8.25 and 11.0 by various schemes are shown in Fig. \ref{fig_RM_viscous}. As the shock wave hits the perturbed interface, it starts to deform and generates vortices due to the baroclinic effect. As time progresses, the $SF_6$, which is heavier, penetrates the air, the lighter fluid, leading to the formation of a spike. For this simulation, viscous fluxes are computed using the implicit gradient \textcolor{black}{$\alpha$-damping} approach to isolate the effect of the inviscid fluxes. Figs. \ref{fig_RM_viscous} indicate that there are no noticeable spurious oscillations for the MIG4 scheme and the interface thickness is thinner than the MEG6 and TENO5. MIG4 scheme also shows improved resolution regarding the roll-up vortices, indicating the schemes' low numerical dissipation in smooth flow regions compared to MEG6. These results indicate that the implicit gradient scheme can capture material interface within a few cells even for longer duration simulations.

\section{ \textcolor{black}{Conclusions and Future work}}\label{sec-5}

This paper presented a gradient-based reconstruction approach for simulations of compressible single and multi-species Navier-Stokes equations. \textcolor{black}{The proposed} methods are accurate and computationally efficient as the higher-order gradients of the flow variables are computed once and reused for both inviscid and viscous fluxes. \textcolor{black}{Furthermore,} superior results are obtained for several benchmark test cases involving shocks, material interfaces\textcolor{black}{, and small-scale} features. The contributions and observations of the paper are summarized as follows:

\begin{itemize}

\item Fourth order \textcolor{black}{$\alpha$-damping} schemes with superior spectral properties are derived for viscous fluxes. Results indicated significant improvements in flow features for the viscous flows depending on the discretization approach's spectral properties and also prevented odd-even decoupling.
\item Unlike interpolation-based schemes, the proposed schemes allow sharing of gradients between viscous and inviscid schemes. Such an approach improved the efficiency of the simulations. The implicit gradient approach with superior spectral properties showed better resolution of the flow structures.
\item Shock capturing is carried out by MP scheme, which is also improved by reusing the gradients. Results obtained by the proposed schemes, MEG and MIG, are superior to the TENO scheme.
\item For the multi-species flows, apart from the improved resolution, the material interface is thinner and is captured within a few cells by the proposed implicit gradient scheme.
\item \textcolor{black}{Finally, A summary of the difference schemes considered in this paper indicating the order of accuracy for each and the best one in terms of the dispersion/dissipation property referring to the relevant figure are shown in the Tables \ref{tab:visc_schema} and \ref{tab:inv_schema}. Table \ref{tab:visc_schema} corresponds to the viscous schemes and the Table \ref{tab:inv_schema} corresponds to the inviscid schemes.}
\end{itemize}

\begin{table}[htbp]
  \centering
  \caption{ \textcolor{black}{Summary of the viscous schemes considered in this paper.}}
    \begin{tabular}{|c|c|c|c|c|}
    \hline
Visocus Scheme & Order & High-frequency damping & Figure & Example \\
    \hline
    $\alpha$-EG & 4th   & Yes   &   Fig. \ref{fig:second_deriv}    & Example \ref {ex:vs}  \\
	\hline
    $\alpha$-IG & 4th   & Yes   &   Fig. \ref{fig:second_deriv}    & Example \ref {ex:vs} \\
    \hline
    Shen's scheme \cite{shen2010large} & 6th   & No    &    Fig. \ref{fig:second_deriv}   & Example \ref {ex:vs} \\
    \hline
    \end{tabular}%
  \label{tab:visc_schema}%
\end{table}%

\begin{table}[htbp]
  \centering
  \caption{ \textcolor{black}{Summary of the nonlinear inviscid schemes considered in this paper.}}
    \begin{tabular}{|c|c|c|c|c|c|}
    \hline
           Nonlinear inviscid scheme & Order & Dispersion error & Dissipation error & Representative examples \\
            & & (Fig. \ref{fig_disp}) & (Fig. \ref{fig_disp}) &  \\
           \hline
           \hline
           MIG   & 4th   & Low   & Medium & Examples \ref{ex:ssl}, \ref{shock-entropy} and \ref{ex:triple} \\
\hline
           MEG   & 4th   & Medium & Medium &  Examples \ref{ex:ssl}, \ref{shock-entropy} and \ref{ex:triple} \\
          \hline
           TENO5 & 5th   & High  & Low   & Examples \ref{ex:ssl}, \ref{shock-entropy} and \ref{ex:triple} \\
          \hline
    \end{tabular}%
  \label{tab:inv_schema}%
\end{table}%

\textcolor{black}
{The following points can improve and extend the proposed approach:}
\begin{enumerate}
\item \textcolor{black}{The current approach is limited to the second-order accuracy for non-linear problems due to the reconstruction of the primitive variables. The primary objective was to reuse the gradients for both viscous and inviscid fluxes. However, genuine high-order accuracy, fourth order, can be obtained by reconstructing the fluxes. An algorithm that can obtain high-order accuracy by flux reconstruction, thereby obtaining high-order accuracy yet sharing the gradients between viscous and inviscid fluxes, is possible and will be presented elsewhere.}

\item \textcolor{black}{This paper shows that it is possible to obtain fourth-order accuracy with the kappa scheme, which is expected to be only third-order accurate as it is a quadratic reconstruction scheme \cite{van2021towards,vanleer1979}. However, even higher-order accuracy, for example, seventh-order, is also possible, and it will be presented elsewhere.}

\item \textcolor{black}{It is also possible to reconstruct conservative variables and share the gradients between viscous and inviscid fluxes. Such an approach may improve (or otherwise) the results for some test cases of single species flow. However, Houim and Kuo \cite{houim2011low} mentioned in their paper that the results would be drastically different depending on the choice of variables (primitive, conservative, characteristic, fluxes, etc.) used for reconstruction or interpolation. It is beyond the scope of the paper to address such differences and is one of the reasons to present only primitive variable reconstruction in this paper.}
\item \textcolor{black}{Extension of the proposed method to curvilinear coordinates such that it is free-stream preserving \cite{Nonomura2010} and applies to complex geometry simulations will be presented elsewhere.}
\item \textcolor{black}{Sharing of gradients is not limited to the current approach but can also be used in kinetic and entropy-preserving schemes (KEEP) \cite{kuya2021high}. In the KEEP approach, the Navier-Stokes equations are written in a split form which means the gradients of the primitive and conservative variables are computed and, therefore, can be reused in the viscous fluxes (or vice versa).} 
\item \textcolor{black}{The author's primary motivation for this approach is to simulate multi-component reacting flows \cite{houim2011low,ziegler2011adaptive}. With the sharing of the gradients between convective and viscous fluxes (including species diffusion fluxes), the gradients of species mass fractions (9 species for H2/air reacting case)  need to be computed only once, making it a very efficient approach to reacting viscous flow simulations.}
\end{enumerate}

\section*{Appendix}

\renewcommand{\thesubsection}{\Alph{subsection}}
{\subsection{Riemann solver}\label{sec-3.3}

As explained in the section (\ref{sec-3.1.1}), Riemann solvers approximate the convective flux after obtaining the reconstructed states at the interface. This section illustrates how the HLLC Riemann solver \cite{batten1997choice,toro1994restoration} approximates convective fluxes. For simplicity, only the HLLC approximations for both single and multi-species flows in the x-direction are illustrated in this section. The HLLC flux in $x$-direction is given by:

\begin{equation}\label{eq:HLLC}
\mathbf{F}^{\rm Riemann}= \mathbf{F}^{HLLC}=\left\{\begin{array}{ll}
\mathbf{F}_{L} & , \text { if } \quad 0 \leq S_{L} \\
\mathbf{F}_{* L} & , \text { if } \quad S_{L} \leq 0 \leq S_{*} \\
\mathbf{F}_{* R} & , \text { if } \quad S_{*} \leq 0 \leq S_{R} \\
\mathbf{F}_{R} & , \text { if } \quad 0 \geq S_{R}
\end{array}\right.
\end{equation}
\begin{equation}
\mathbf{F}_{* K}=\mathbf{F}_{K}+S_{K}\left(\mathbf{Q}_{* K}-\mathbf{Q}_{K}\right)
\end{equation}
where $L$ and $R$ are the left and right states respectively. With $K$ = $L$ or $R$, the star state for single-species flow is defined as:

\begin{equation}\label{eqn-hllc-single}
\mathbf{Q}_{* K}=\left(\frac{S_{\mathrm{K}}-u_{\mathrm{K}}}{S_{\mathrm{K}}-S_{*}}\right)\left[\begin{array}{c}
\rho_{K} \\
\rho_{\mathrm{K}} S_{*} \\
\rho_{K} v_{K} \\
E_{k}+\left(S_{*}-u_{K}\right)\left(\rho_{K} S_{*}+\frac{p_{K}}{S_{K}-u_{K}}\right)
\end{array}\right]
\end{equation}

For two-species  flow with five-equation model is defined as:

\begin{equation}\label{eqn-hllc-multi}
\mathbf{Q}_{* \mathrm{K}}=\left(\frac{S_{\mathrm{K}}-u_{\mathrm{K}}}{S_{\mathrm{K}}-S_{*}}\right)\left(\begin{array}{c}
\left(\alpha_{1} \rho_{1}\right)_{\mathrm{K}} \\
\left(\alpha_{2} \rho_{2}\right)_{\mathrm{K}} \\
\rho_{\mathrm{K}} S_{*} \\
\rho_{\mathrm{K}} v_{\mathrm{K}} \\
E_{\mathrm{K}}+\left(S_{*}-u_{\mathrm{K}}\right)\left(\rho_{\mathrm{K}} S_{*}+\frac{p_{\mathrm{K}}}{S_{\mathrm{K}}-u_{\mathrm{K}}}\right) \\
\alpha_{1_{\mathrm{K}}}
\end{array}\right)
\end{equation}
 In the above expressions, the waves speeds  $S_L$ and $S_R$ can be obtained as suggested by Einfeldt \cite{einfeldt1988godunov}, $ S_L = min(u_{L}-c_L, \tilde{u}-\tilde{c}) \ \text{and} \ S_R = max(u_{R}+c_R, \tilde{u} +\tilde{c})$, where $\tilde {u}$ and $\tilde {c}$ are the Roe averages from the left and right states. Batten et al. \cite{batten1997choice} provided a closed form expression for $S_*$ which is as follows:

\begin{align}
 \label{eqn:HLLCmiddlewaveestimate}
 S_* = \frac{p_R - p_L + \rho_Lu_{nL}(S_L - u_{nL}) - \rho_Ru_{nR}(S_R - u_{nR})}{\rho_L(S_L - u_{nL}) -\rho_R(S_R - u_{nR})}.
\end{align}
The HLLC solver is also used for the computation of the non-conservative equation in the multi-species  flows to ensure consistency with the other conservative equations. The approximated velocity components at any cell interface for the approximation of source term in \textcolor{black}{Equation (\ref{eqn-source})} is as follows, for instance, the x component of the velocity is computed as:
\begin{equation}
\begin{aligned}
{{u}_{i+\half}} &=\frac{1+\operatorname{sgn}\left(S_{*}\right)}{2}\left[u_{L}+s_{-}\left(\frac{S_{L}-u_{L}}{S_{L}-S_{*}}-1\right)\right] +\frac{1-\operatorname{sgn}\left(S_{*}\right)}{2}\left[u_{R}+s_{+}\left(\frac{S_{R}-u_{R}}{S_{R}-S_{*}}-1\right)\right]
\end{aligned}
\end{equation}
where
\begin{equation}
s_{-}=\min \left(0, s_{L}\right), \quad s_{+}=\max \left(0, s_{R}\right)
\end{equation}
Similarly, the $y-$ component of the velocity can also be computed. Finally, the source term can be evaluated within a computational cell $i,j$ using,
\begin{equation}\label{eqn-sourceterm}
\begin{aligned}
\left(\alpha_{1} \nabla \cdot \mathbf{u}\right)_{i, j} &=\left(\alpha_{1}\right)_{i, j}\left[ \frac{1}{\Delta x}\left(u_{i+\frac{1}{2}, j}-u_{i-\frac{1}{2}, j}\right)+\frac{1}{\Delta y}\left(v_{i, j+\frac{1}{2}}-v_{i, j-\frac{1}{2}}\right)\right]
\end{aligned}
\end{equation}

\bibliographystyle{elsarticle-num}
\bibliography{bleep_ref}

\end{document}